\RequirePackage{fix-cm}
\documentclass[smallextended]{svjour3}       % onecolumn (second format)

\smartqed  

\usepackage{graphicx}
\usepackage{footnote}
\usepackage[multiple]{footmisc}
\usepackage{amsmath,amssymb,amsfonts}
\usepackage{algorithmic}
\usepackage{graphicx}
\usepackage{textcomp}
\usepackage{xcolor}
\usepackage[round,sort]{natbib}
\usepackage{pifont}
\usepackage[most]{tcolorbox}
\usepackage[T1]{fontenc}
\usepackage{booktabs}
\usepackage{placeins}
\usepackage{subfig}
\usepackage{enumitem}
\usepackage[flushleft]{threeparttable}
\usepackage[colorlinks=true]{hyperref}

\usepackage{multirow}

\hypersetup{
linkcolor=blue
,citecolor=blue
,filecolor=blue
,urlcolor=blue
,menucolor=blue
,runcolor=blue
,linkbordercolor=blue
,citebordercolor=teal
,filebordercolor=blue
,urlbordercolor=blue
,menubordercolor=blue
,runbordercolor=blue
}

\begin{document}

\title{Studying the Characteristics of AIOps Projects on GitHub}

\author{Roozbeh Aghili \and
        Heng Li \and
        Foutse Khomh 
}

\institute{Roozbeh Aghili, Heng Li, Foutse Khomh \at
              Department of Computer Engineering and Software Engineering \\
              Polytechnique Montreal\\
              Montreal, QC, Canada\\
              \email{\{roozbeh.aghili, heng.li, foutse.khomh\}@polymtl.ca}        
}

\date{Received: date / Accepted: date}
% The correct dates will be entered by the editor

\maketitle

\begin{abstract}
Artificial Intelligence for IT Operations (AIOps) leverages AI approaches to handle the massive amount of data generated during the operations of software systems. Prior works have proposed various AIOps solutions to support different tasks in system operations and maintenance, such as anomaly detection. In this study, we conduct an in-depth analysis of open-source AIOps projects to understand the characteristics of AIOps in practice. We first carefully identify a set of AIOps projects from GitHub and analyze their repository metrics (e.g., the used programming languages). Then, we qualitatively examine the projects to understand their input data, analysis techniques, and goals. Finally, we assess the quality of these projects using different quality metrics, such as the number of bugs. To provide context, we also sample two sets of baseline projects from GitHub: a random sample of machine learning projects and a random sample of general-purposed projects. By comparing different metrics between our identified AIOps projects and these baselines, we derive meaningful insights. Our results reveal a recent and growing interest in AIOps solutions. However, the quality metrics indicate that AIOps projects suffer from more issues than our baseline projects. We also pinpoint the most common issues in AIOps approaches and discuss potential solutions to address these challenges. Our findings offer valuable guidance to researchers and practitioners, enabling them to comprehend the current state of AIOps practices and shed light on different ways of improving AIOps' weaker aspects. To the best of our knowledge, this work marks the first attempt to characterize open-source AIOps projects.

% \PACS{PACS code1 \and PACS code2 \and more}
% \subclass{MSC code1 \and MSC code2 \and more}
\end{abstract}
\section{Introduction}
\label{intro}
With the prevalence of %expansion of the 
generated data in large-scale systems, monitoring this data and transforming it into practical insights is becoming a complex challenge. Artificial Intelligence for IT Operations (AIOps) has been introduced to cope with this challenge. It combines big data, machine learning (ML) approaches, and other advanced analysis techniques (e.g., statistical analysis)
% (e.g., \Foutse{such as?}) 
to analyze
% \Foutse{analyse or understand?}
system performance patterns to be able to
% \Foutse{with the aim to?} 
improve service quality and reduce operational costs~\citep{dang2019aiops,gartnerwebsite}. By utilizing AIOps techniques, organizations are now able to collect and combine different sources of system data and use them to perform various tasks (e.g., anomaly detection or failure prediction) in their DevOps or operations environment~\citep{dang2019aiops,gartnerwebsite}. %, root cause analysis, failure prediction, and fix the failures before actually happening.

A considerable amount of research has been performed on the topic of AIOps. % since its introduction~\citep{gartnerwebsite}.
Prior work has proposed AIOps solutions for various maintenance and operations tasks, such as predicting node failures~\citep{lin2018predicting, li2020predicting}, predicting task or job failures~\citep{gao2020task, el2017learning, rosa2015catching}, anomaly detection~\citep{he2018identifying, lim2014identifying}, and self healing~\citep{ding2014mining, ding2012healing}.
However, no work has systematically studied AIOps practices in real-world projects (e.g., public GitHub projects).

Studying AIOps practices in real-world projects is important and has several benefits, including (1) helping researchers and practitioners understand the current status of AIOps solutions and the characteristics of AIOps projects; (2) providing guidance for researchers and practitioners to adopt best-performing
% \Foutse{similar to what? do you mean 'best performing' for example?} 
AIOps solutions for their application scenarios; 
%identifying the existed gap and issues and recommend improvement opportunities; 
and (3) identifying problems in AIOps practices and shedding lights on future research opportunities.

\begin{comment}
\begin{itemize}
\item Help researchers and practitioners understand the current status of AIOps solutions.
\item Identify the existed gap and issues and recommend improvement opportunities.
\item Provide insights for future work.
\end{itemize}
\end{comment}

Therefore, this work identifies and studies a set of AIOps projects publicly available on GitHub. We also compare the selected AIOps projects with two baselines: traditional ML projects and General-purpose projects. Our goal is to understand the characteristics of these AIOps projects in the context of the baseline projects. Specifically, we first investigate the overall characteristics of these AIOps projects in terms of their GitHub metrics (RQ1), then we dig deeper into the individual projects to understand ``what'' goals these projects aim to achieve (RQ2), and finally, we examine how well these goals are achieved in terms of the code quality of these projects (RQ3). Our research questions are as follows.

% \heng{First paragraph: The context (operation data is becoming large and complex), definition, and importance of AIOps.}

% \heng{Second paragraph: Prior work has proposed AIOps solutions for ... Briefly summary of prior work.}

% \heng{Third paragraph: However, no work has studied AIOps practices in real-world projects. The benefits of studying real-world AIOps Projects: 1) help researchers and practitioners understand the current status; 2) identify gap and improvement opportunities; 3) provide insights for future work.}

% \heng{Four paragraph: Therefore, this work identifies and studies a set of AIOps projects from GitHub... Briefly summarize our work... Specifically, we work aims to answer the following three research questions (RQs).}

\begin{description}
\item[RQ1] \textbf{\textit{What are the characteristics of AIOps projects in terms of their GitHub metrics?}}
% \heng{Brief summary of approach and results}
We analyze the GitHub metrics
% \Foutse{can we be specific on this? current description is quite vague and uses different terms to refer to the same information!...repository statistics...GitHub metrics...features...it is preferable to define it clearly and use the term consistently!}
of AIOps and baseline projects
% \Foutse{what of the baselines?} 
to understand the current status of AIOps projects and also compare them with baselines in terms of their GitHub metrics, such as the programming languages and the number of stars.
% \Foutse{what is feature now?}. 
We observe that AIOps solutions are being developed with a faster growth rate compared to the baselines.
% \heng{we cannot say more AIOps projects are being developed, as the absolute increase is obviously not more for AIOps. Instead we should say ``faster growth''} 
% \Foutse{this is a github metric? or a defined one? we are VAGUE!} 
AIOps projects also have a higher distribution of popularity metrics (e.g., number of stars and forks), and also more pull requests and issues compared to baselines.
% \Foutse{more pull requests doesn't mean more challenges necessary...there are many potential confounding factors to this!}.

\item[RQ2] \textbf{\textit{What are the characteristics of AIOps projects in terms of their input data, analysis techniques, and goals?}}
% \heng{Brief summary of approach and results}
In order to understand the characteristics of AIOps projects (i.e., their input data, analysis techniques, and goals), we manually investigate each project. %We use open card sorting \citep{rugg1997sorting, spencer2009card} to derive the high-level categories for input, techniques, and goals.
% and discovering the main categories 
% \Foutse{what categories? this is vague! did we do ground theory? may be better to explain that!}
% , we identify the projects’ primary input, techniques, and goals.
We find that monitoring data (e.g., logs and performance metrics) is the most used input data, classical machine learning techniques are the most adopted analysis techniques, and anomaly detection is the primary goal of many AIOps projects.
% \Foutse{did we look if this is stable overtime? maybe some other type of data/ approaches have been picking up recently!} \Roozbeh{no we didn't.}\heng{it would be interesting to have a discussion point in RQ2 for the trend of these aspects. Not necessarily quantitative, qualitative discussion would be fine.}

\item[RQ3] \textbf{\textit{What is the code quality of AIOps projects?}}
% \heng{Brief summary of approach and results}
We further analyze the source code of AIOps and baseline projects to identify any interesting patterns related to their quality. We find that AIOps projects have a higher issue 
% \Foutse{can we be more specific? is this bugs, tech debt..etc} 
rate, specifically in terms of bugs, code smells, and technical debt, than the baselines. 

\end{description}

We share our replication package on GitHub~\footnote{\url{https://github.com/roozbehaghili/studying_aiops_github}} so that future work can reproduce or extend our study. Our work makes several important contributions:

\begin{enumerate}
    \item As the first study on AIOps practices in real-world projects, our work helps practitioners and researchers understand the status of AIOps from a practical point of view.
    \item Our qualitative analysis of the input data, analysis techniques, and goals of the AIOps projects can help practitioners and researchers consider and adopt AIOps solutions that fit into their specific application scenarios.
    \item Our work identifies problems in AIOps practices (e.g., code quality) and sheds light on future research opportunities in AIOps.
\end{enumerate}
 
The rest of the paper is organized as follows. Section~\ref{sec:experiment-setup} describes the experiment setup of our study, including the collection and preparation of the AIOps project data used for answering our research questions. Section~\ref{sec:rqs&results} presents our approach and results for answering the research questions. Section~\ref{sec:discussion} provides further discussions of our results. Section~\ref{sec:threats} discusses the threats to the validity of our findings. Section~\ref{sec:related_work} summarizes prior research related to our work, and finally, Section~\ref{sec:conclusion} concludes our paper.

\section{Experiment Setup} \label{sec:experiment-setup}
% \Foutse{add an introductory sentence!}
This section describes our approach for collecting the AIOps and baseline projects. We first present the overview of our study, then describe the steps for collecting and verifying AIOps and baseline projects, respectively.
% \heng{use projects consistently}
\subsection{\textbf{Overview of our study}}
\begin{figure*}
  \centering
  \includegraphics[width=1\textwidth]{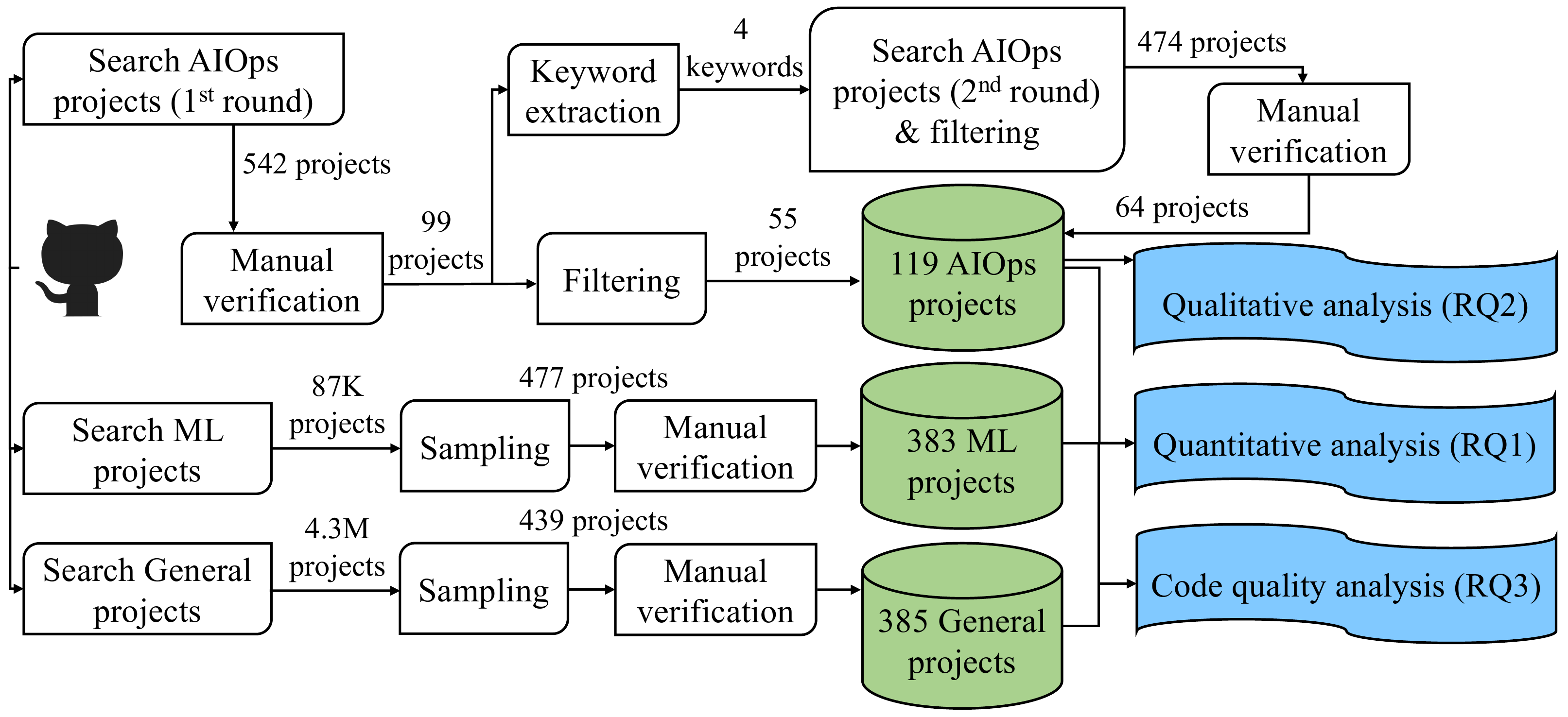}
\caption{An overview of our study.}
\label{fig:overview}       
\end{figure*}

Figure~\ref{fig:overview} presents an overview of our approach to study the characteristics of AIOps projects. We use GitHub as the main source to extract the needed data for our analysis. As of January 2023, GitHub has over 100 million registered developers and over 372 million repositories~\citep{githubb}. GitHub is also considered the largest hosting service for open-source software systems~\citep{li2020exploratory}. Hence, projects found on GitHub are likely to reflect the diversity of existing AIOps projects. Many existing studies also extract the needed information for their analysis from GitHub~\citep[e.g.,][]{openja2022technical, majidi2022empirical, dakhel2023dev2vec, foalem2023studying}.
We start by searching the projects with the keyword ``AIOps''. % and the projects under the ``AIOps'' topic\footnote{https://github.com/topics/aiops}.
Then, through manual verification (e.g., removing non-AIOps projects), keyword expansion (i.e., through pattern mining), second-round search and manual verification, and threshold-based filtering (e.g., by the number of stars), we collect a total of 119 AIOps projects that are used to answer our research questions. 
%Overall, we first study the repositories with the keyword AIOps, then perform pattern mining ~\citep{han2007frequent} to find other related keywords. After that, we filter the projects based on some of the GitHub metrics (i.e., number of stars and forks). Next, we study the repositories with the four other keywords that we have found. 
In order to better understand the characteristics of AIOps projects in a bigger context, we also compare our identified AIOps projects with two baselines:
% 1) \heng{number}
\begin{enumerate}
\item randomly sampled machine learning (ML) projects;
\item randomly sampled general projects
% \Foutse{isn't this Python only?}.\Roozbeh{no}\heng{maybe briefly explain why not limiting to Python: because we don't limit AIOps projects to a particular programming language}
\end{enumerate}
% (ML) projects, and2) \heng{number} randomly sampled general projects. 
%We also create two baselines to compare our results with them, machine learning and general baselines.
We choose the ML baseline because most of the AIOps projects leverage ML techniques. We choose the General baseline to compare our AIOps projects with general software applications on GitHub. Finally, we perform qualitative and quantitative analyses on the collected data to answer our research questions. 
Below, we describe the details of our data collection. 
The detailed approaches for answering our research questions are presented in Section~\ref{sec:rqs&results}.

%\subsection{\textbf{Studied Projects}}
\subsection{\textbf{Collecting AIOps Projects}}

Through two rounds of searching AIOps-related keywords on GitHub, we collect a total of 1016 candidate projects. Through filtering and manual verification, we end up with 119 of them as our final set of AIOps projects. The methodology we use to select these projects follows the systematic approach recommended by Basili et al.~\citeyearpar{basili1986experimentation} and is described in the following sections.

%\subsubsection{AIOps Keyword}
\subsubsection{Search AIOps projects (first round)} \label{sec:first-round-search}
In the initial step of identifying suitable projects, we utilize the GitHub interface to search for repositories specifically labeled as AIOps projects. To do this, we employ the keyword ``AIOps'' and search across four key sections of each repository: the repository name, the ``about'' section, the ``topics'' section, and the contents of the ``readme'' file. After searching for the keyword ``AIOps'', we find a total of 542 repositories that match our criteria. These 542 repositories represent all the available projects on GitHub that have been labeled as AIOps projects.

\subsubsection{Manual verification (first round)} \label{sec:manual-first-round}
Based on existing definitions of AIOps~\citep{gartnerwebsite,dang2019aiops}, we consider AIOps projects as: \textit{any project that uses IT Operations-related data, utilizes advanced analysis technologies such as machine-learning or statistical analysis to reach valuable insights or enhance the system’s quality by actions such as monitoring and automation.}

Therefore, not all the 542 discovered repositories are good AIOps candidates and suitable for our study. We hence select our subject projects based on three criteria: 
\begin{enumerate}
\item The project should be about AIOps, 
% \heng{we need a concrete definition of AIOps projects either in introduction or here}
not similar topics such as Machine Learning Operations (MLOps)\footnote{A set of practices to maintain and deploy machine learning models.} or Development and Operations (DevOps)\footnote{A set of practices that aim to shorten the system development life cycle while preserving high quality.}. Therefore, we delete the projects that are mainly about other topics. 
\item The projects should contain sufficient code. Therefore, we delete the repositories that do not have any code or only contain a few lines of code. We also delete projects that are a collection of papers, slides, or other repositories. However, we accept the projects that have created a dataset so that other developers and researchers can use it in their work.  
\item The projects should not be toy projects: we delete the projects that are homework assignments or university projects.
% \Foutse{clarify the cut off for 'toy' project!}\Roozbeh{I've clarified it in the next sentence!}\heng{made it more direct}
\end{enumerate}

To select the desirable projects based on the explained criteria, the first two authors of the paper (i.e., coders) independently perform a coding process, adding a YES (AIOps projects) or NO (non-AIOps projects) tag to each project. We perform the coding process as below.

\textbf{Step 1: Coding.} Each coder studies and analyzes all the 542 repositories and independently decides if each project should be added to the final list of projects.

\textbf{Step 2: Discussion.} The coders share their responses and discuss their approaches for selecting a project. The discussion session's goal is to reach the same understanding of the inclusion criteria among the coders.

\textbf{Step 3: Revision.} Based on the discussion, each coder revises their responses from step 1. 

\textbf{Step 4: Resolving disagreements.} In the last step, the coders discuss any conflicts that may remain and try to resolve them. If an agreement can not be reached, the third author would analyze the project and a vote is performed. % to get the final decision.

After performing the manual verification process, we obtain a total of 99 candidate AIOps projects from the 542 projects derived from the search results, which corresponds to a selection rate of 18\%. 

%\subsubsection{Pattern Mining}
\subsubsection{Keyword extraction (pattern mining)}
Expanding our search to gain a more comprehensive view of real-world AIOps projects is necessary as the term ``AIOps'' was only introduced in 2018~\citep{gartnerwebsite}. As a result, some projects might have existed prior to the introduction of this terminology, implementing AIOps solutions without explicitly using the exact term. To achieve a broader scope, we extend our search to include additional projects that may not explicitly label themselves as AIOps repositories but are, in fact, implementing AIOps solutions. To achieve this, we extract all the topics associated with each of the 99 AIOps projects resulting from the previous step. These topics can be found in the ``topics'' section of each GitHub repository. 

We then use frequent pattern mining~\citep{han2007frequent}, a method aimed at discovering associations and patterns within a given dataset. Specifically, we utilize the frequent pattern growth technique~\citep{han2004mining, han2000mining} to identify the most common topics among GitHub repositories. To conduct this analysis, we set the support parameter to 2, indicating that a pattern should appear in at least two projects for consideration. Through this approach, we identify a total of 194 patterns among the topics present in the repositories. Next, we perform a discussion session involving all three authors to decide which patterns hold potential for identifying additional AIOps projects. From this discussion, we narrow down our selection to four pairs of two-item patterns: ``anomaly detection'' and ``log analysis'', ``log analysis'' and ``machine learning'', ``anomaly detection'' and ``machine learning'', as well as ``machine learning'' and ``metrics''. All the selected keyword pairs are among the most frequently used topics in the projects. We use these four sets of keywords to find more AIOps repositories. 

\subsubsection{Search AIOps projects (second round)} 
Using the four pairs of keywords obtained in the previous step, we conduct the second round of search on the GitHub interface to identify more projects related to AIOps. We follow the same process as described in Section~\ref{sec:first-round-search}. After completing this second-round search and removing any duplicated projects that were already identified in the first-round search results, we find a total of 474 unique projects.

\subsubsection{Filtering}
Based on the knowledge that we have gained from our first-round manual verification, we apply a filtering phase to remove the toy projects. To achieve this, we employ two filtering criteria based on the number of stars and forks for each project. Specifically, we only consider projects that have both stars and forks greater than or equal to 1 (stars: >=1 \& forks: >=1). The purpose of removing toy projects is to have relatively mature projects and not soil our results with small repositories~\citep{munaiah2017curating}. Given the limited number of AIOps projects on GitHub, adding stricter filtering criteria would result in much fewer projects. Therefore, we choose a low-bar filtering approach to reduce the manual effort of analyzing all the projects with stars or forks of 0.

We utilize the filtering process for both the projects from the first and second rounds of searches. In the first round, we manually verify all the projects before applying the filtering step. This deliberate approach allows us to gain a comprehensive understanding of the status of the AIOps projects. Following the filtering process, the initially verified 99 projects are reduced to 55. For the projects from the second round, we opt to apply the filtering before the manual verification process. This decision saves unnecessary manual effort by excluding projects that do not meet the filtering criteria from the beginning.  

\subsubsection{Manual verification (second round)}
As detailed in Section~\ref{sec:manual-first-round}, not all of the repositories obtained from our initial search are suitable for our study. To carefully select the projects that align with our research objectives, we carry out a manual verification process, repeating the steps outlined in Section~\ref{sec:manual-first-round}. During this verification, the coders examine all 474 projects that resulted from the expanded keyword search to determine their suitability as real AIOps projects. After steps of separate coding, discussion, revision, and resolving disagreements, we identify 64 projects from the expanded keyword search. This corresponds to a selection rate of 14\%, 
Finally, we combine the 55 projects identified using the ``AIOps'' keyword with the 64 projects discovered through the expanded keywords.
The final set of 119 AIOps projects is used to answer our research questions. The approaches for answering our research questions are detailed in Section~\ref{sec:rqs&results}.

\subsubsection{Measuring the reliability of our manual verification
\label{sec:reliability}
}
Reliability is vital to ensure the validity of the coding results \citep{artstein2008inter}. The coding results are reliable if there exists a certain level of agreement between the coders, known as inter-coder agreement. In this study, we use Cohen’s kappa \citep{cohen1960coefficient} to measure the reliability of the agreements between two coders. Cohen’s kappa is one of the most common approaches to measuring the inter-coder agreement \citep{artstein2008inter}. Table~\ref{tab:cohen} indicates the relation between the value of Cohen’s kappa and the level of agreement \citep{mchugh2012interrater}.

\begin{table}[!ht]
    \centering
    \caption{Interpretation of Cohen’s kappa.}
    \resizebox{0.6\textwidth}{!}
    {
    \small
    \begin{tabular}{cc}
        \toprule
        Value of Cohen’s k &  Level of
        Agreement  \\

        \midrule
        0-.20 & None\\
        .21-.39 & Minimal \\
        .40-.59 & Weak \\
        .60-.79 & Moderate \\
        .80-.90 & Strong \\
        .90-1 & Almost Perfect 
    \end{tabular}
    }
    \label{tab:cohen}
\end{table}

Our manual verification for choosing the proper AIOps projects achieves a Cohen’s kappa of 0.68 before the discussion session. After the discussion session between the coders, the kappa score increased to 0.84. As shown in Table~\ref{tab:cohen}, kappa $\geq$ 0.80 indicates a strong agreement.

%\subsection{\textbf{Baselines}}
\subsection{\textbf{Collecting Baseline Projects}}
\label{sec:Baselines}

To understand how AIOps projects differ from traditional software projects, %only reporting the AIOps results is not sufficient. Therefore, 
we create two baselines and compare the AIOps projects with them. We select Machine Learning (ML) projects as our first baseline and General projects as our second baseline. %\\

\subsubsection{Machine Learning projects}
For our initial baseline, we choose ML projects. We select ML for the first baseline because AIOps can be considered as an application domain of machine learning.
In similar studies, researchers typically search for specific frameworks to gather ML projects on GitHub. For example, \citep{zhang2018empirical} select their project set by searching the keyword ``TensorFlow'', while \citep{islam2019comprehensive} gather their projects using various keywords such as ``TensorFlow'' and ``Keras''. However, as our focus is not limited to any specific framework, we use more general keywords. To gather our set of ML baseline projects, we utilize two keywords: ``machine learning'' and ``deep learning.'' This approach allows us to cast a broader net and capture a comprehensive range of relevant ML projects for comparison and analysis.

To compare the AIOps projects with the baselines in a similar context, we apply a similar filtering process to the baseline projects. Like the AIOps projects, we only extract ML projects that have at least one star and one fork. Additionally, we take into account that ML projects generally have a longer history than AIOps projects. To address this difference, we apply the same date range for the creation of the ML baseline projects as observed in the AIOps set. Specifically, the earliest and latest creation dates in the AIOps projects are 2012/12/25 and 2022/10/27, respectively. Thus, we apply the same date range to filter the ML baseline projects.
Following the search and removal of duplicate projects in two queries, we obtain a total of 87,276 unique repositories for the ML baseline. 
Due to time and computational resource limitations, extracting and analyzing all these projects becomes impractical. Hence, following prior work (e.g., Chen et al.~\citeyearpar{chen2020comprehensive} and Zhang et al~\citeyearpar{zhang2019empirical}), a sample of 383 projects is needed to represent the pool of 87,275 ML repositories with a confidence level of 95\% and a confidence interval of 5\%.  %\\

\subsubsection{General projects}
As our second baseline, we choose general projects from GitHub, meaning that we do not focus on any particular topic in our search. We also do not limit the General baseline to a specific programming language (e.g., Python) since AIOps projects are not limited to a single language as well. 
In this way, the General baseline captures the general characteristics of all GitHub projects. We then apply the same filtering phase as we did for the ML baseline, but without indicating any specific topic.
After completing the filtering, we obtain 4,358,342 public and available repositories for the General baseline. Similar to the ML baseline, a sample of 385 projects is required to represent the pool of 4,358,342 General repositories with a confidence level of 95\% and confidence interval of 5\%.

\subsubsection{Manual verification}
In order to maintain consistency with our methodology for selecting AIOps projects, we apply manual verification to our two baseline sets as well. The selection of baseline projects is based on two key criteria:

\begin{enumerate}
\item The projects should contain sufficient code and be mature. Therefore, we delete the repositories that do not have code and are a collection of papers, slides, or other repositories.
\item The projects should not be toy projects: we delete the projects that are homework assignments or university projects.
\end{enumerate}

To construct our baseline sets, we randomly extract 500 projects from both ML and General pool of repositories. We then divide each set of baseline projects into two parts of 20\% and 80\% portions. Similar to Section~\ref{sec:manual-first-round}, the two coders independently label 20\% of each baseline to ensure a reliable assessment. We then measure the reliability of our coding using Cohen’s kappa~\citep{cohen1960coefficient}. The results of our manual verification for choosing the ML and General baselines indicate a strong agreement between the coders, with Cohen's kappa scores of 0.81 and 0.91, respectively. As the measurement shows a strong agreement between coders, the rest of baseline projects (the 80\% portions) are labeled by the first coder. 

We continue our manual labeling process until we obtain the statistical representative set for each baseline. In the case of ML projects, we thoroughly analyze 477 projects, ultimately selecting 383 of them to create our final set of ML baseline. Similarly, for the General projects, we thoroughly analyze 439 projects, resulting in the selection of 385 projects as our final set of General baseline.

\bigskip
As shown in Figure~\ref{fig:overview}, we compare our AIOps projects with the baseline projects in RQ1 where we study the repositories statistics of the projects, as well as in RQ3 where we study the code quality of the projects. In RQ2, we perform a qualitative study for the AIOps projects only, since the research question is specifically about AIOps projects (i.e., the input data, analysis techniques, and goals of AIOps solutions).

\section{Research Questions and Results} \label{sec:rqs&results}
This section presents the details of our research questions (RQs) and the results. We organize each RQ by its motivation, approach, and results. % of each research question.

%\subsection{\textbf{What are the characteristics of AIOps projects?}}

\subsection{\textbf{RQ1. What are the characteristics of AIOps projects in terms of their GitHub metrics?}}

\subsubsection{\textbf{Motivation}}
Prior studies proposed AIOps solutions that leverage AI technologies to support various software operation efforts~\citep{li2022cdx, li2020predicting, di2021prometheus}. 
However, no work has investigated real-world AIOps projects and their characteristics. 
Thus, this RQ bridges the gap to study the characteristics of AIOps projects and compare them with the baseline projects in terms of their GitHub metrics. With this comparison, we can understand the similar and different patterns between the characteristics of AIOps projects and the baselines.
Our findings can help AIOps researchers and practitioners understand the state of AIOps in practice.

\subsubsection{\textbf{Approach}}

% \heng{Try to organize each RQ approach using small titles like in this RQ.}

In this RQ, we analyze the repository characteristics of AIOps projects and compare them with the baseline projects. 
We use 
%different GitHub metrics to obtain the repository statistics of these projects. 
%These metrics are the number of stars, forks, commits, contributors, releases, pull requests, issues, size, creation date, last updated date, the status of being archived, and language. We write a script to extract the metrics of both AIOps and baseline projects. We use 
GitHub REST API~\citep{GitHubSearchAPI:2022} to retrieve the repository characteristics of these projects. %the data. 
In particular, we analyze the characteristics of AIOps projects and the baseline projects from three perspectives: growth of repositories, programming languages, and repository metrics.

\begin{figure*}
\centering
\includegraphics[width=0.75\textwidth]{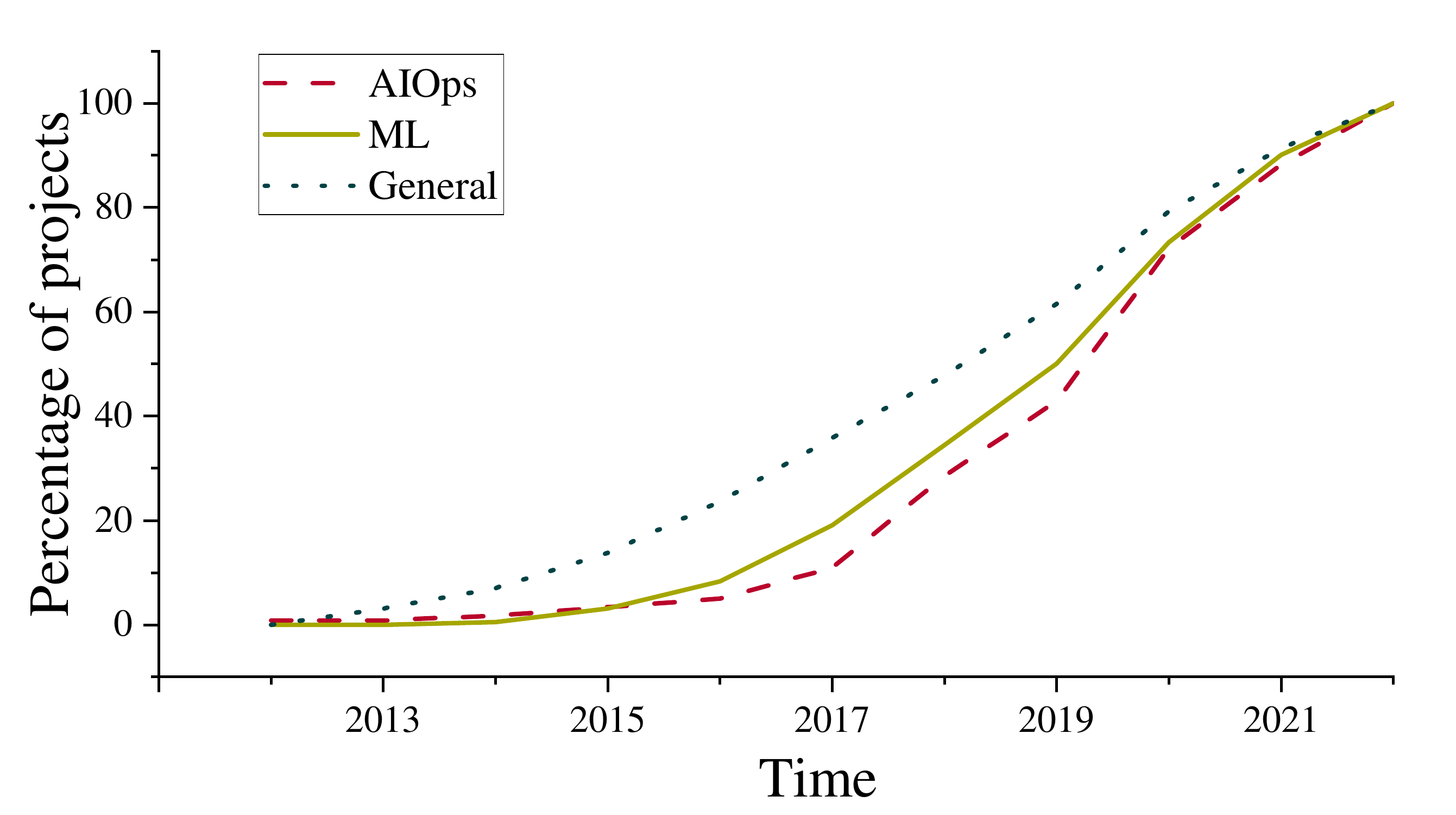}
\caption{The cumulative distribution of the creation time of AIOps and baseline projects.}
\label{fig:creation-date}    
\end{figure*}

\begin{table}[!t]
    \centering
    \caption{The top-5 languages of AIOps and baseline projects.}
    \resizebox{1\textwidth}{!}
    {
    \small
    \begin{tabular}{ll|ll|ll}
        \toprule
        \multicolumn{2}{c|}{\textbf{AIOps}} &  \multicolumn{2}{c|}{\textbf{ML}} &
        \multicolumn{2}{c}{\textbf{General}} \\
        Language & Usage $(\%)$ & 
        Language & Usage $(\%)$ & 
        Language & Usage $(\%)$ \\

        \midrule
        Python & 71.4 & Python & 81.7 & Python & 21.6\\
        Java & 10.1 & HTML & 2.6 & JavaScript & 16.4 \\
        Go & 3.4 & R & 1.8 & Java & 8.1 \\
        HTML & 2.5 & C++ & 1.8 & TypeScript & 4.7 \\
        JavaScript & 1.7 & JavaScript & 1.8 & PHP & 4.4 
    \end{tabular}
    }
    \label{tab:languages}
\end{table}

\noindent\textbf{Growth of repositories.}
To understand the evolution of the population of AIOps projects, we analyze the distribution of the AIOps projects based on their creation time. We also compare the creation time distribution with that of the baseline projects. 
% Please recall that (see Section~\ref{sec:experiment-setup}), for the baseline projects, we only consider the ones created in the same time periods as the AIOps projects.

\noindent\textbf{Programming languages.}
Developers may use different programming languages for AIOps projects. Understanding the distribution of the programming languages can provide insights for future work to support AIOps project development. Each project may use multiple programming languages. In this work, we extract and present the primary language of each project.

\noindent\textbf{Repository metrics.}
We study the repository metrics of the AIOps projects, including the number of stars, forks, commits, contributors, releases, pull requests, issues, size, 
% age, creation date, last updated date, 
and the status of being archived. %, and languages
% \heng{make sure we only keep the ones that are used in the results}.  
%We write a script to extract the metrics of both AIOps and baseline projects.
% Most of the metrics are directly retrieved using the GitHub REST API. 
For the pull requests, we sum the open and closed pull requests. % and show them as pull requests. 
Similarly, for the issues, we sum the open and closed issues. % and represent them as issues.  
% We also define the age of a project as the difference between the last update date and the creation date. We use this metric to qualify the maturity of the repositories.

\noindent\textbf{Statistical tests.}\label{sec:statistics}
We further perform statistical tests to evaluate the statistically significant difference between metrics for AIOps and baseline projects. We first conduct the \textit{Shapiro-Wilk} test~\citep{shapiro1965analysis} to test the normality of our metrics. Using the widely accepted 0.05 significance threshold, the Shapiro-Wilk test shows that the GitHub metrics of AIOps and baseline projects do not follow a normal distribution. Since our data is not normally distributed, we select nonparametric tests. We use \textit{Mann-Whitney U} test~\citep{mann1947test} to compare our samples. We also use \textit{Cliff’s delta} test~\citep{cliff1993dominance} to test the effect size between our samples. Regarding Mann-Whiteny U test, we use the significance threshold of 0.05. Regarding Cliff’s delta test, we use the scale presented by Romano et al.~\citeyearpar{romano2006exploring}, that effect of |d| = 0.147 is small, |d| = 0.33 is medium, and |d| = 0.474 is large.

\begin{table}
\centering
\caption{Detailed results of \textit{Mann–Whitney U} and \textit{Cliff's delta} tests on projects' GitHub metrics.}
\begin{threeparttable}
   \begin{tabular}{l|ll|ll}
\hline
  \multirow{2}{*}{Metric} & \multicolumn{2}{c|}{\textbf{AIOps vs. ML}} & \multicolumn{2}{c}{\textbf{AIOps vs. General}} \\ 
  \cline{2-5}
  & \multicolumn{1}{l|}{p-value} & effect size & \multicolumn{1}{l|}{p-value} & effect size \\ \hline
Stars & \multicolumn{1}{l|}{0.00} & ** & \multicolumn{1}{l|}{0.00} & *** \\
Forks & \multicolumn{1}{l|}{0.00} & ** & \multicolumn{1}{l|}{0.00} & *** \\
Commits & \multicolumn{1}{l|}{0.34} & - & \multicolumn{1}{l|}{0.92} & - \\
Contributors & \multicolumn{1}{l|}{0.00} & ** & \multicolumn{1}{l|}{0.02} & * \\
Releases & \multicolumn{1}{l|}{0.04} & * & \multicolumn{1}{l|}{0.87} & - \\
Pull requests & \multicolumn{1}{l|}{0.00} & ** & \multicolumn{1}{l|}{0.59} & - \\
Issues & \multicolumn{1}{l|}{0.00} & ** & \multicolumn{1}{l|}{0.01} & * \\
Size & \multicolumn{1}{l|}{0.02} & * & \multicolumn{1}{l|}{0.00} & *** \\ \midrule
\end{tabular}
    \begin{tablenotes}
  \small
  \item \textit{Mann–Whitney U} results are shown in \textit{p-value} columns. If the sets have statistically different distributions, the \textit{Cliff's delta} results are shown in \textit{effect size} columns.

  *: negligible effect
  **: small effect
  ***: medium effect
     \end{tablenotes}
\end{threeparttable}
\label{tab:github_metrics_statistics}
\end{table}

\subsubsection{\textbf{Results}}
\label{rq1_results}
% \heng{Organize all the results like the first paragraph: first use a bold sentence to present the key message, then discuss the results to support the key message. The first RQ can have three key messages based on the three results.}

\noindent\textbf{Compared to the ML and General baselines, AIOps projects are relatively new and exhibit rapid growth in recent years.}
Figure~\ref{fig:creation-date} represents the percentage of projects created in and before each year for the AIOps and baseline projects. %based on the creation date of each repository. 
As mentioned in Section \ref{sec:Baselines}, the creation date of all the projects is between 2012/12/25 and 2022/10/27. 
As shown in Figure~\ref{fig:creation-date}, in the first few years (from 2012 to 2017), the number of AIOps projects is very small. In recent years (from 2017 to 2022), the AIOps projects experience a faster growth compared to the ML and General baselines. As can be seen, the ML projects also exhibit a faster growth than the General baseline projects.
%This figure indicates the speed of growth in ML projects (especially in the last 5 years) is much more than the general ones, and the AIOps repositories even evolve faster.

\begin{figure*}[]
  \centering
  \subfloat[]{\includegraphics[width=0.45\textwidth]{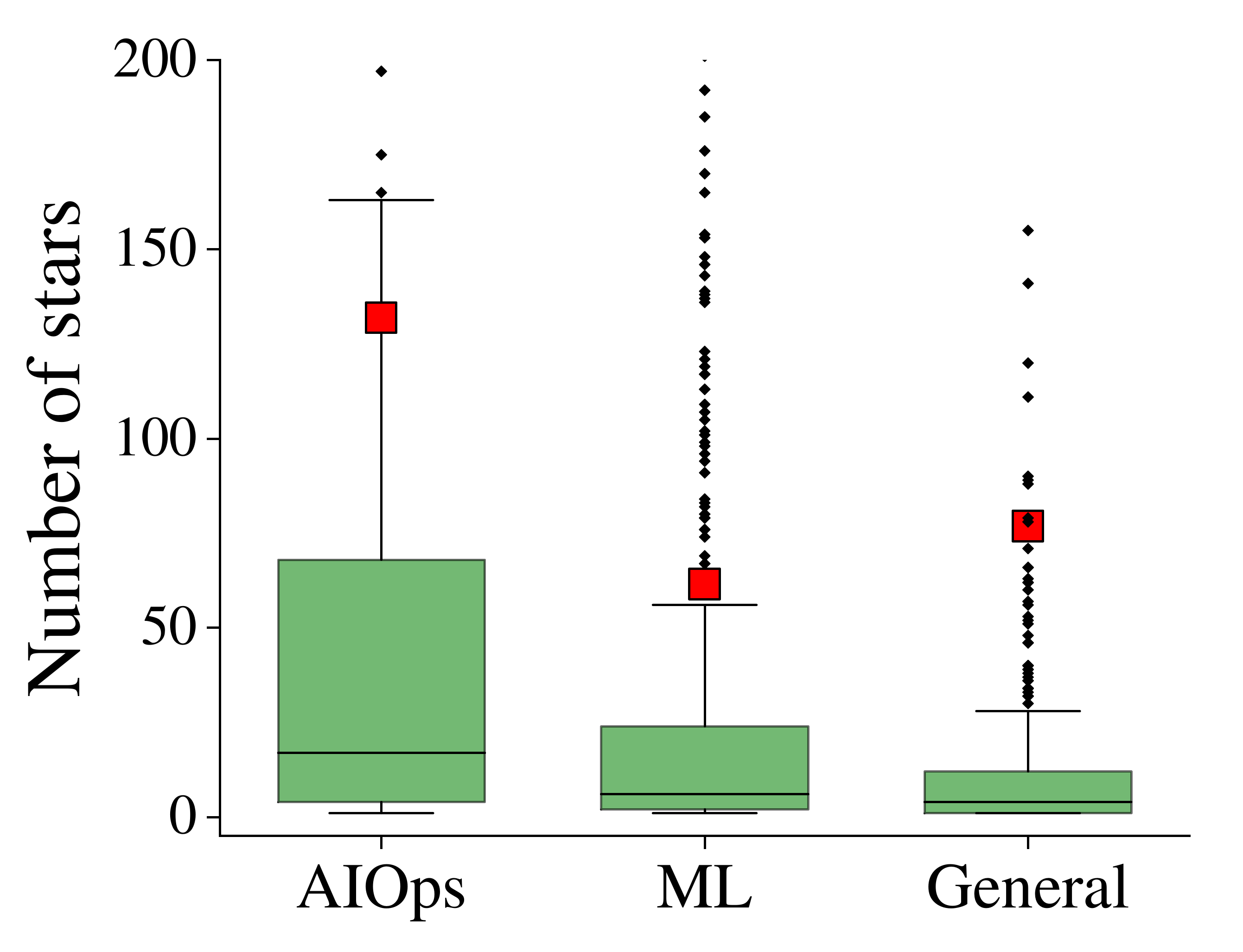}\label{fig:stars}}
  \subfloat[]{\includegraphics[width=0.45\textwidth]{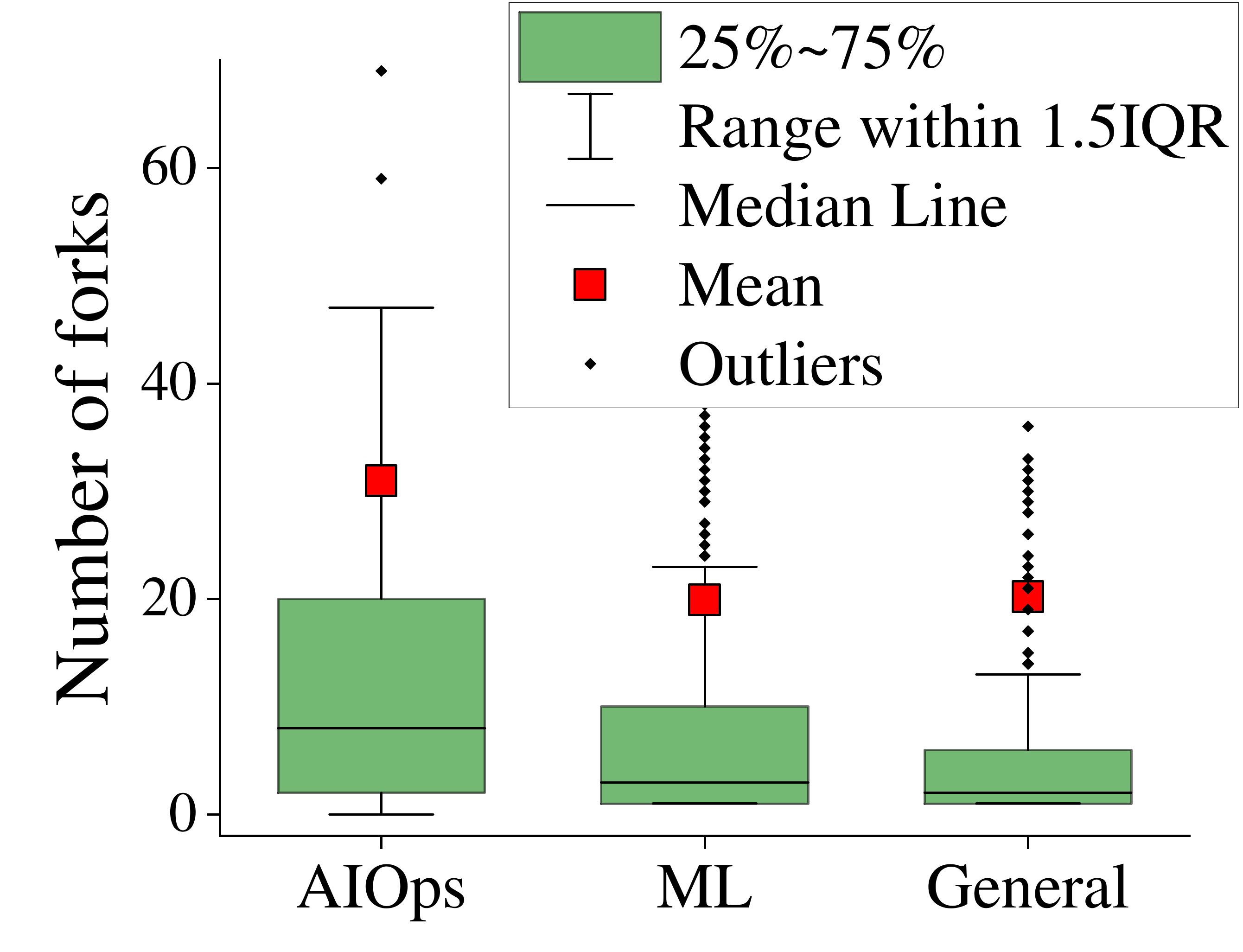}\label{fig:forks}} \\
  \subfloat[]{\includegraphics[width=0.45\textwidth]{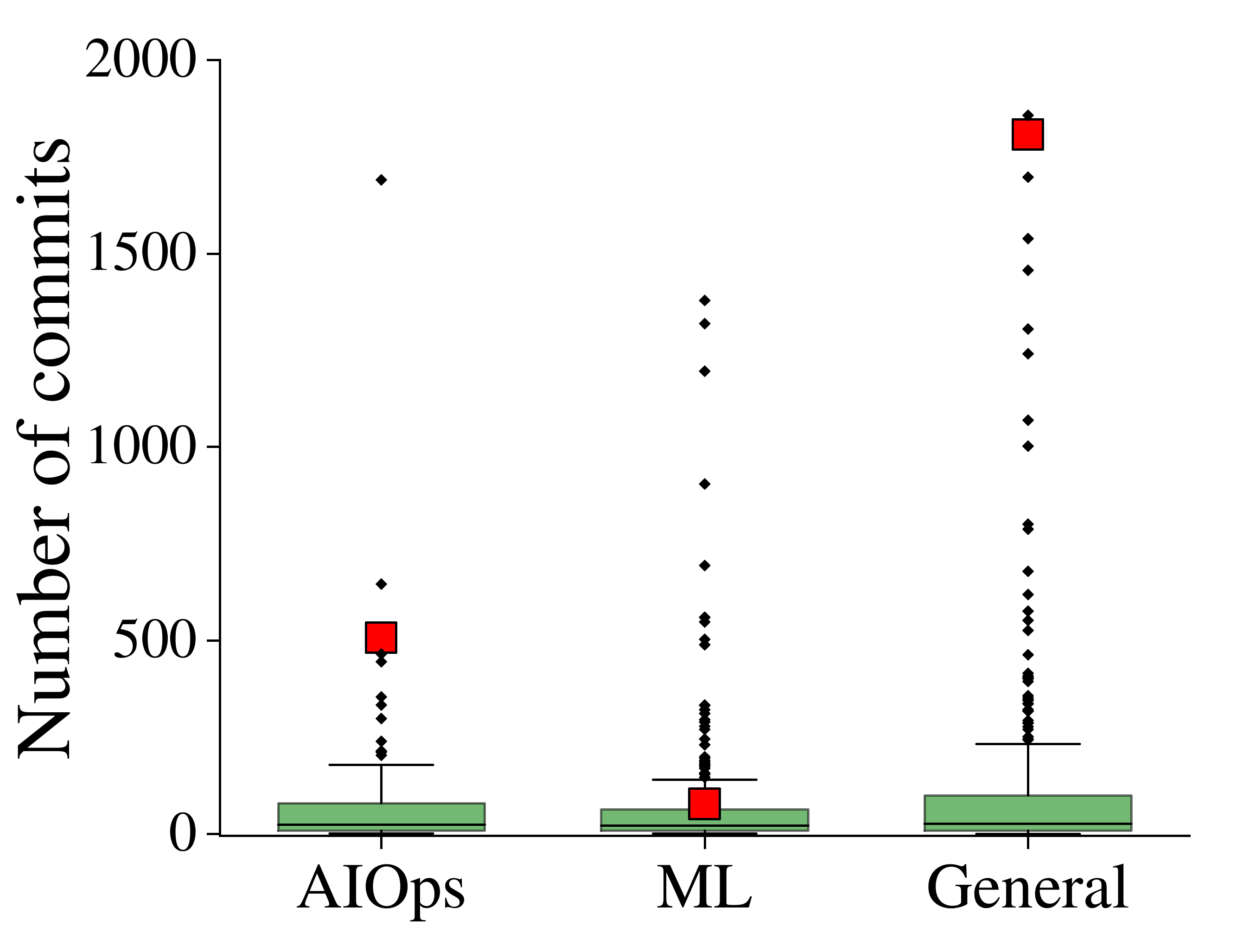}\label{fig:commits}}
  \subfloat[]{\includegraphics[width=0.45\textwidth]{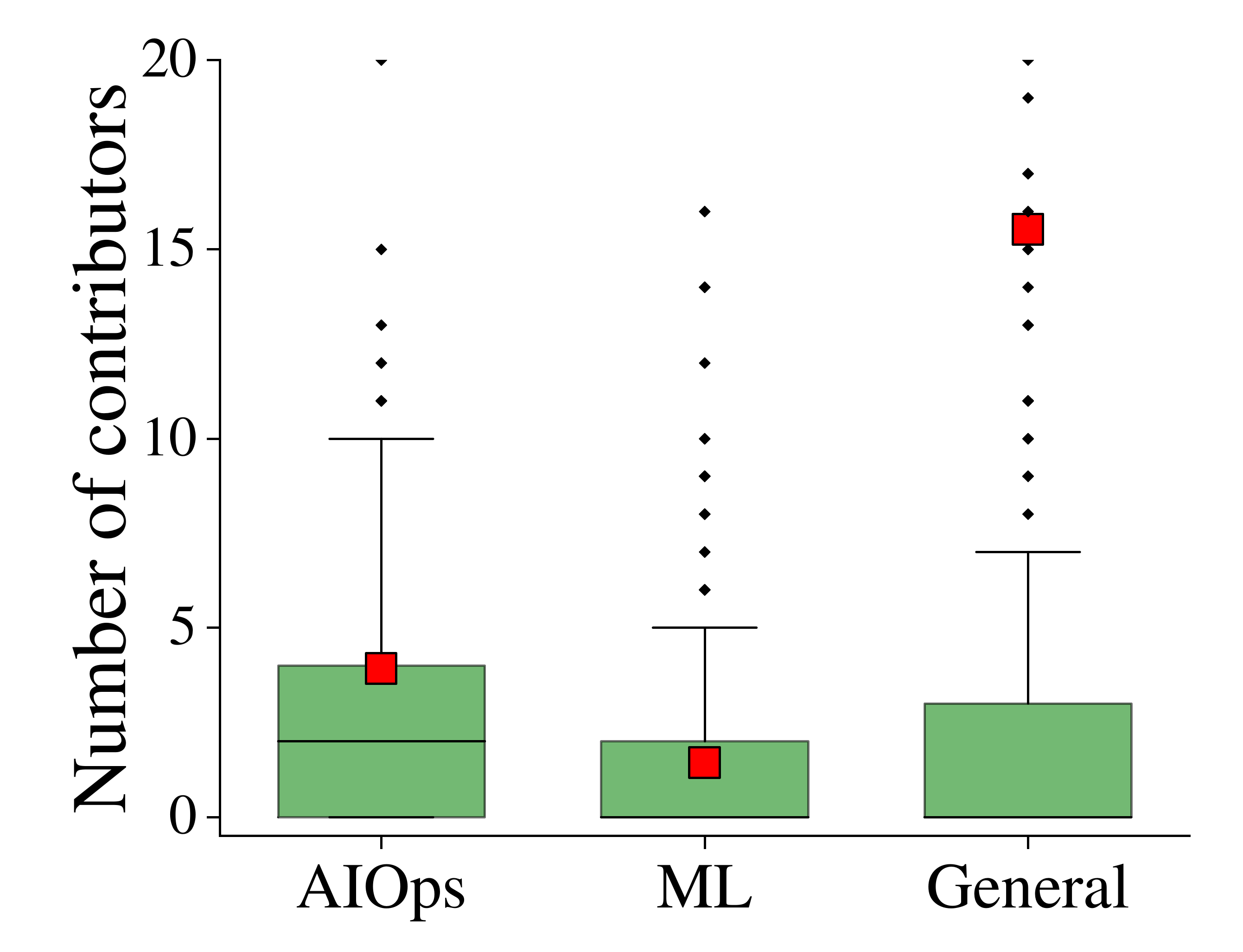}\label{fig:contributors}} \\
  \subfloat[]{\includegraphics[width=0.45\textwidth]{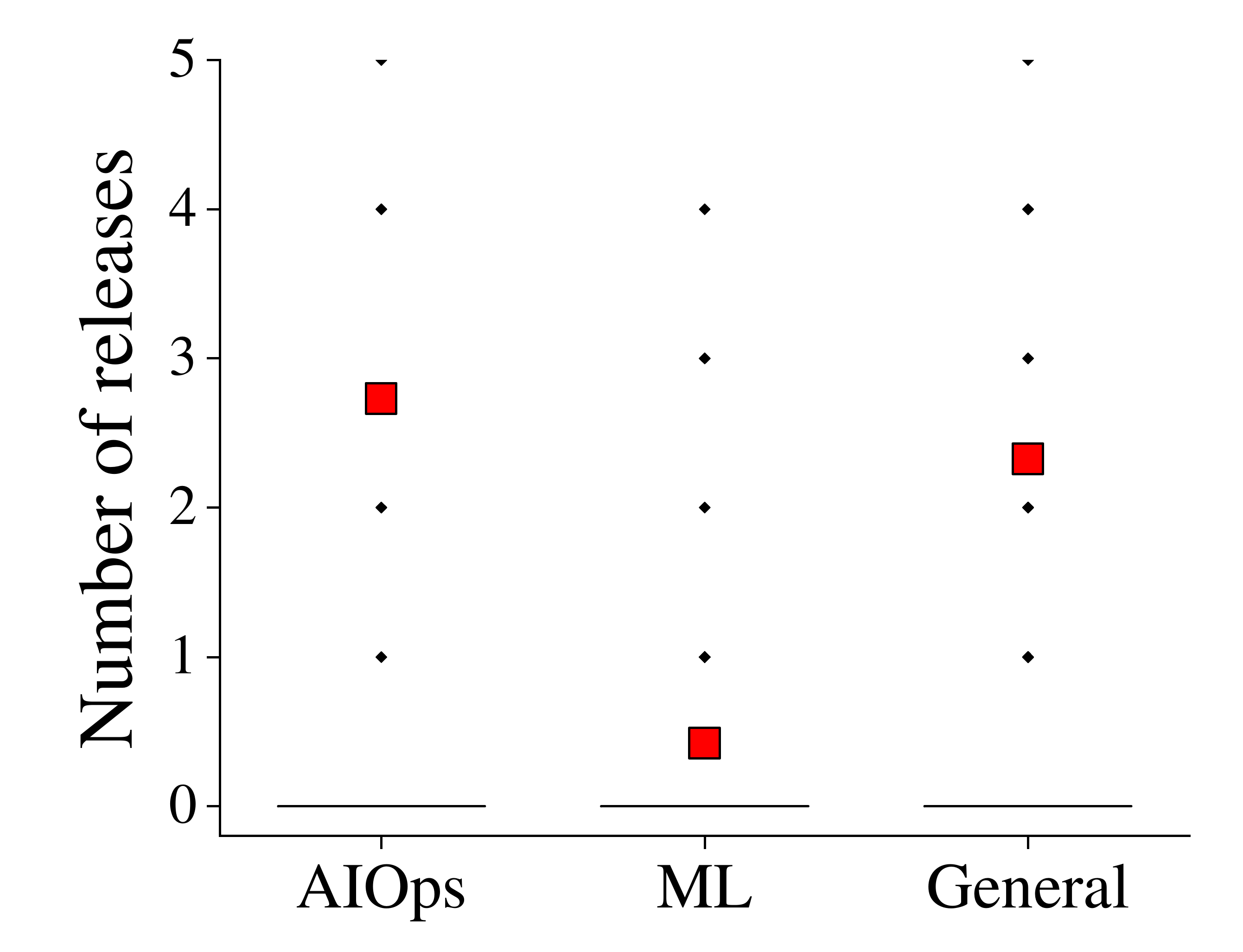}\label{fig:releases}}
  \subfloat[]{\includegraphics[width=0.45\textwidth]{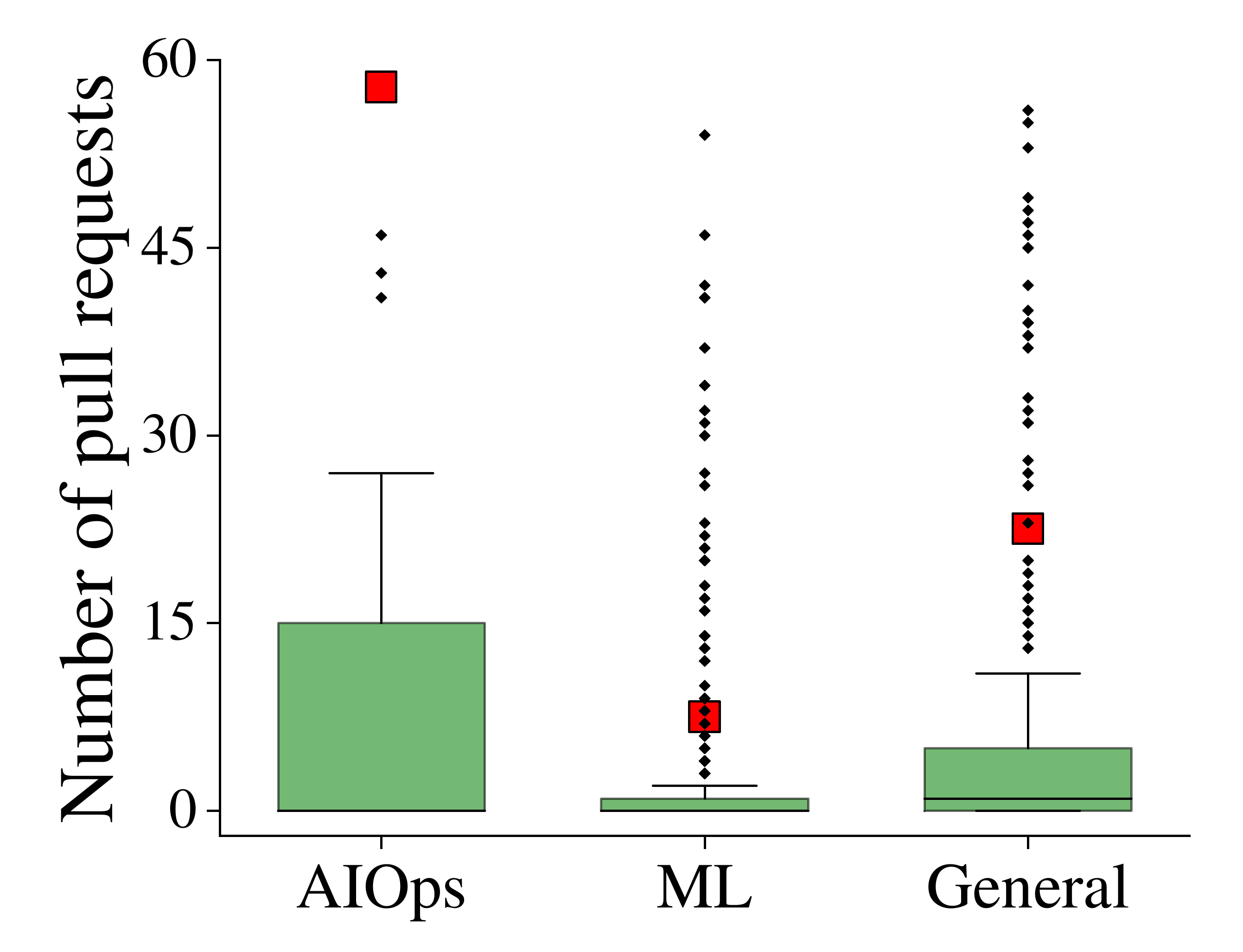}\label{fig:pull_requests}} \\
  \subfloat[]{\includegraphics[width=0.45\textwidth]{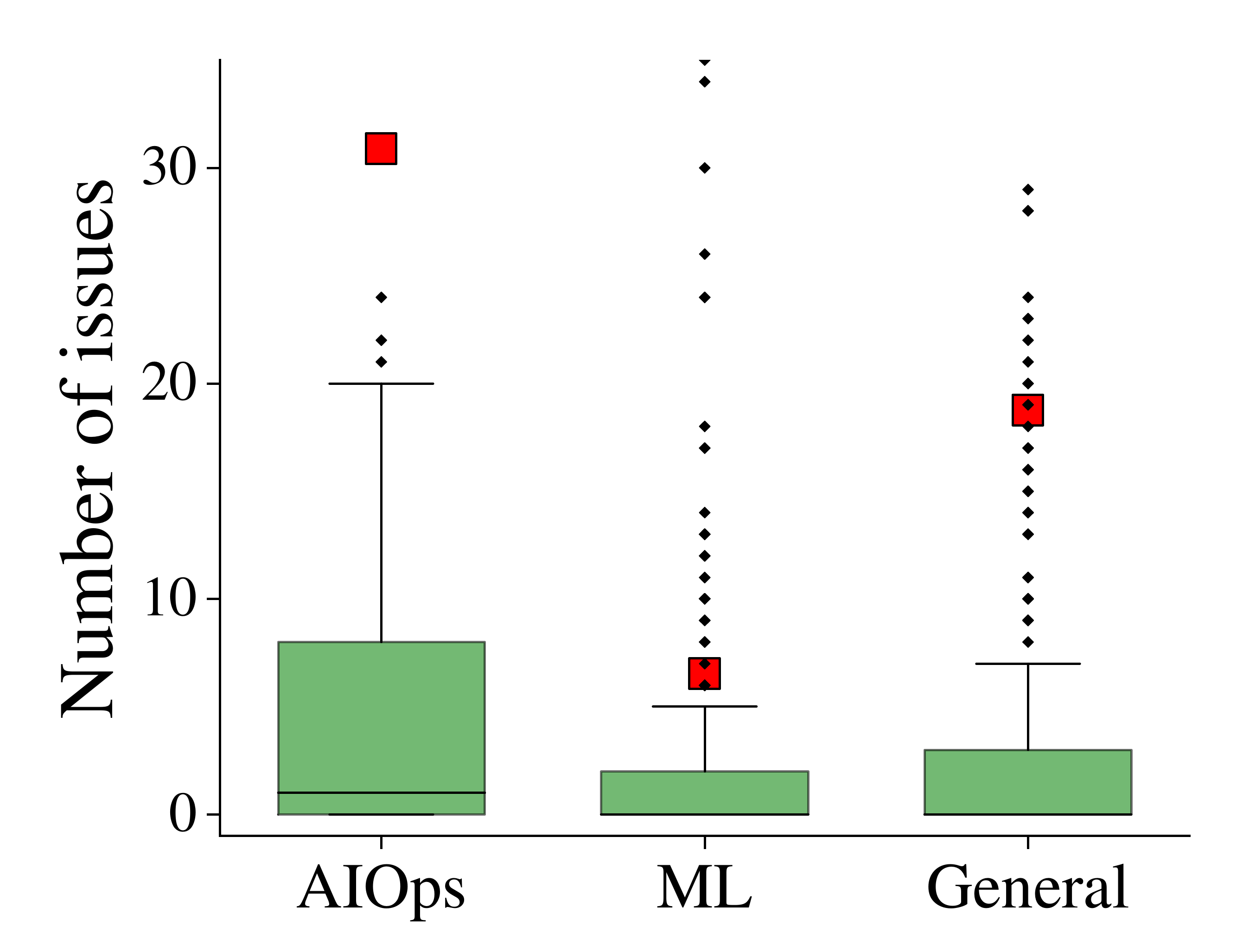}\label{fig:issues}}
  \subfloat[]{\includegraphics[width=0.45\textwidth]{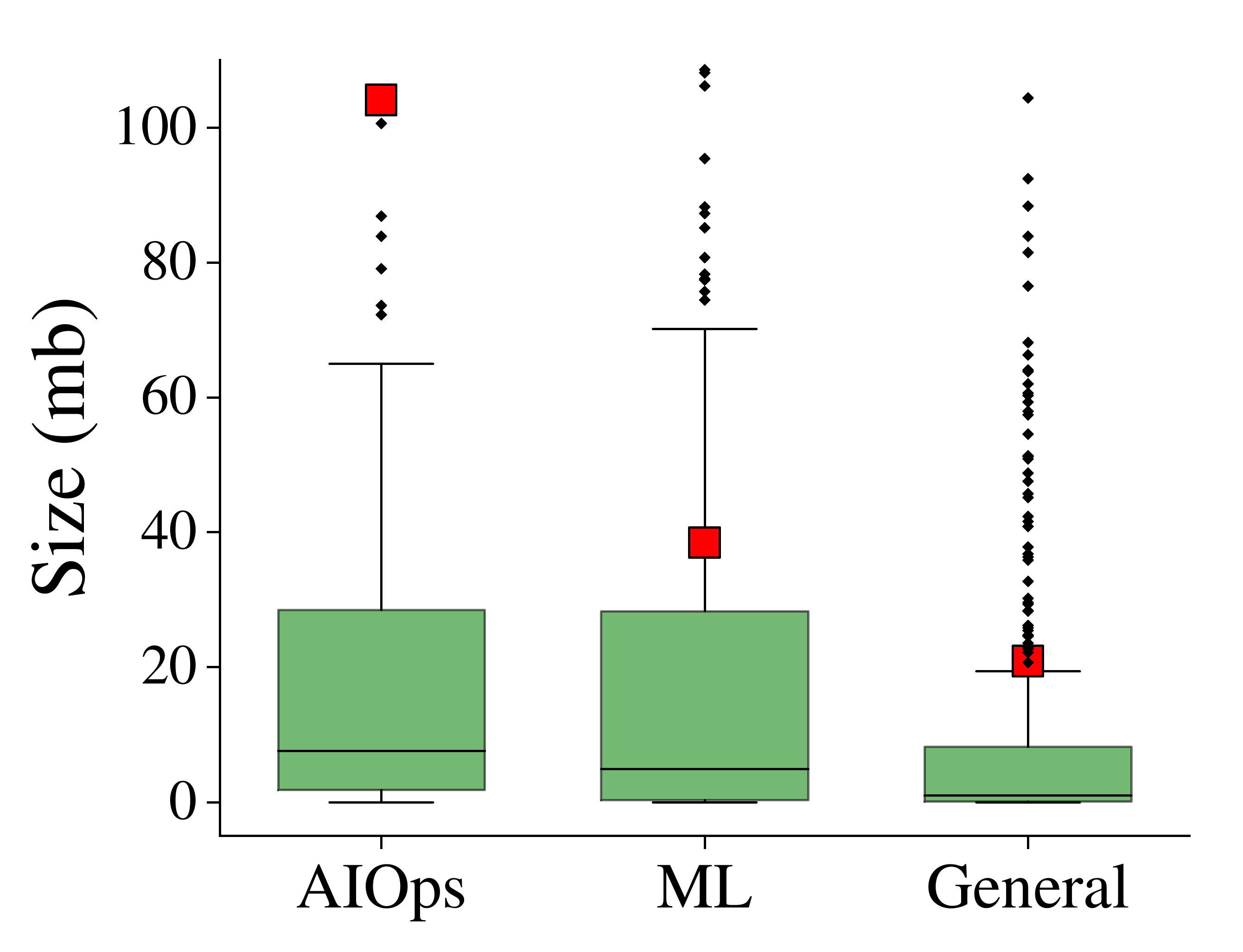}\label{fig:size}}
  \caption{Box plots of GitHub metrics for AIOps and baseline projects. The median number of 0 for the releases, pull requests, and issues indicate that more than half of the projects do not have any release, pull request, or issue.}
\label{fig:github_box_plot}
 \vspace{-1em}
\end{figure*}

\noindent\textbf{Like in ML projects, Python is the dominant programming language in the AIOps projects; however, unlike in ML projects, Java is also a major programming language used in AIOps projects.}
% \heng{this is the second key message. Add details of the results to support the key message.}
As shown in Table \ref{tab:languages}, Python is the most used language among all the groups; however, the usage of Python is much higher in AIOps and ML repositories than in the General baseline (71.4\% and 81.7\% in contrast to 21.6\%). Another interesting finding is the relatively high usage of Java in AIOps projects (10.1\%), while Java is not among the top-5 popular languages in the ML baseline.

% \begin{figure*}[h!]
%   \centering
%   \subfloat[]{\includegraphics[width=0.24\textwidth]{figures/figure_files/stars.pdf}\label{fig:stars}}
%   \subfloat[]{\includegraphics[width=0.24\textwidth]{figures/figure_files/forks.pdf}\label{fig:forks}}
%   \subfloat[]{\includegraphics[width=0.24\textwidth]{figures/figure_files/commits.pdf}\label{fig:commits}}
%   \subfloat[]{\includegraphics[width=0.24\textwidth]{figures/figure_files/contributors.pdf}\label{fig:contributors}} \\
%   \subfloat[]{\includegraphics[width=0.24\textwidth]{figures/figure_files/pull_requests.pdf}\label{fig:pull_requests}}
%   \subfloat[]{\includegraphics[width=0.24\textwidth]{figures/figure_files/issues.pdf}\label{fig:issues}}
%   \subfloat[]{\includegraphics[width=0.24\textwidth]{figures/figure_files/size.pdf}\label{fig:size}}
%   \subfloat[]{\includegraphics[width=0.24\textwidth]{figures/figure_files/age.pdf}\label{fig:age}}
%   \caption{Distribution of GitHub metrics: (a) number of stars, (b) number of forks, (c) number of commits, (d) number of contributors, (e) number of pull requests, (f) number of issues, (g) the size of projects, (h) the age of projects. The range is changed duo to the amount of outliers. 
%   }
%   \label{fig:github_box_plot}
% \end{figure*}

\noindent\textbf{On average, AIOps projects are more popular and active than the baselines.}
 Figure~\ref{fig:github_box_plot} represents the box plots of GitHub metrics for the AIOps projects and the baselines. Table~\ref{tab:github_metrics_statistics} shows the p-value and effect size of the GitHub metrics of AIOps projects compared to the baselines.

Considering the number of stars and forks, AIOps projects are more popular than the baselines, as both median and mean values in AIOps projects are higher than the baselines. In terms of the  median value, AIOps projects have 4 times more stars and forks than the General baseline (median values of stars in AIOps and General projects are 17 and 4, and median values of forks in AIOps and General projects are 8 and 2). Regarding the statistical test results in Table~\ref{tab:github_metrics_statistics}, the p-value of stars and forks indicate that AIOps projects have statistically different distributions from ML and General baselines. For both of the metrics, the effect size is small compared to the ML baseline and is medium compared to the General baseline.

Regarding the number of commits, as shown in the statistical test results in Table~\ref{tab:github_metrics_statistics}, there is not a statistically significant difference between AIOps and baselines. Regarding the number of contributors, it seems that AIOps projects are more collaborative (median of 2 in AIOps and 0 in baselines). Statistical results also corroborate this finding, with the effect size of small compared to ML baseline and negligible (with a statistically significant difference) compared to General baseline. Regarding the number of releases, there is not a significant difference between AIOps and General baseline; however, AIOps projects have a statistically significant difference compared to ML baseline.

Comparing the number of pull requests and issues, as shown in Figures~\ref{fig:pull_requests} and~\ref{fig:issues}, we notice that AIOps projects experience more pull requests and issues. Statistical results also confirm this finding, as p-values of pull requests and issues are less than 0.05 (except pull requests for General baseline). This may be explained in three ways. First, the AIOps projects are more active and popular (i.e., more developers proactively develop them), leading to a more significant number of pull requests and issues. The second interpretation might be that AIOps projects are in the first stages of formation and not mature enough, having more defects and flaws and more developers trying to fix these problems. The third explanation is that AIOps projects are on average larger than the two baseline projects (as shown in Figure~\ref{fig:size}), which may lead to more pull requests and issues.

Comparing the size of the projects, AIOps projects tend to be larger than General baseline as the statistical tests indicate a medium effect size. AIOps projects also are larger than ML projects; however, the difference is not as large as General baseline, as the statistical tests result in a negligible (with a statistically significant difference) effect size. Comparing the median values, the size of AIOps projects is 7 times larger than General baseline (7.6 MB for AIOps projects compared to 1.0 MB for General baseline).

We further analyze the percentage of projects that have been archived. The results indicate that only 1.7\% of AIOps projects have been archived, while this amount for ML and General baseline is 1.8\% and 4.2\%, respectively.

Overall, taking into consideration all the mentioned metrics, AIOps projects seem to be more active and popular than both baselines.

\newenvironment{mybox}[1]{%
    \begin{tcolorbox}[title={Summary of RQ1}]%
    }{ 
    \end{tcolorbox}
}
\begin{mybox}{}
On average, AIOps projects are receiving more attention than the ML and General baselines. The primary language used in them is Python, followed by Java. They are growing faster than the baselines in recent years, demonstrating the growing needs and active practices in this area.  
The size of AIOps projects is larger than the baselines, and focusing on other GitHub metrics such as number of stars, forks, and releases, AIOps projects seem to be more popular and active than both baselines. 
\end{mybox}
%\subsection{\textbf{RQ2. What are the input data characteristics, analysis techniques, and analysis goals in AIOps projects?}}
\subsection{\textbf{RQ2. What are the characteristics of AIOps projects in terms of their input data, analysis techniques, and goals?}}
\label{sec:rq2}

\subsubsection{\textbf{Motivation}}
AIOps researchers and practitioners leverage different techniques to analyze different types of operational data and achieve different goals. However, it is unclear how real AIOps projects leverage data and technologies to achieve the goals. 
In this RQ, we qualitatively analyze our set of AIOps projects to understand the characteristics of these projects' input data, analysis techniques, and goals.
Our results can help researchers and practitioners further understand the status of AIOps practices and the characteristics of AIOps projects. %, understand the current trends, and identify future directions to fill the missing areas.
Our results can also provide insights for future work to provide support for different AIOps application scenarios. 

\subsubsection{\textbf{Approach}}
\begin{figure*}
\centering
\includegraphics[width=0.7\textwidth]{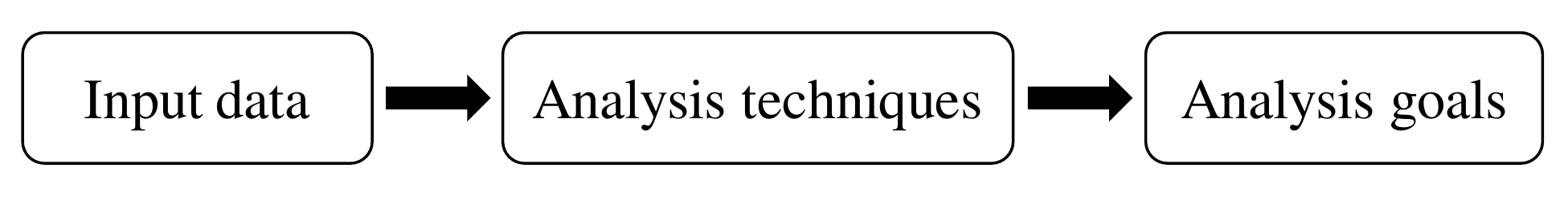}
\caption{The three key concepts of our manual analysis.}
\label{fig:flow}    
\end{figure*}

%In order to categorize the , 
We manually examine each AIOps project to understand its input data, analysis techniques, and goals.
For each project, we manually investigate four sources of information; the ``about'' section, the ``readme'' file, %code comments, 
the source code, and the additional documentations if available.
%Not all projects have all four sources of information; however, using a combination of the available information, we were able to extract the needed information to answer this research question.
Figure~\ref{fig:flow} illustrates the three key concepts of our manual analysis (input data, analysis techniques, and goals) and their relationship.

%Below, we first define the three key concepts of our manual analysis: the input data, the analysis techniques, and the goals of the AIOps projects. Then, we describe the details of our coding process.

\begin{itemize}
    \item \textbf{Input data:}
    The types of data (e.g., log data) that an AIOps project takes as inputs to achieve its objectives.
    
    \item \textbf{Analysis techniques:}
    The main analysis techniques (e.g., machine learning techniques) that an AIOps project adopts to analyze the input data and achieve its objectives.
    
    \item \textbf{Goals:}
    The objectives (e.g., anomaly detection) that an AIOps project aims to achieve through its input data and analysis techniques.
\end{itemize}

\noindent\textbf{Manual coding process.}
We use open coding approach~\citep{khandkar2009open} 
% \Roozbeh{maybe we should use Grounded theory instead of open coding?}\heng{open coding is fine} 
to extract the information related to the three key concepts shown in Figure~\ref{fig:flow}. %  the extracted data. 
Open coding is widely used among software engineering researchers to conclude a high-level abstraction from lower-level data~\citep{wohlin2015towards, stol2016grounded}. 
To label the projects, the first two authors of the paper (i.e., coders) jointly perform a coding process, determining each project's input data, analysis techniques, and goals. We perform a five-step coding process as follows.

\begin{figure*}
\centering
\includegraphics[width=0.75\textwidth]{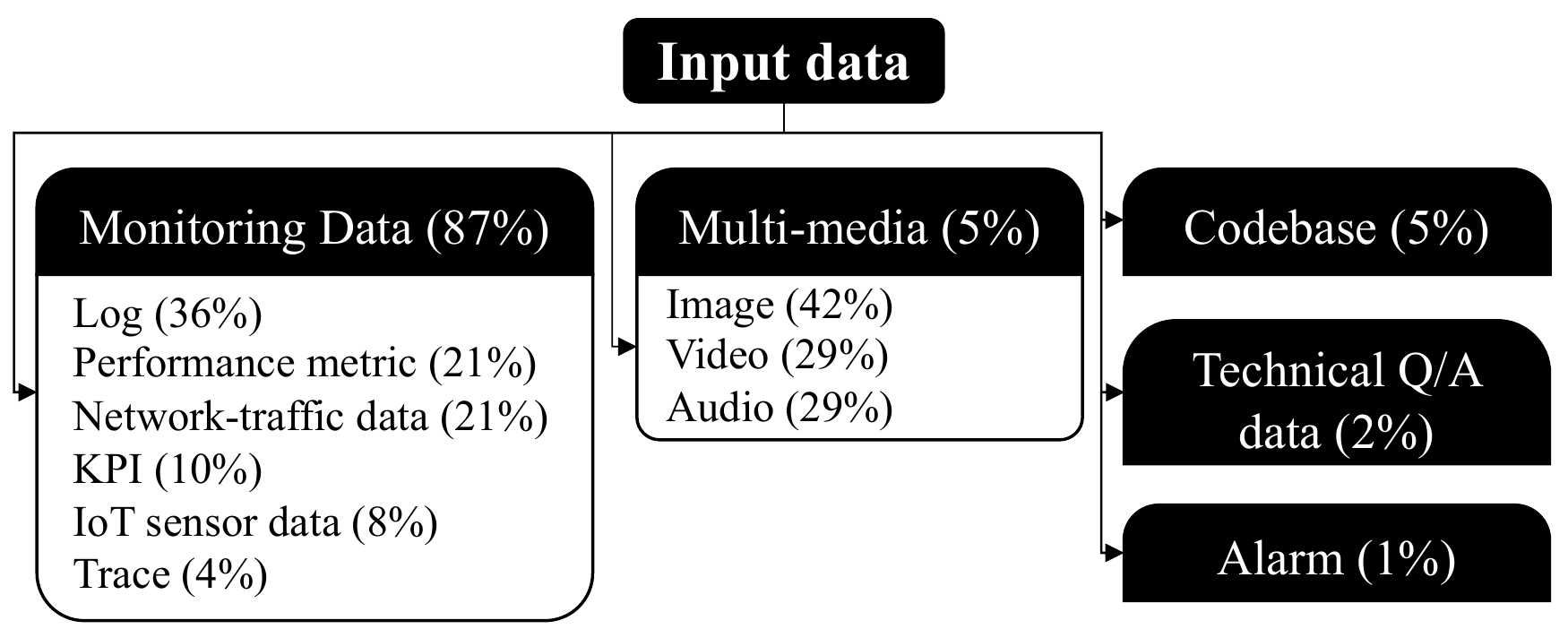}
\caption{The categorization of the input data used in AIOps projects. The high level categories are highlighted in dark.}
\label{fig:inputs}    
\end{figure*}

\begin{figure*}
\centering
\includegraphics[width=1\textwidth]{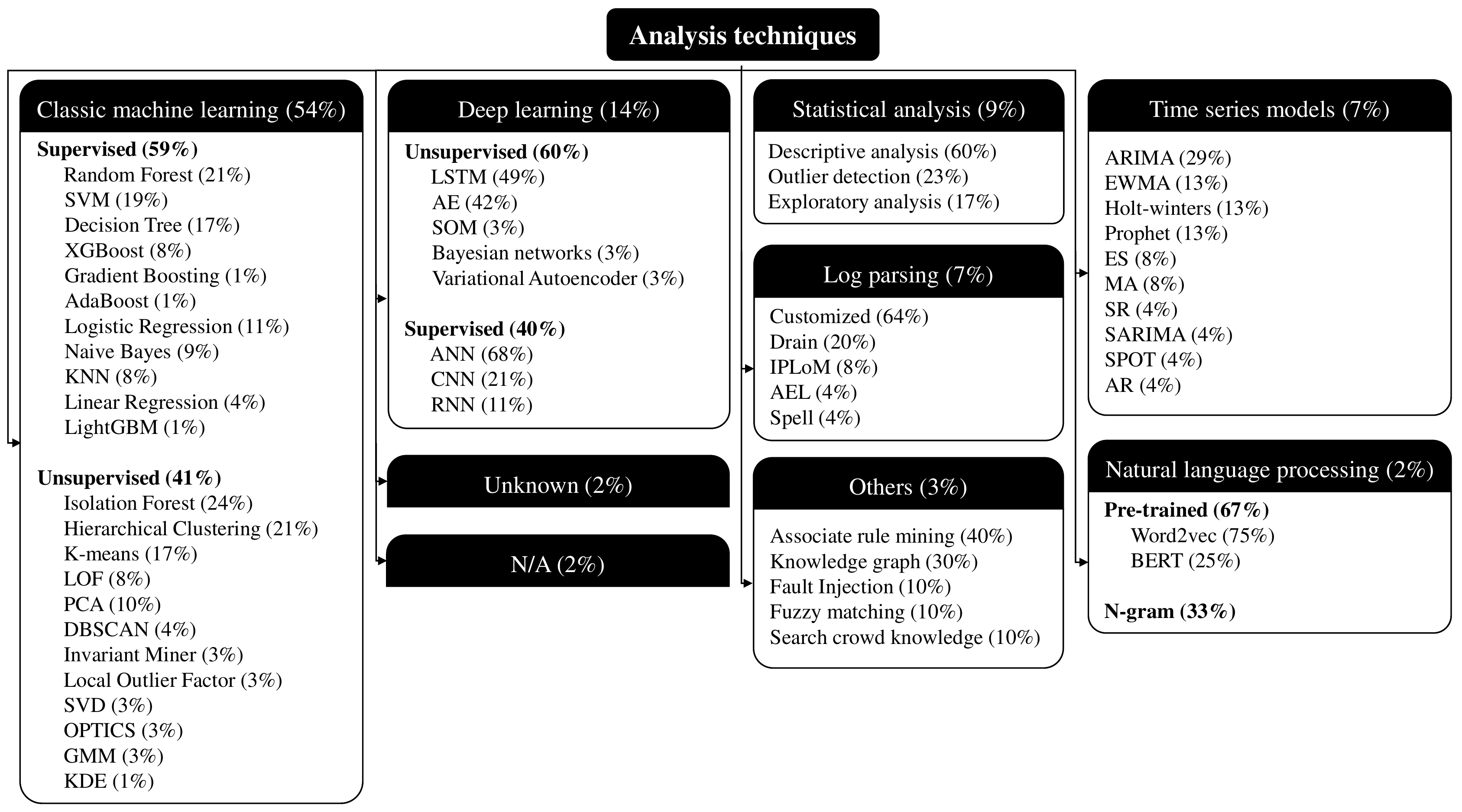}
\caption[caption]{The categorization of the analysis techniques used in AIOps projects. The high level categories are highlighted in dark.\footnotemark}
\label{fig:techniques}    
\end{figure*}

\begin{figure*}
\centering
\includegraphics[width=0.8\textwidth]{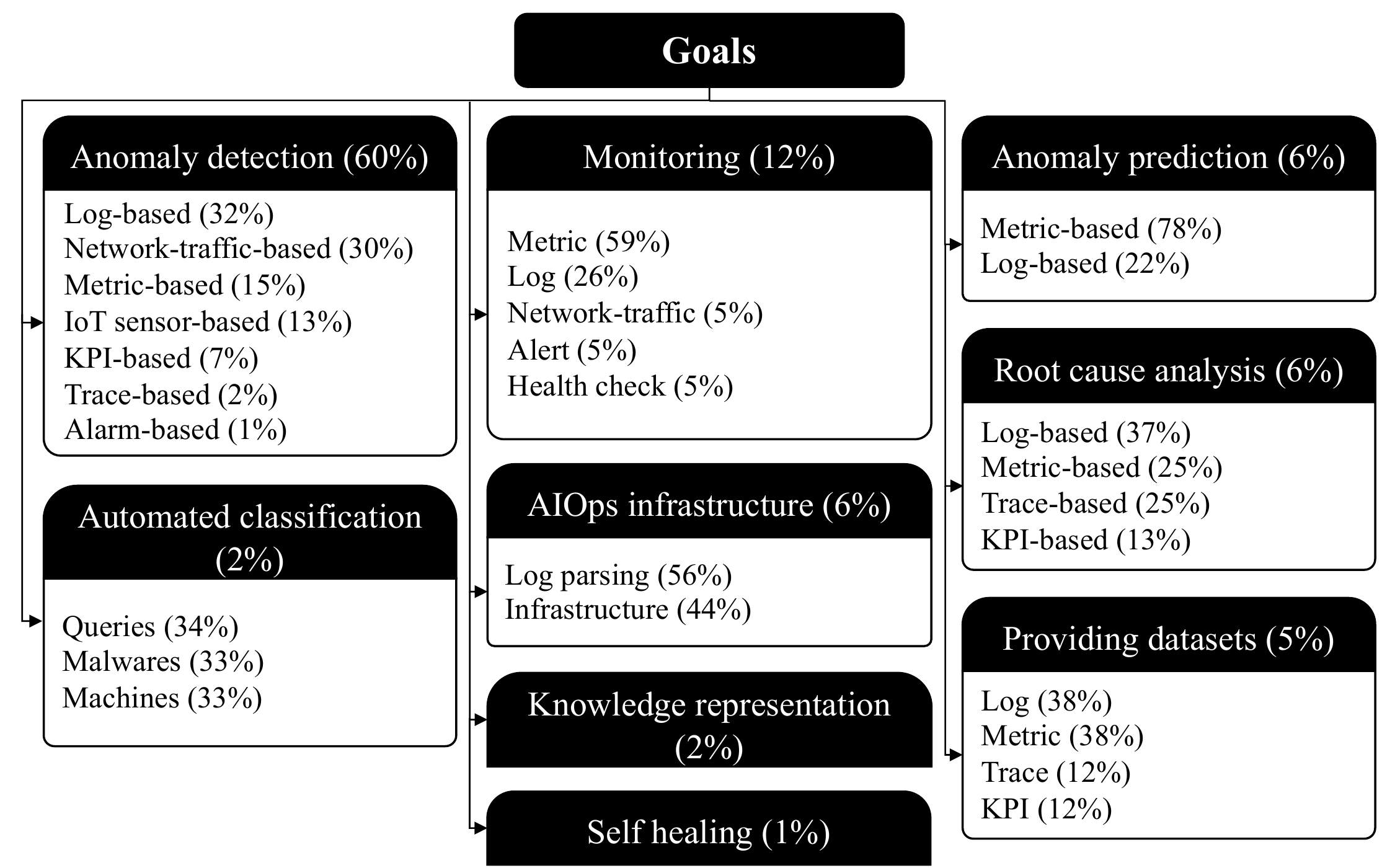}
% \caption{The categorization of the goals of the AIOps projects. The high level categories are highlighted in dark.}
\caption[Caption]{The categorization of the goals of the AIOps projects. The high level categories are highlighted in dark.}
\label{fig:goals}    
\end{figure*}

\textbf{Step 1: Coding.} Each coder analyzes the 97 AIOps projects and assign labels for each concept (input data, techniques, and goals) of each project. % to the three key concepts for each project. %the three mentioned parts. 
Multiple labels can be assigned to a concept of a project.
This step takes a few days for each coder to complete.

\textbf{Step 2: Discussion.} The coders share their responses and discuss the created labels. The main goal of the discussion session is to obtain a common understanding of the labels for the input data, techniques, and goals. Based on the separate labels of the coders, we join related labels together and take apart some high-level labels into smaller ones. After this session, we finalize the labels for each concept (input data, techniques, and goals).

\footnotetext{LSTM approaches can be used in a supervised or unsupervised manner~\citep{chen2021experience}. In our set of projects and primarily for anomaly detection, projects use LSTM in the unsupervised form.}

\textbf{Step 3: Revision.} Based on the results of the discussion session and the agreed-upon labels, each coder revises his responses from Step 1. 

\textbf{Step 4: Resolving disagreements.} The coders compare their final results from step 3 and discuss any conflict that may remain. The coders try to resolve the conflicts, but if an agreement can not be reached, the third author analyzes the project, and the final decision is made.

\textbf{Step 5: Final revision.} In the final stage, we create a mind map from all the produced %themes and 
labels. We then discuss the labels and form an hierarchy, change some labels’ names for clarity, and merge some small categories to be cohesive.

% \heng{Can we have reliability metrics for this manual analysis?} \Roozbeh{It should be a bit hard since there are many categories (it's not a typical 2-category cohen) and we changed all labels many many times and in different ways. Let's send it without it and if the reviewers mentioned it, we will add it.}

\subsubsection{\textbf{Results}}
\label{rq2_results}
%\textbf{Logs are the most used input data, machine learning approaches are the most popular techniques, and anomaly detection is the most critical goal of AIOps projects.}

%We categorize the input data, techniques, and goals of each of the 97 AIOps projects. As shown in Fig \ref{fig:flow}, each project uses its input data, then performs the desired analysis techniques on the input to achieve the project’s goals. Hence, we divide this subsection into three segments and talk about them separately. 
Figures~\ref{fig:inputs},~\ref{fig:techniques}, and~\ref{fig:goals} present our categorization of the input data, analysis techniques, and goals of the AIOps projects, % from our manual code process,
respectively. We further define our coding labels in Tables~\ref{tab:input_definitions},~\ref{tab:techniques_definitions}, and~\ref{tab:goals_definitions}.
It is important to note that each project may have multiple input data, analysis techniques, or goals. \\

\noindent\textbf{Input data: Monitoring data (e.g., logs, performance metrics, and network-traffic data) is the dominant type of input data of the AIOps projects, with logs being the most commonly used input data type.}
The categorization of the input data is illustrated in Figure~\ref{fig:inputs}.
We divide the input data of AIOps projects into five main categories: monitoring data, multi-media, codebase, technical question-answer (QA) data, and alarm. We define these categories and provide an example
% \heng{an example?} 
for each in Table~\ref{tab:input_definitions}. Among them, monitoring data is the most popular, used in 87\% of the projects. The monitoring data is divided into 6 sub-categories: log, performance metric, network-traffic data, Key Performance Indicator (KPI), Internet of Things (IoT) sensor data, and trace. Among them, log data is the most commonly used. These data types could be used for different purposes such as extensive monitoring, debugging, performance analysis, test analysis, and business analytics~\citep{chen2021survey, svoboda2015network}. In this paper, we use the umbrella term of ``monitoring data'' to name these data types.

We find two interesting input data types for the AIOps projects: network traffic and IoT sensor data. A considerable proportion of the projects (21\%) use network traffic data as their input. Also, 8\% of projects use IoT sensor data. 

\begin{table*}[h!]
\caption{Different types of input data, their definitions, and examples.}
% \heng{The difference between performance metrics and KPI is not clear. I suggest merge them.}}\Roozbeh{I tried to better define them. If it's still not good, we may merge them.}
\centering
\small
    \resizebox{1\textwidth}{!}
    {
    \begin{threeparttable}
    \begin{tabular}{lll}
        \midrule
        Input data & Definition & Repository example \\
        \midrule
        \textbf{Monitoring Data} & \textbf{Different types of data that} \\
        & \textbf{record the runtime status of a system.} \\
        
        \rule{0pt}{4ex} 
        \quad Log & \quad System-generated data that records & Repository~\href{https://github.com/logpai/loglizer}{(58811148)} uses log data as  \\
        & \quad runtime events that have happened. & input to perform anomaly detection. \\

        \rule{0pt}{4ex} 
        \quad Performance metric & \quad Quantitative measurements used to & Repository~\href{https://github.com/jixinpu/aiopstools}{(160285839)} uses \\
        & \quad track the performance of a system. & different performance metrics such as \\
        & \quad These metrics are often operational, & CPU and memory usage to perform \\
        & \quad such as CPU usage. & various tasks such as anomaly \\
        & & detection and time series forecasting. \\

        \rule{0pt}{4ex} 
        \quad Network-traffic data & \quad Monitoring data that records  & Repository~\href{https://github.com/alexamanpreet/Network-Log-and-Traffic-Analysis}{(79239275)} uses network- \\
        & \quad network activities. &  traffic data to identify malicious \\
        & & behaviors and attacks. \\
        
        \rule{0pt}{4ex} 
        \quad KPI & \quad Measurements related to key business & Repository~\href{https://github.com/logpai/Log3C}{(142442484)} uses different \\
        & \quad goals of a system. These metrics are & KPIs in time intervals \\ 
        & \quad often strategic.  & to identify impactful system problems. \\

        \rule{0pt}{4ex} 
        \quad IoT sensor data & \quad Data collected by devices connected & Repository~\href{https://github.com/kaiwaehner/ksql-udf-deep-learning-mqtt-iot}{(142325304)} uses real-time \\
        & \quad to an IoT network. & IoT sensor data to detect anomalies.\\

        \rule{0pt}{4ex} 
        \quad Trace & \quad A specialized use of logging to record & Repository~\href{https://github.com/CloudWise-OpenSource/GAIA-DataSet}{(397983735)} gathers a \\
        & \quad information about a system's & dataset of traces that can be used \\
        & \quad execution with comprehensive details. & to analyze operations problems.\\
        \midrule
        
        \textbf{Multi-media} & \textbf{Different types of multi-media data, } & Repository~\href{https://github.com/whylabs/whylogs}{(287642401)} generates \\
        & \textbf{including images, videos, and} & summaries of data types including \\
        & \textbf{audios.} & image data for monitoring purposes \\
        \midrule
        
        \textbf{Codebase} & \textbf{Source code and configuration} & Repository~\href{https://github.com/keikoproj/active-monitor}{(201529303)} can be \\
        & \textbf{files of software systems.} & installed on Kubernetes source code \\
        & \textbf{} & and provide self-monitoring and \\
        & & self-healing. \\
        \midrule
        
        \textbf{Technical Q/A data} & \textbf{Data collected from} & Repository~\href{https://github.com/OS-ABC/AIOps-Event-Graph-WebData}{(345320486)} extracts \\
        & \textbf{technical Q/A websites such as } & information from Stack Overflow to \\
        & \textbf{Stack Overflow.} & find fast solutions for faults in their \\
        & \textbf{} & platform. \\
        \midrule
        
        \textbf{Alarm} & \textbf{Alarms generated} & Repository~\href{https://github.com/jixinpu/aiopstools}{(160285839)} uses alarms \\
        & \textbf{during system run time.} & to find the association rules between \\
        & & them. \\
        \midrule

 \end{tabular}
    \begin{tablenotes}
    \small
      \item To find a GitHub repository with its ID, one can either click the hyperlink or use the link \textit{https://api.github.com/repositories/\{ID\}} where \textit{\{ID\}} is replaced by a specific repository ID.
    \end{tablenotes}
    \end{threeparttable}
    }
    \label{tab:input_definitions}
\end{table*}

\begin{table*}[h!]
\centering
\small
\caption{Different types of analysis techniques, their definitions, and examples.}
%  \heng{Keep the first letter cases of the second-level categories consistent}
    \resizebox{1\textwidth}{!}
    {
    \begin{threeparttable}
    \begin{tabular}{lll}
        \midrule
        Analysis techniques & Definition & Repository example \\
        \midrule
        \textbf{Classic machine} & \textbf{Classic machine learning techniques such as}  \\
        \textbf{learning} & \textbf{Logistic Regression and Decision Tree.} & \\
        
        \rule{0pt}{4ex} 
        \quad Supervised-learning & \quad A leaning technique that uses & Repository~\href{https://github.com/shimo85/2019AIOps_ai}{(165321356)} uses various \\
        & \quad labeled datasets. & classic supervised approaches such as \\
        & & Random Forest and Decision Tree to \\
        & & detect anomalies in KPI data. \\
        
        \rule{0pt}{4ex} 
        \quad Unsupervised-learning & \quad An approach that system learns & Repository~\href{https://github.com/logpai/loglizer}{(58811148)} implements \\
        & \quad without using labeled datasets. &  multiple unsupervised approaches \\
        & & such as Isolation Forest and Invariant \\
        & & Miner to detect anomalies. \\
        
        \midrule
        \textbf{Deep learning} & \textbf{A subfield of machine learning that} & \\
        & \textbf{uses artificial neural networks with multiple} & \\
        & \textbf{layers (i.e., deep neural networks).} & \\
        
        \rule{0pt}{4ex}
        \quad Supervised-learning & \quad A leaning technique that uses & Repository~\href{https://github.com/Shauqi/Attack-and-Anomaly-Detection-in-IoT-Sensors-in-IoT-Sites-Using-Machine-Learning-Approaches}{(187774599)} uses Artificial \\
        & \quad labeled datasets. & Neural Network to find anomalous \\
        & & behavior on IoT sensor data. \\
        
        \rule{0pt}{4ex}
        \quad Unsupervised-learning & \quad A technique that learns & Repository~\href{https://github.com/donglee-afar/logdeep}{(246569386)} uses Long \\
        & \quad patterns from unlabelled data. & Short-Term Memory to detect \\
        & & anomalies from log data. \\
        
        \midrule
        \textbf{Statistical analysis} & \textbf{Analytical techniques to} & \\
        & \textbf{understand, analyze and interpret } & \\
        & \textbf{the input data.} & \\
        
        \rule{0pt}{4ex}
        \quad Descriptive analysis & \quad Describing the features of data and & Repository~\href{https://github.com/IBM-Cloud/iot-device-phone-simulator}{(114942949)} uses descriptive \\
        & \quad summarize data in a quantitatively & analysis such as measuring minimum and \\
        & \quad manner. & maximum values and plotting scatter \\
        & & figures to interpret their IoT sensor data. \\
        
        \rule{0pt}{4ex}
        \quad Statistical outlier & \quad Applying statistical tests such as & Repository~\href{https://github.com/ThirdEyeData/Anomaly-Detections-Apache-Spark}{(156308650)} uses z-score \\
        \quad detection & \quad z-score to identify outlier values. & test to detect anomalies on CPU usage \\
        & & data. \\
        
        \rule{0pt}{4ex}
        \quad Exploratory analysis & \quad Exploring data to identify new & Repository~\href{https://github.com/BBVA/Tarkin}{(123162193)} uses exploratory \\
        & \quad connections, inspect missing data, & analysis such as plotting and describing \\
        & \quad or check hypotheses. & their log data to inspect and understand \\
        & & their input data. \\

        \midrule
        \textbf{Time series model} & \textbf{Techniques that aim to model time} & Repository~\href{https://github.com/dreamhomes/TroubleShooter}{(302842095)} applies different \\
        & \textbf{series data, mainly used for finding} & time series models such as Auto \\
        & \textbf{trends and forecasting.} & Regression and Holt-winters to detect \\ 
        & & anomalies and analyze root causes. \\

        \midrule
        \textbf{Log parsing} & \textbf{Techniques that analyze and extract} & Repository~\href{https://github.com/ixalodecte/AI-Log-Analyzer}{(345575577)} uses Drain to do \\
        & \textbf{information from log data.} & log parsing. It then uses the extracted \\
        &  & information to perform anomaly detection. \\
        
        \midrule
        \textbf{Natural language} & \textbf{Techniques that aim to analyze} & Repository~\href{https://github.com/Ohou-csu/AIOps-Learning-and-Exploration}{(316407231)} leverages \\
        \textbf{processing} & \textbf{and model text data.} & Bidirectional Encoder Representations \\
        & & from Transformers (BERT), a pre-trained \\
        & & language model as one of their techniques \\
        & & to perform anomaly prediction. \\
        
        \midrule
        \textbf{Others} & \textbf{Other techniques that could not be} & Repository~\href{https://github.com/NCBI-Hackathons/Semantic-search-log-analysis-pipeline}{(146802240)} uses fuzzy \\
        & \textbf{categorized in previous categories.} & matching to classify web queries. \\
        
        \midrule
        \textbf{Unknown} & \textbf{Projects for which we could not find} &  \\
        & \textbf{any specific techniques.} & \\
        
        \midrule
        \textbf{N/A} & \textbf{Projects that only provide} & Repository~\href{https://github.com/NetManAIOps/KPI-Anomaly-Detection}{(238914477)} is a dataset of \\
        & \textbf{datasets or do not use any} & KPI data and does not use any \\
        & \textbf{analysis techniques.} & analysis techniques. \\
        \midrule

 \end{tabular}
    \begin{tablenotes}
    \small
      \item To find a GitHub repository with its ID, one can either click the hyperlink or use the link \textit{https://api.github.com/repositories/\{ID\}} where \textit{\{ID\}} is replaced by a specific repository ID.
    \end{tablenotes}
    \end{threeparttable}
    }
    \label{tab:techniques_definitions}
\end{table*}

\begin{table*}
\centering
\small
\caption{Different types of goals, their definitions, and examples.}
    \resizebox{1\textwidth}{!}
    {
    \begin{threeparttable}
    \begin{tabular}{lll}
        \midrule
        Goal & Definition & Repository example \\
        \midrule
        
        \textbf{Anomaly detection} & \textbf{Identifying anomalies that deviate} & Repository~\href{https://github.com/alexfrancow/A-Detector}{(134266587)} uses log data \\
        & \textbf{from the normal behavior.} & and analyze them to find anomalous \\
        & & behavior. It then display anomalies \\
        & & using dynamic graphics. \\
        \midrule
        
        \textbf{Monitoring} & \textbf{Collecting and observing the} & Repository~\href{https://github.com/opendistro-for-elasticsearch/anomaly-detection}{(221989665)} provides data \\
        & \textbf{real-time stream of data to} & monitoring and alerting.\\
        & \textbf{understand system runtime status.} & \\
        \midrule
        
        \textbf{Anomaly prediction} & \textbf{Analyzing historical data to } & Repository~\href{https://github.com/geekidharsh/predicting-harddrive-failures-using-ml}{(169132015)} uses metric \\
        & \textbf{forecast future anomalies.} & data of hard drives to predict the \\
        & & failures in the near future. \\
        \midrule
        
        \textbf{Root cause analysis} & \textbf{Identifying the root causes of} & Repository~\href{https://github.com/afritzler/oopsie}{(238914477)} analyzes the \\
        & \textbf{faults or problems.} & logs of Kubernetes containers to find \\
        & & the root causes of issues. \\
        \midrule

        \textbf{AIOps infrastructure} & \textbf{Providing infrastructure support} & Repository~\href{https://github.com/nailo2c/spellpy}{(244678163)} provides \\
        & \textbf{or utility functions such as log} & automated parsing of raw logs. \\
        & \textbf{parsing.} & \\
        \midrule
        
        \textbf{Providing datasets} & \textbf{Providing datasets to be used} & Repository~\href{https://github.com/logpai/loghub}{(60705895)} collects various \\
        & \textbf{in other AIOps projects.} & system log datasets that can be used \\
        & & for log analysis.\\
        \midrule

        \textbf{Knowledge representation} & \textbf{Extracting and summarizing} & Repository~\href{https://github.com/OS-ABC/AIOps-Event-Graph-WebData}{(345320486)} extracts \\
        & \textbf{knowledge from datasets} & information from Stack Overflow to  \\
        & \textbf{or websites.} & find fast solutions for faults in their \\
        & & platform. \\
        \midrule
        
        \textbf{Automated classification} & \textbf{Classifying different input data} & Repository~\href{https://github.com/NCBI-Hackathons/Semantic-search-log-analysis-pipeline}{(238914477)} classifies web \\
        & \textbf{instance based on their similarities.} & queries to find similar important \\
        & & information and trends. \\
        \midrule  
        
        \textbf{Self healing} & \textbf{Conducting health checks and} & Repository~\href{https://github.com/keikoproj/active-monitor}{(238914477)} aims to provide \\
        & \textbf{automatically fixing the issues.} & software systems with monitoring and \\
        & & self-healing. \\
        \midrule

 \end{tabular}
    \begin{tablenotes}
    \small
      \item To find a GitHub repository with its ID, one can either click the hyperlink or use the link \textit{https://api.github.com/repositories/\{ID\}} where \textit{\{ID\}} is replaced by a specific repository ID.
    \end{tablenotes}
    \end{threeparttable}
    }
    \label{tab:goals_definitions}
\end{table*}

% \heng{Briefly describe other categories (multi-media, source code, etc.) A brief definition of each category is fine.}
The second popular category of input data for AIOps approaches is multi-media (image, video, or audio) (5\%), followed by codebase (5\%), technical Q/A data (2\%), and alarm (1\%). We define the input of a project as ``codebase'' if it analyzes the source code or configuration files of other software. We define the input as ``technical Q/A data'' if the project analyzes data from Q/A websites like Stack Overflow.

\bigskip
\noindent\textbf{Analysis techniques: Classical machine learning models are the most commonly used analysis techniques, followed far behind by deep learning, statistical analysis, and time series models.}
We present our categorization of the applied analysis techniques of AIOps projects in detail in Figure~\ref{fig:techniques}. 
We derive 9 high-level categories: classic machine learning (54\%), deep learning (14\%), statistical analysis (9\%), time series models (7\%), log parsing (7\%), others (3\%), natural language processing (2\%), unknown (2\%), and N/A (2\%). Definitions and examples for each category can be found in Table~\ref{tab:techniques_definitions}.
% \textcolor{red}{We categorize a project as ``unknown'' if we can not find the techniques used in the project.
% We assign a project with an ``N/A'' label if it does not have any type of analysis. Mainly, the projects that are datasets are categorized with this label.} \heng{this can be kept here}\Roozbeh{Again, didn't understand.}\heng{this is fine, can be removed}

Both the classic machine learning and deep learning categories are further divided into supervised and unsupervised learning algorithms. % and report all the used algorithms we find. 
Regarding classic machine learning approaches, supervised algorithms are used more than unsupervised ones (59\% compared to 41\%). The top-3 supervised algorithms are Random Forest (23\%), Support Vector Machine (SVM) (19\%), and Decision Tree (17\%). The top-3 unsupervised algorithms are Isolation Forest (24\%), Hierarchical Clustering (21\%), and K-means (13\%). Regarding the deep learning approaches, unsupervised techniques are more popular than supervised algorithms (60\% compared to 40\%). Long Short-Term Memory (LSTM), AutoEncoder (AE), and Self-Organizing Map (SOM) are the most used unsupervised algorithms, while Artificial Neural Networks (ANN) is the dominant supervised technique. 

Overall, considering supervised and unsupervised usage in classical machine learning and deep learning approaches, 37\% of the projects in our study employ supervised learning, while 31\% utilize unsupervised approaches. Hence, the difference between the adoption of supervised and unsupervised methods is not substantial. Therefore, we could not find a strong correlation with Dang et al.~\citeyearpar{dang2019aiops}, where they state that in \textit{many} AIOps cases, only unsupervised machine learning models are practical. However, we find that some projects in our study explore a combination of supervised and unsupervised models, conducting comparisons to assess their performance for specific tasks. While unsupervised techniques outperforms the supervised approaches in scenarios where data labeling is limited or new patterns of data can emerge, there are also instances where supervised approaches achieve better results. For example, as mentioned by He et al.~\citeyearpar{he2016experience} and Chen et al.~\citeyearpar{chen2021experience}, supervised machine learning algorithms usually outperform in anomaly detection use cases.

The following three common techniques are  statistical analysis, time series models, and log parsing. We find three major statistical analysis techniques: descriptive analysis, outlier detection, and exploratory analysis. We categorize the technique of a project as descriptive analysis if it analyzes the data numerically, and we group it as exploratory if it uses visualization to analyze the data. Usually, the exploratory analysis will be done after performing the descriptive analysis. 

As most of the input data is time series (e.g., logs, performance metrics, network traffic data), it is not surprising that some projects use time series techniques to model their data. AutoRegressive Integrated Moving Average (ARIMA) is the most common time-series technique followed by Exponentially Weighted Moving Average (EWMA). 

As shown in Figure~\ref{fig:inputs}, logs are the most-used input data in AIOps projects. To handle logs, projects either use log parsing techniques or Natural Language Processing (NLP) approaches. Regarding log parsing techniques, developers tend to write their customized version of log parsers (with 64\%). After coming up with customized log parsers, the most common log parser that is used among AIOps projects is Drain (with 20\%). Regarding NLP approaches, only 2\% of AIOps projects use them, with Word2vec and BERT as the most common techniques. with the recent developments in the NLP field, for example generating language models such as BERT~\citep{devlin2018bert} and CodeBERT~\citep{feng2020codebert}, we believe AIOps solutions can also benefit more from NLP techniques. We discuss this point in more detail in Section~\ref{sec:discussion}.

\begin{figure*}
\centering
\includegraphics[width=0.9\textwidth]{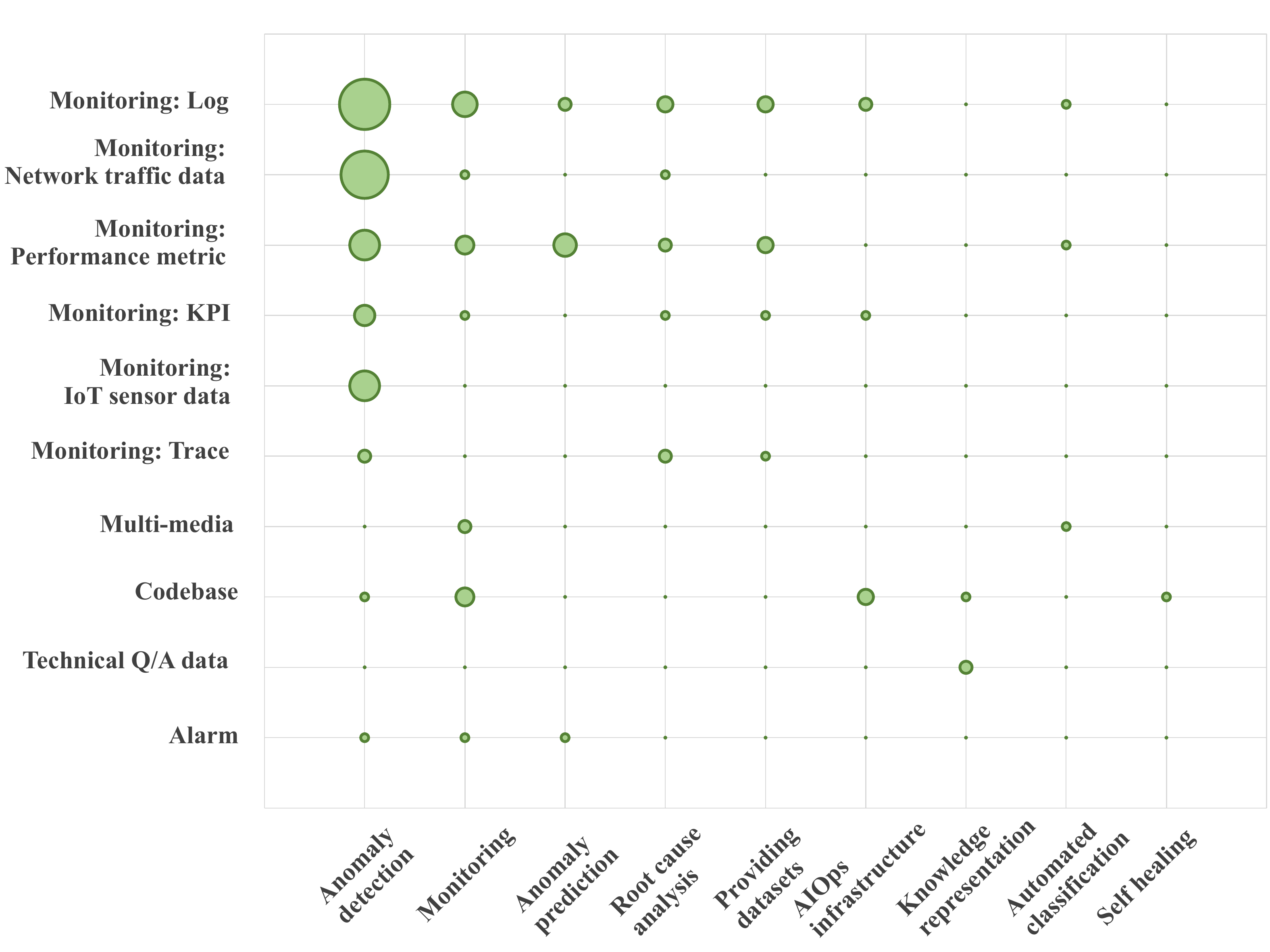}
\caption{The relation between the input data and goals of the AIOps projects. The sizes of the circles are proportional to the number of projects that use a certain input for a certain goal.}
\label{fig:bubble}    
\end{figure*}

\bigskip
\noindent\textbf{Goals:
% \heng{let's use ``goal'', not ``analysis goals'', as some goals are not analysis (like monitoring)?}
Anomaly detection is the most popular goal of the studied AIOps projects, followed by monitoring, anomaly prediction, root cause analysis, and AIOps infrastructure.}
We find 9 categories for the goals of the AIOps projects which are shown in Figure~\ref{fig:goals} and are described in Table~\ref{tab:goals_definitions}. Anomaly detection (60\%) and monitoring (12\%) are the most common reasons for using AIOps solutions. Anomaly prediction, root cause analysis, and AIOps infrastructure are the next main goals of our set of projects. We categorize the projects as ``providing datasets'' if they mainly provide public AIOps data that can be used by other practitioners or researchers. We categorize projects as ``AIOps infrastructure'' if they provide infrastructural support such as log parsing. 

We find that for tasks such as anomaly detection and root cause analysis, developers usually use logs as their main source. However, for monitoring and anomaly prediction tasks, the main source of data is performance metrics. We also find that only 1\% of projects do self-healing as their final goal. It means, in the other projects, after achieving the final goal, for example, anomaly detection, an agent (e.g., developer) should decide what to do with the founded anomalies. This not completely automated procedure will lead to a loss of time and resources. We discuss this point in more detail in Section~\ref{sec:discussion}.
% \heng{Discuss the combinations of the input data and goals (the heatmap).}

The relation between the input data and the goals of the AIOps projects is shown in Figure~\ref{fig:bubble}. 
As shown in this figure, log data and network traffic data are the most used data types for achieving the goal of anomaly detection, while performance metric data is the most used data type for anomaly prediction. Log data and performance metric data are also the most common data sources for the goal of root cause analysis. 
%As can be seen, the mixture of ``anomaly detection’’ using ``log’’ and ``network traffic data’’ are the most popular combinations of inputs and goals among the AIOps projects.
%We have also shown the data category they are using to reach these goals. 
%For example, the network traffic data and logs are the most used data to do anomaly detection, while for monitoring, the projects mainly use performance metrics.

\newenvironment{mybox2}[1]{%
    \begin{tcolorbox}[title={Summary of RQ2}]%
    }{Logs are the most commonly used input data in the studied AIOps projects, followed by performance metrics and network-traffic data. Classical machine learning techniques are the most used analysis techniques, followed (far behind) by deep learning, statistical analysis, time series models, and log parsing. The most popular goals of the AIOps projects are anomaly detection, followed by monitoring, anomaly prediction, root cause analysis, and AIOps infrastructure. 
    %and anomaly detection is the most critical goal of AIOps projects.
    \end{tcolorbox}
}
\begin{mybox2}{}

\end{mybox2}

\subsection{\textbf{RQ3. What is the code quality of AIOps projects?}}

\subsubsection{\textbf{Motivation}}
% \Foutse{this motivation is weird! what do you mean by recognising?may be better to argue that it is vital to ensure the quality of AIops projects to avoid relying on wrong signals about the health of the IT infrastructure...etc..}
Ensuring the quality of AIOps projects is vital for the software operation tasks they are designed for. Otherwise, the insights derived from their analysis would not be reliable. Thus, in this RQ, we aim to understand the code quality of the AIOps projects. We analyze different metrics in order to come to a comprehensive conclusion. We also compare the code quality of AIOps projects with our baselines to find similar patterns or differences.
Our results can provide insights for future work to improve the quality of AIOps projects.

\begin{table*}
\centering
\caption{The code quality metrics and their definitions.}
\resizebox{1\textwidth}{!}
    {
    \begin{threeparttable}
    \begin{tabular}{ll}
        \midrule
        Metric & Definition \\
        \midrule
        
        \textbf{Size} & \textbf{The metrics that represents the size of a project.} \\
        
        \rule{0pt}{4ex} 
        \quad Lines of Code (LOC) & \quad Number of lines that contain at least one character which is not a whitespace, \\
        & \quad a tabulation, or part of a comment. \\
        
        \rule{0pt}{4ex} 
        \quad Comment lines & \quad Number of lines containing comments. Non-significant comments (empty comment \\
        & \quad lines, comment lines only having special characters, etc.) are not considered. \\
        
        \rule{0pt}{4ex} 
        \quad Density of comments & \quad The amount of lines of comments compared to lines of code. It is calculated\\
        & \quad based on the following formula. \\
        & \quad Density of comments = Comment lines / (Lines of code + Comment lines) * 100 \\
        
        \midrule
        \textbf{Reliability} & \textbf{The issues that make the code behave differently as the developer was} \\
        & \textbf{intended.} \\
        
        \rule{0pt}{4ex} 
        \quad Number of bugs & \quad Total number of bugs in a project. A bug is defined as an issue that represents \\
        & \quad something wrong in the code. \\
        
        \midrule
        \textbf{Maintainability} & \textbf{The issues that make the code more difficult to update than it should.} \\
        
        \rule{0pt}{4ex} 
        \quad Number of code smells & \quad Total number of code smells in a project. A code smell is a violation of design \\
        & \quad patterns that may negatively impact software quality. \citep{rasool2015review}\\
        
        \rule{0pt}{4ex} 
        \quad Technical debt time & \quad The estimated time required to fix all the code smells. If the values are written \\
        & \quad in days, an 8-hour day is assumed. \\
        
        \midrule
        \textbf{Security} & \textbf{The issues that make potential weaknesses in terms of security. These} \\
        & \textbf{issues might be benefited by hackers.} \\
        
        \rule{0pt}{4ex} 
        \quad Number of vulnerabilities & \quad Total number of vulnerabilities in a project. A vulnebarity is a piece of code that \\
        & \quad could be exploit by a hacker. \\
        
        \rule{0pt}{4ex} 
        \quad Number of security hotspots & \quad Total number of security hotspots in a project. A security hotspot is the pieces of \\
        & \quad code that are security-sensitive. \\
        \midrule
        
 \end{tabular}
    \begin{tablenotes}
    \small
      \item All the definitions are extracted from SonarQube. A more detailed definition could be found on \href{https://docs.sonarqube.org/latest/user-guide/metric-definitions/}{SonarQube documentations}.
    \end{tablenotes}
    \end{threeparttable}
    }
    \label{tab:sonar_definition}
\end{table*}

\subsubsection{\textbf{Approach}}

%Several tools have been designed to automate the source code analysis and evaluate the code quality. Among them, we select
We use SonarQube\footnote{\url{https://www.sonarqube.org/}}, a static code analysis tool that supports a large number of programming languages, including Python, Java, JavaScript, and Go, to statically measure the quality of the AIOps projects and the baselines, as AIOps projects are developed by different languages (RQ1). Many recent studies have utilized or evaluated SonarQube for code quality measurement~\citep{businge2019studying, saarimaki2019accuracy, lomio2021fault}.

\noindent \textbf{Code quality metrics.}
To understand the code quality of AIOps projects, we measure the metrics of each studied project along four dimensions: size, reliability, maintainability, and security. The detailed list of the measured metrics include: Lines of Code (LOC), comment lines, density of comments, number of bugs, number of code smells, number of vulnerabilities, number of security hotspots, and technical debt time. All these quality metrics are extracted using SonarQube. We define these metrics in Table~\ref{tab:sonar_definition}.

SonarQube also assigns a severity level to each of the issues. It categorizes the severities into four groups; Minor, Major, Critical, and Blocker, from the lowest to the highest severity\footnote{\url{https://docs.sonarqube.org/latest/user-guide/rules/}}.
% \heng{order them by severity}

% \heng{This paragraph is related to the content in ``Finding the most violated rules and rule categories''. Please consider merging them} 
\noindent \textbf{Finding the most violated rules and rule categories.} To provide insights into the primary code quality issues among the AIOps projects, we count each issue’s assigned \textit{rules} and \textit{rule categories} (i.e., \textit{tags}) and report the most repeated ones. SonarQube evaluates the source code against its set of rules to detect specific issues. Also, rule categories are a way to categorize trivial rules into higher-level concepts. SonarQube rules include code smells, bugs, vulnerabilities, and security hotspots. Each rule is related to a specific defined issue, and different rules can be part of one rule category.

To find the most violated issues (i.e., rules and rule categories), it is essential to consider two aspects: the occurrence of an issue in each project and the percentage of projects that have that issue. Hence, we calculate the weight of violated issues (``W'' in Tables~\ref{tab:sonarqube_rules} and~\ref{tab:sonarqube_tags}) to find the most violated issues.
Each project can have multiple violated issues. We calculate the weight of issue \textit{i} in project \textit{j} using the following formula,
\[w_{ij} = \frac{n_{ij}}{n_{pj}}\]
where $n_{ij}$ is the frequency of issue \textit{i} in project \textit{j} and $n_{pj}$ is the total number of issues in project \textit{j}. Then, to calculate the weight of issue \textit{i} in all the projects, we calculate its average in all the projects.
% \heng{i think better remove 100 here, as you have the percentage symbols \% in Table 9 and 10}
\[w_{i} = \frac{\sum_{j=1}^{n}w_{ij}}{n}\]

% \heng{I think better use rule categories instead of tags (the term is too specific to SonarQube}
% \heng{generalize?} 
% \heng{Briefly describe the approach for the results of the specific rules.}

% The severity of bugs, code smells, vulnerabilities, and security hotspots can be categorized into four groups based on the impact (the effect of that issue on the system) and the likelihood (the probability of causing serious problems) of happening. These four groups are Minor, Major, Critical, and Blocker. We have also defined them in table \ref{tab:severity}.

%In order to analyze the source code of projects, we follow the next steps.
% \heng{Maybe move this part as the last part to have a more logical order.}
\noindent\textbf{Preparing project data and running SonarQube.} We execute the following three steps to analyze the quality of each of the AIOps and baseline projects.
% \heng{Organize through three steps: 1) Collecting project data (clone from GitHub); 2) Prepossessing project data (for Python and Java; what about other languages, if no preprocessing needed for them, need to mention it); 3) Execute SonarQube (may briefly explain some setup/configuration).}

%\begin{itemize}

%\item 
\textbf{Step 1. Clone the projects from GitHub.} %In the first step of the process, 
We clone the GitHub repositories on the local machine so that we can analyze their source code.

%\item 
\textbf{Step 2. Preprocess the project data.} For two of the popular languages among our projects (i.e., Python and Java), we have to perform a preprocessing phase before performing the source code analysis using SonarQube. We do not need to perform this phase for other languages. % since it is unnecessary.

%\begin{enumerate}
    %\item 
    \textit{Python code.} SonarQube does not support the \textit{.ipynb} format which is the file format of Jupyter notebooks. %, a widespread way of writing in Python. Since Python is one of the most favorite languages among both AIOps projects and the baselines, we need to overcome this limitation. 
    To overcome this limitation, we first convert the \textit{.ipynb} files in each project to the \textit{.py} format using the \textit{nbconvert} library~\citep{nbconvert} in Python.
    % \heng{through ... approach} \roozbeh{I used a library. Should I name the library?}\heng{yes, say using the ... Python library}. 
    We then use SonarQube to analyze the resulting \textit{.py} files. % to , which is able to analyze them.
    
    %\item 
    \textit{Java code.} To analyze the Java code, SonarQube requires the compiled files (\textit{.class} files). However, in most cases, developers do not upload the \textit{.class} files in their GitHub repositories and only put the \textit{.java} files. 
    To address this issue, we compile the \textit{.java} files into \textit{.class} files.
    %In order to overcome this bottleneck, we first search in all projects to find if they have java files or not. If they have, we try to find the build automation tool that the project is using. We focus on the most popular tools, namely Gradle, Apache Maven, and Apache ANT.
    We primarily leverage the build automation tool (i.e., Gradle, Apache Maven, and Apache ANT) of the project to compile the project files. For example, for Maven projects, we use \textit{mvn compile}. If we could not find a build automation tool, we use \textit{javac} to compile the \textit{.java} files.

    %After finding the build automation tool, we use the corresponded command to build the projects. Hence, for Gradle projects, we use \textit{gradle classes}; for Maven projects, we use \textit{mvn compile}; and for ANT projects, we use \textit{ant build}. If a project was not classified as one of the above groups, we run the \textit{javac}, which is the primary way to build a java file. Succeeding the creation of source code files, we pass them to the SonarQube for analysis.
%\end{enumerate}

%\item 
\textbf{Step 3. Execute SonarQube.} We write a script that %first performs the described preprocessing phase and then 
sends the source code of each project to a SonarQube web server that is installed locally. 
After SonarQube finishes the analysis, we extract the results.

%\end{itemize}

We are able to analyze the source code of 770 out of the 887 projects (87\%) in all three groups of projects (AIOps projects and the two baselines). The success ratio is 93\%, 92\%, and 79\% for AIOps, ML, and General projects, respectively.
% \heng{mention the success rates for each of the three)}. 
The main reasons of failures %in the analysis of projects 
include removed GitHub repositories %, failures in converting the \textit{.ipynb} notebooks to \textit{.py} python code,
% repositories without source code, 
and failures in compiling the Java projects. Analyzing the source code of projects at this scale is rarely reported in the literature, and performing the SonarQube analysis experiments took several weeks.

% \newcommand{\cmark}{\ding{51}}
% \newcommand{\xmark}{\ding{55}}

% \begin{table}[]
% \centering
% \begin{tabular}{|l|l|l|}
% \hline
% & Impact & Likelihood \\ \hline
% Minor & \xmark \par & \xmark\\ \hline
% Major & \xmark & \cmark \\ \hline
% Critical & \cmark & \xmark \\ \hline
% Blocker & \cmark & \cmark \\ \hline
% \end{tabular}
% \caption{The assignment of severities for bugs, code smells, and hotspot securities.}
% \label{tab:severity}
% \end{table}

\noindent\textbf{ Statistical tests.}
Similar to the first RQ, we perform statistical tests to validate our results and ensure their statistical significance. We use \textit{Mann-Whitney U} test~\citep{mann1947test} to compare the distribution of the metrics between AIOps and baseline projects. We also use \textit{Cliff’s delta} test~\citep{cliff1993dominance} to measure the effect size of the difference between the distributions. We use the same approach and scale (i.e., effect of |d| = 0.147 is small, |d| = 0.33 is medium, and |d| = 0.474 is large) as described in Section~\ref{sec:statistics}.

\subsubsection{\textbf{Results}}
\label{sec:rq3_results}
\begin{figure*}[]
  \centering
  \subfloat[]{\includegraphics[width=0.45\textwidth]{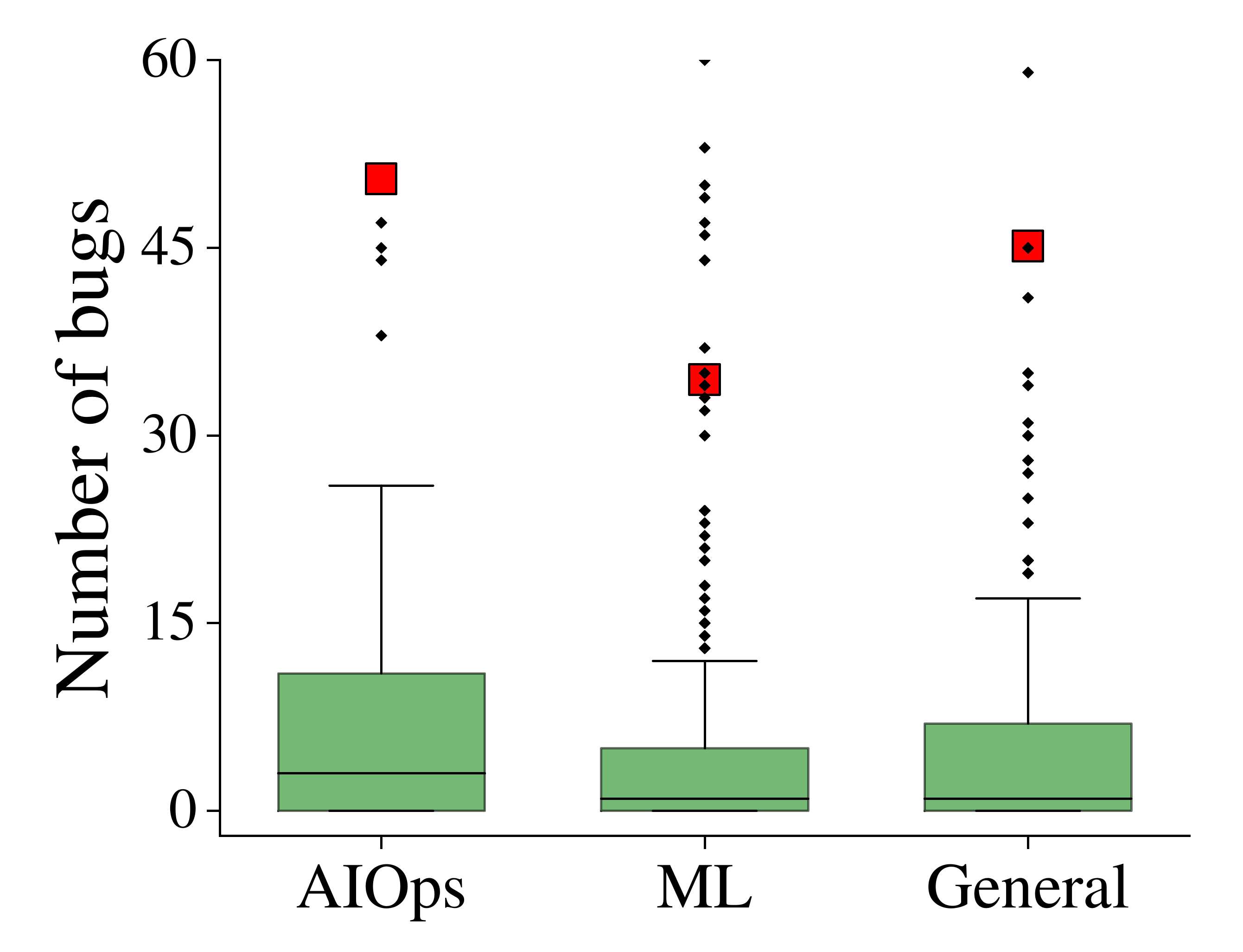}\label{fig:bugs}}
  \subfloat[]{\includegraphics[width=0.45\textwidth]{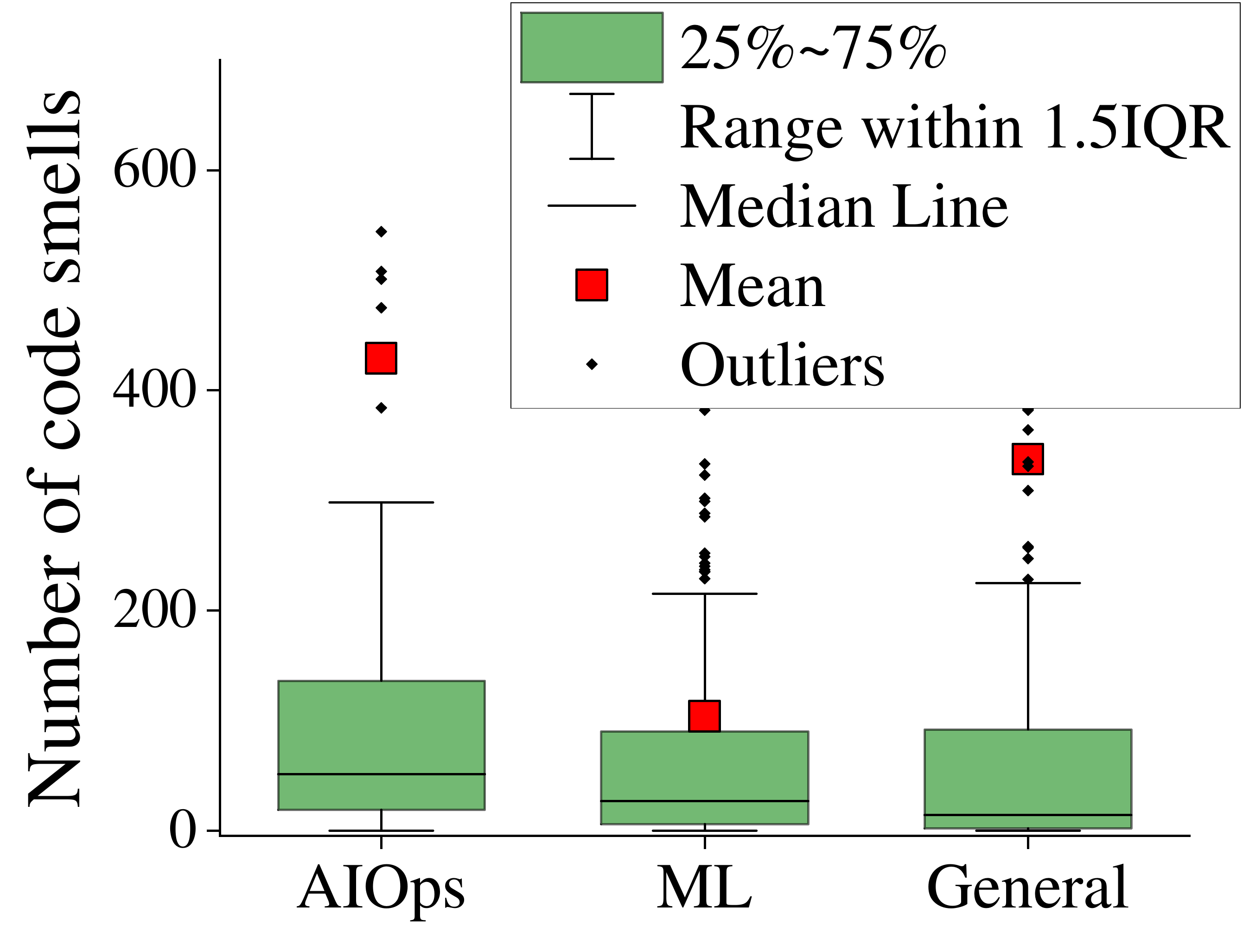}\label{fig:smells}}\\
  \subfloat[]{\includegraphics[width=0.45\textwidth]{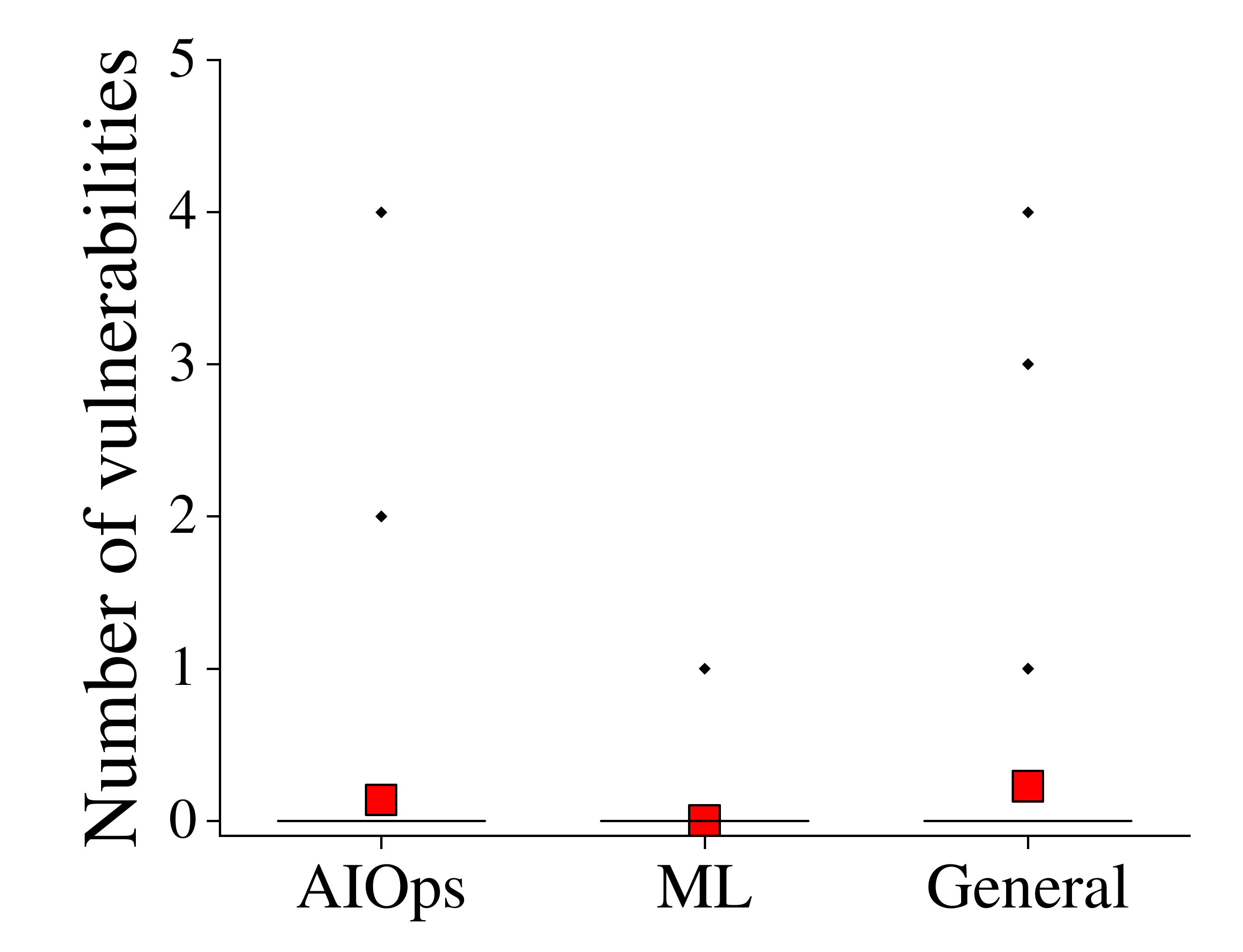}\label{fig:vulnerabilities}}
  \subfloat[]{\includegraphics[width=0.45\textwidth]{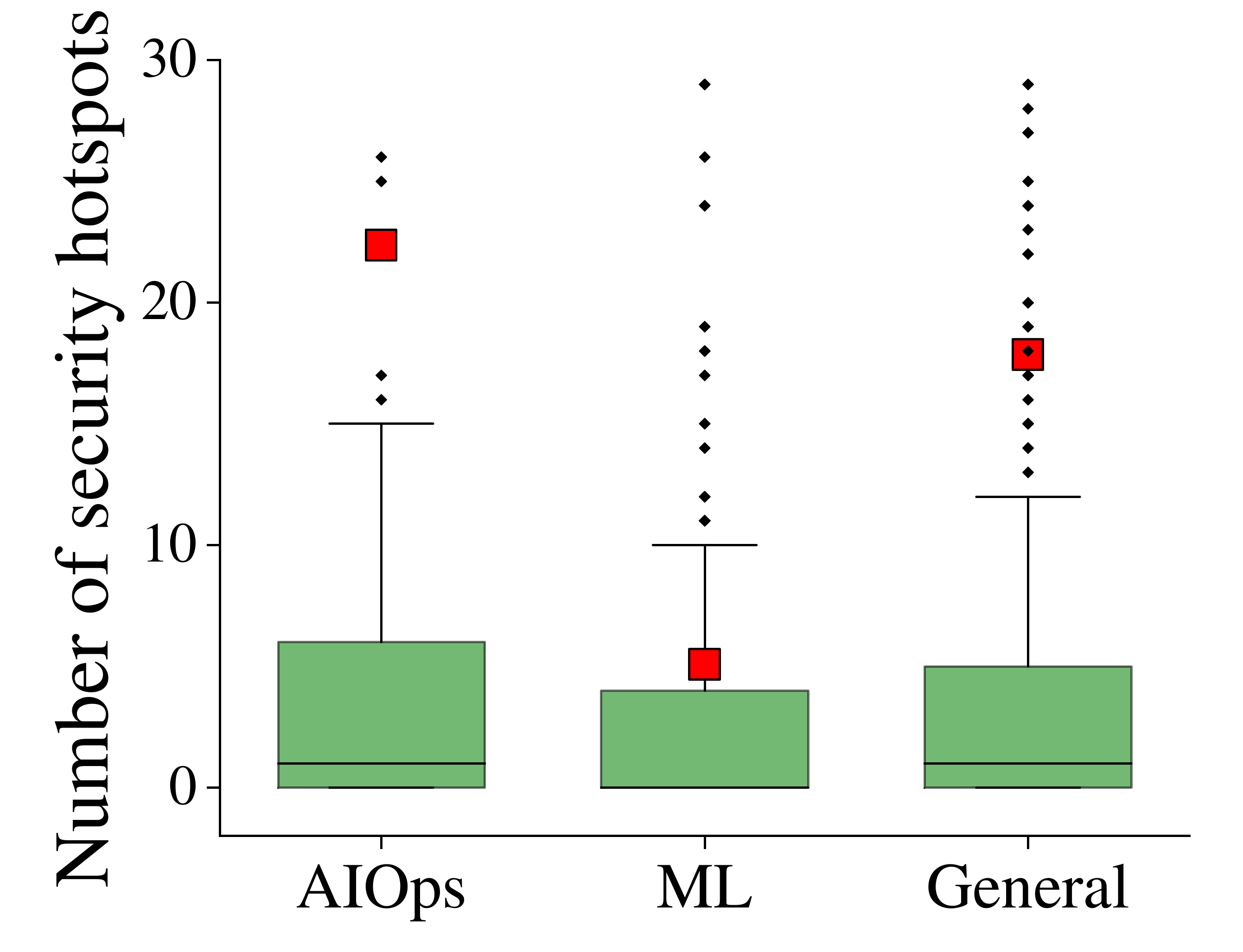}\label{fig:security_hotspots}}\\
  \subfloat[]{\includegraphics[width=0.45\textwidth]{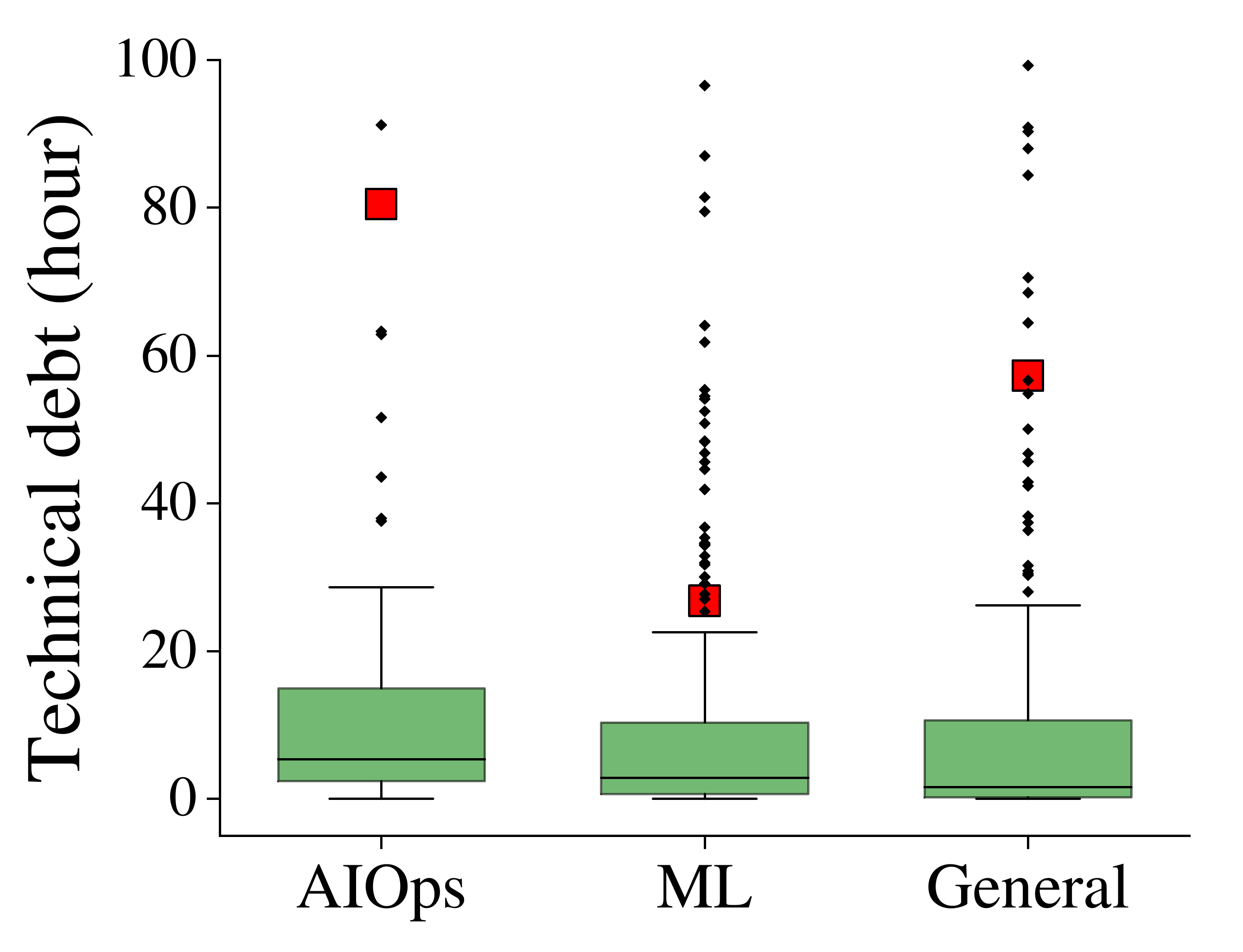}\label{fig:debt}}
  \subfloat[]{\includegraphics[width=0.45\textwidth]{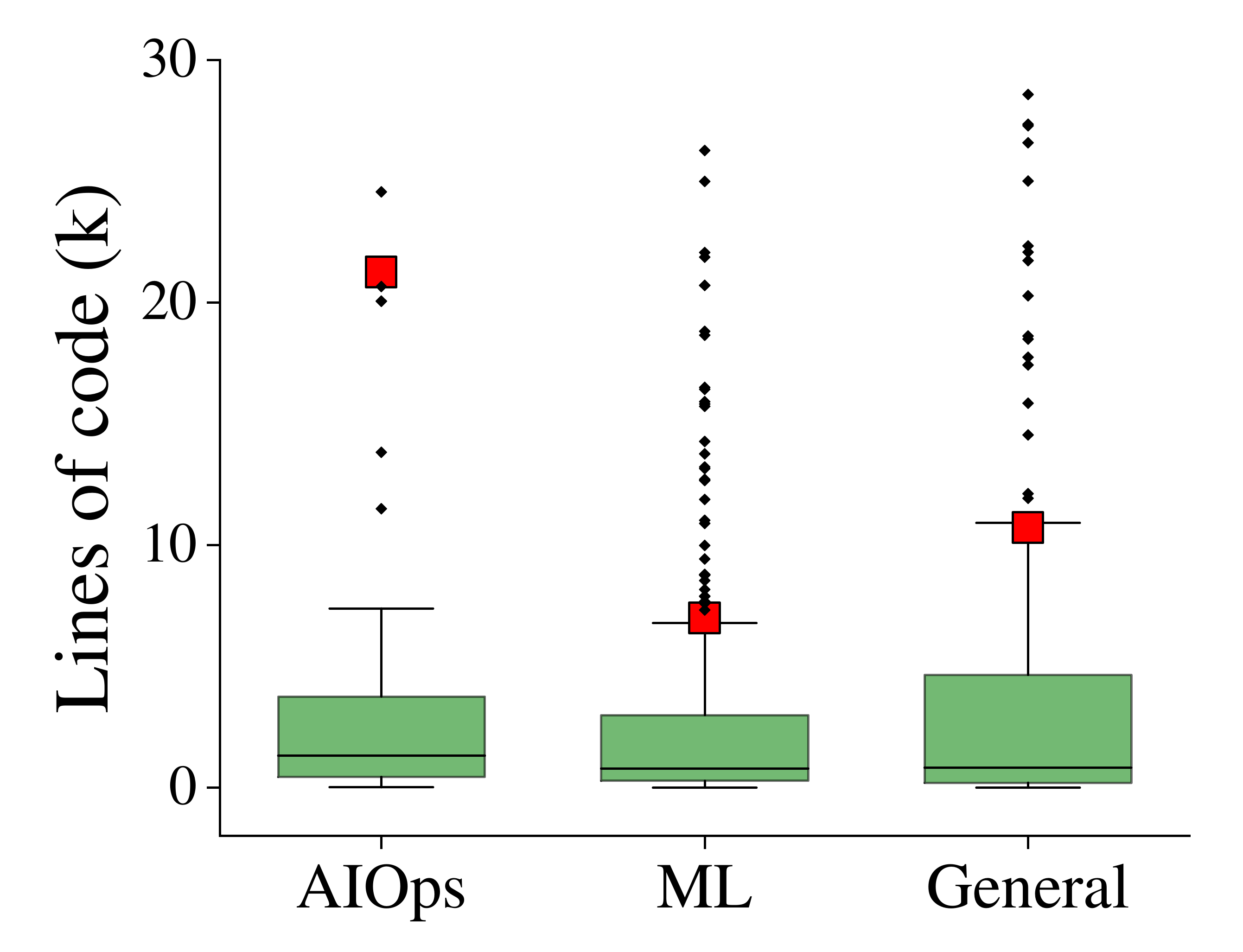}\label{fig:loc}}\\
  \subfloat[]{\includegraphics[width=0.45\textwidth]{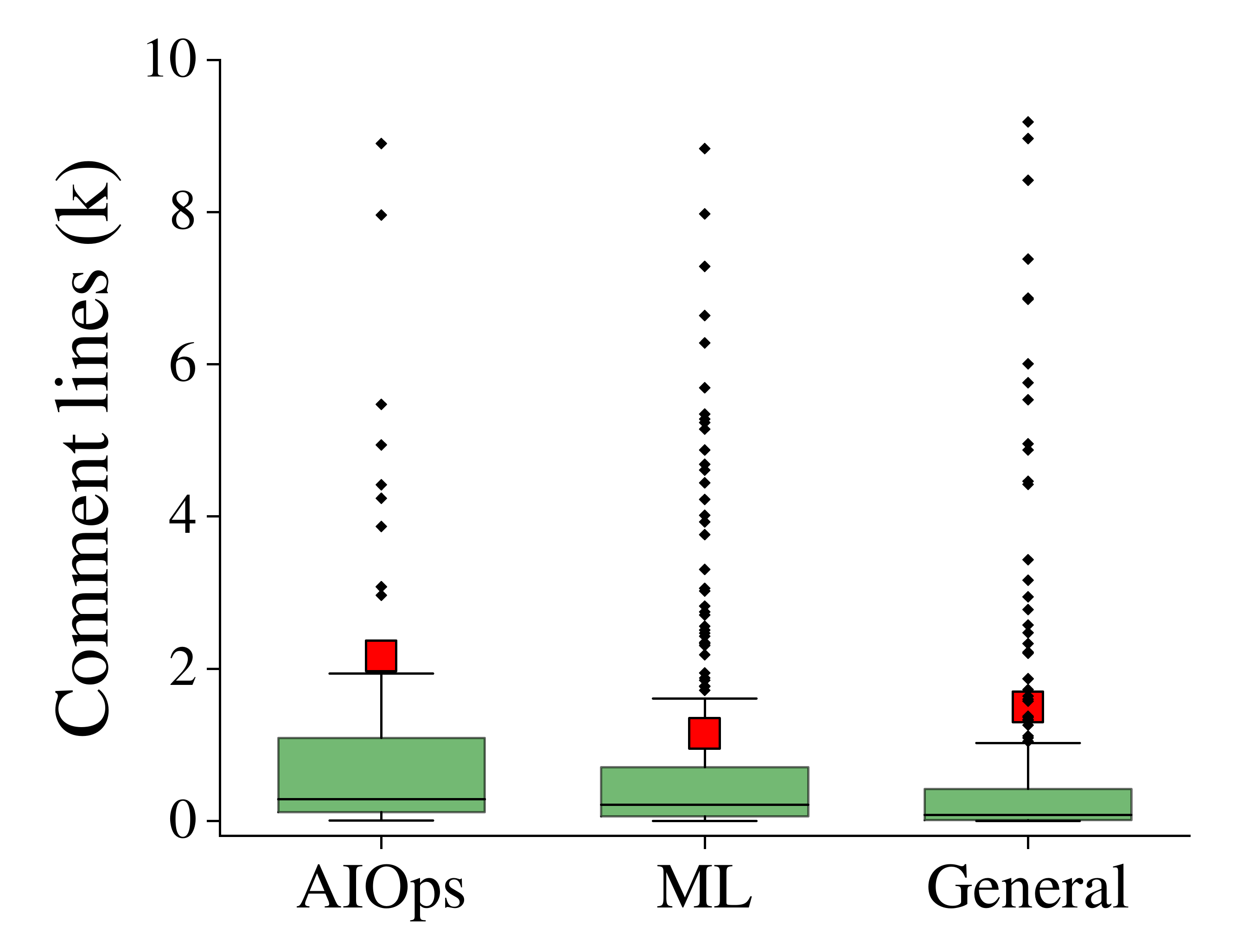}\label{fig:comment_lines}}
  \subfloat[]{\includegraphics[width=0.45\textwidth]{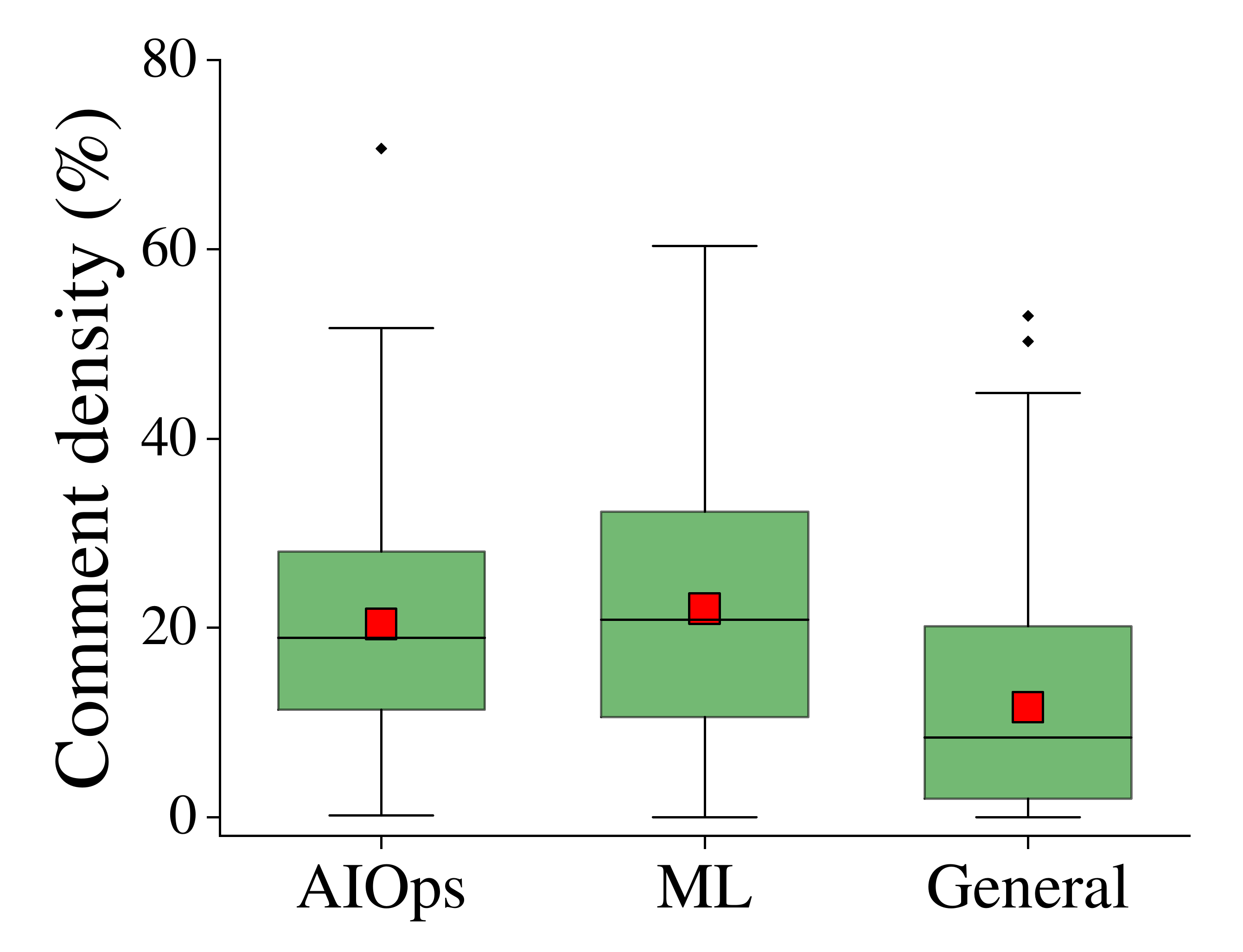}\label{fig:comment_density}}
  \caption{Box plots of code quality metrics for AIOps and baseline projects.}
\label{fig:sonarqube_box_plot}
 \vspace{-1em}
\end{figure*}

\begin{figure*}[]
  \centering
  \subfloat[]{\includegraphics[width=0.45\textwidth]{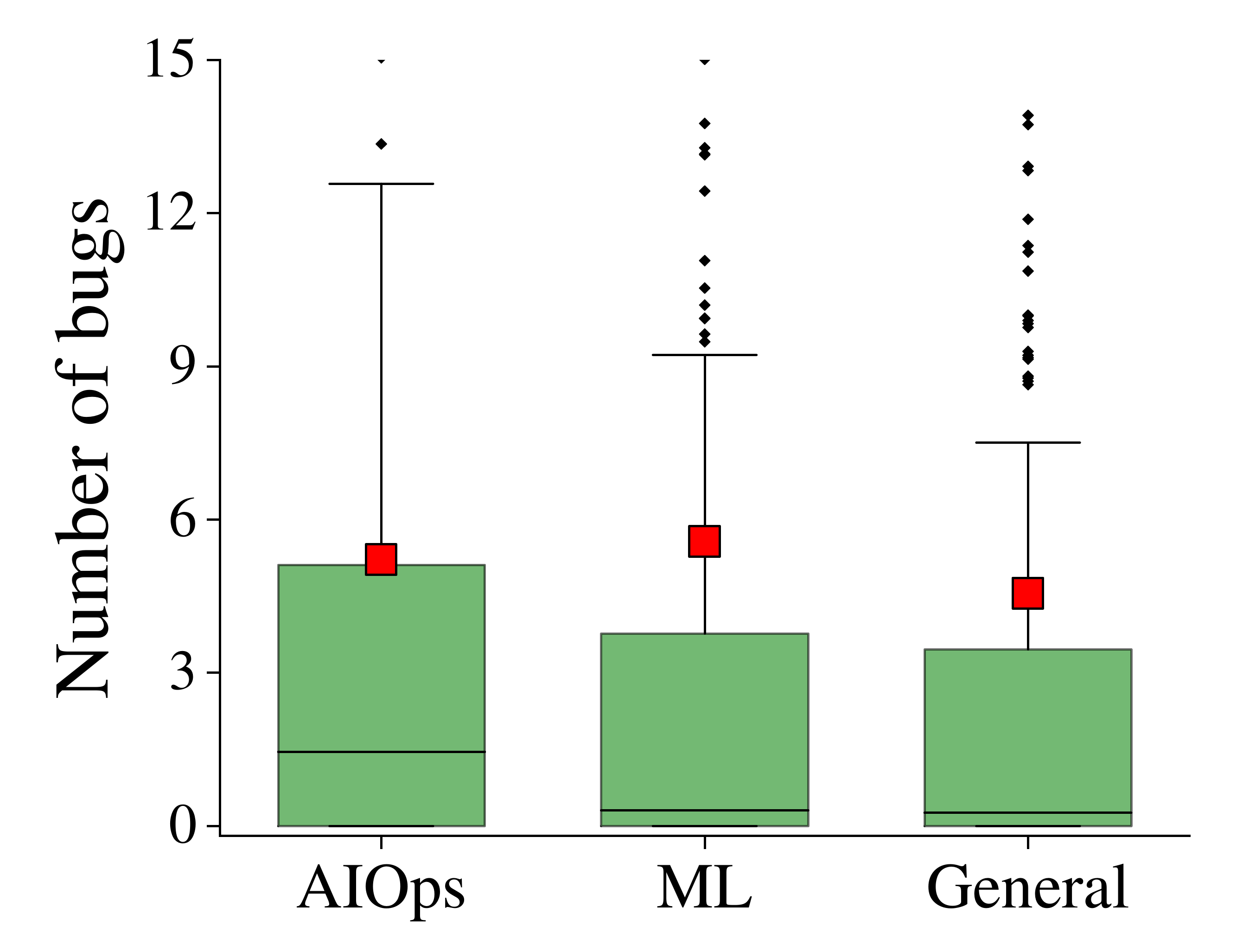}\label{fig:norm_bugs}}
  \subfloat[]{\includegraphics[width=0.45\textwidth]{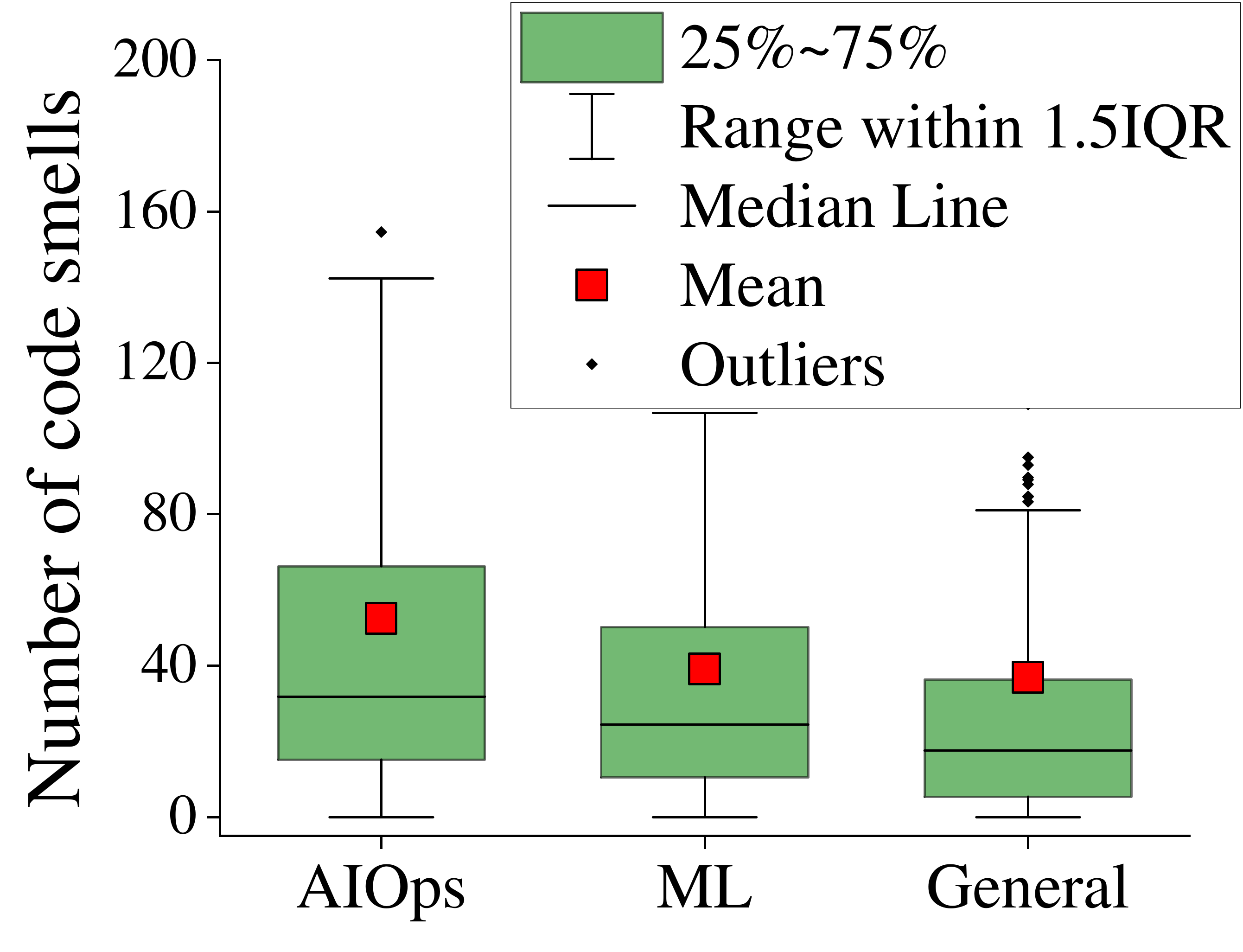}\label{fig:norm_smells}}\\
  \subfloat[]{\includegraphics[width=0.45\textwidth]{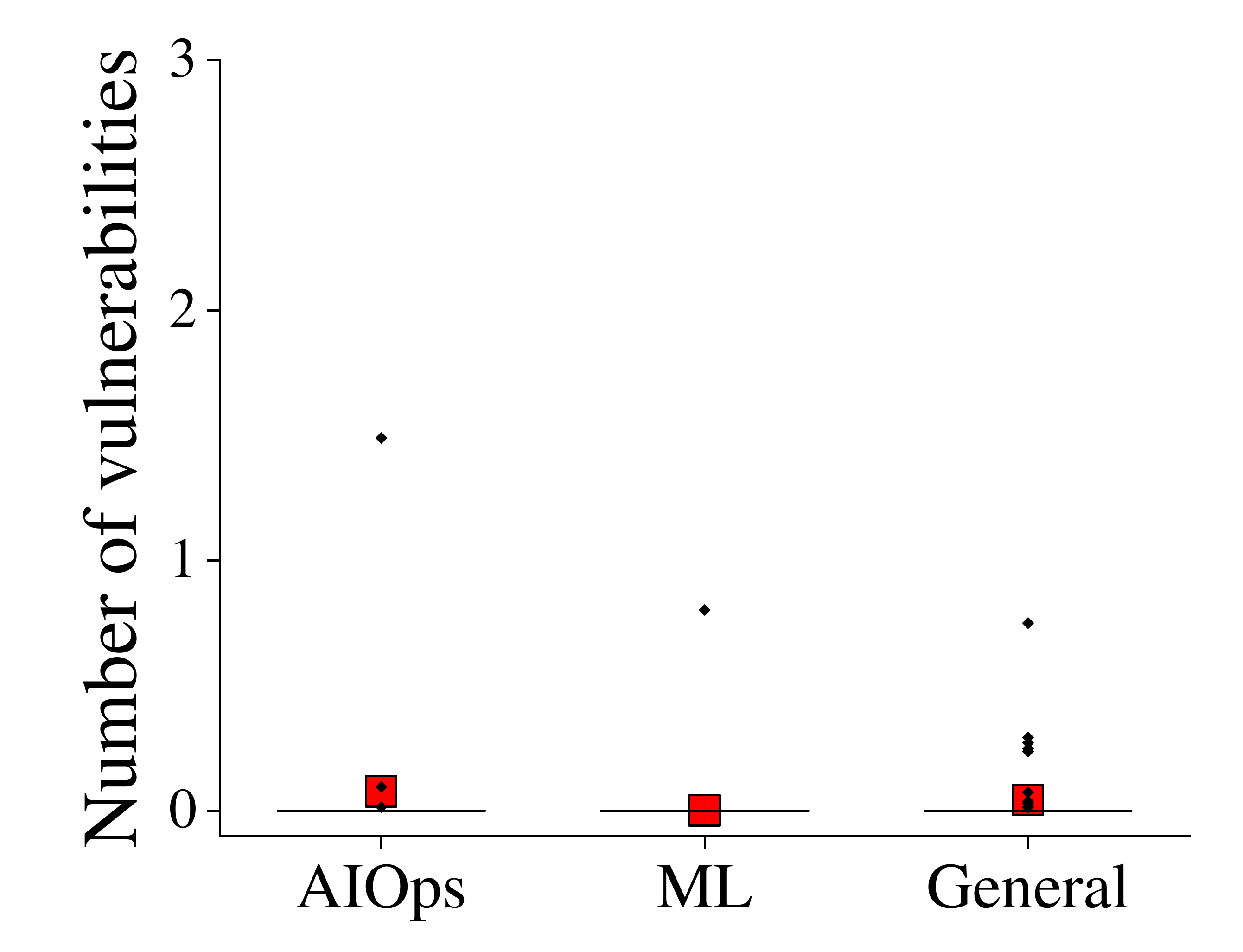}\label{fig:norm_vulnerabilities}}
  \subfloat[]{\includegraphics[width=0.45\textwidth]{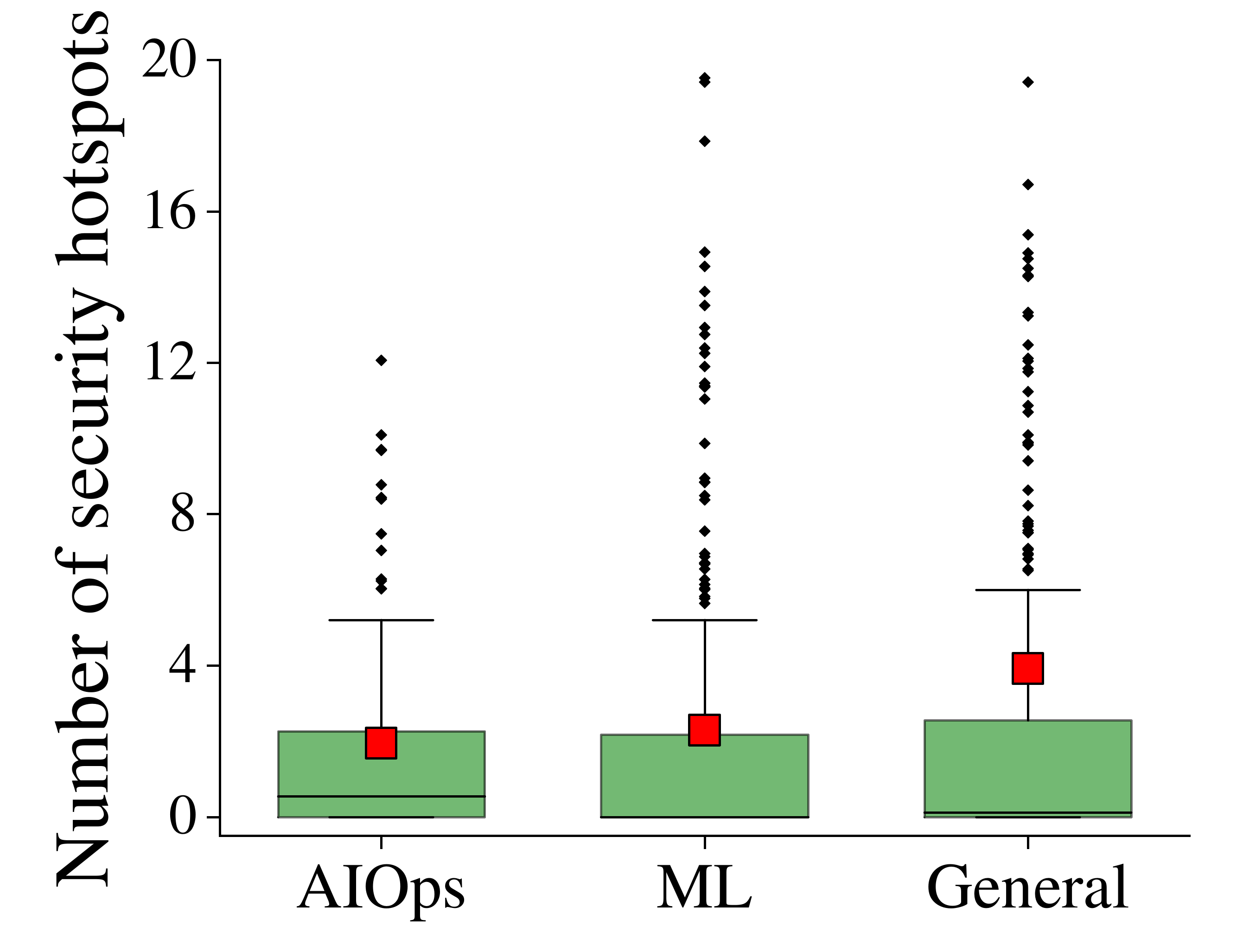}\label{fig:norm_security_hotspots}}\\
  \subfloat[]{\includegraphics[width=0.45\textwidth]{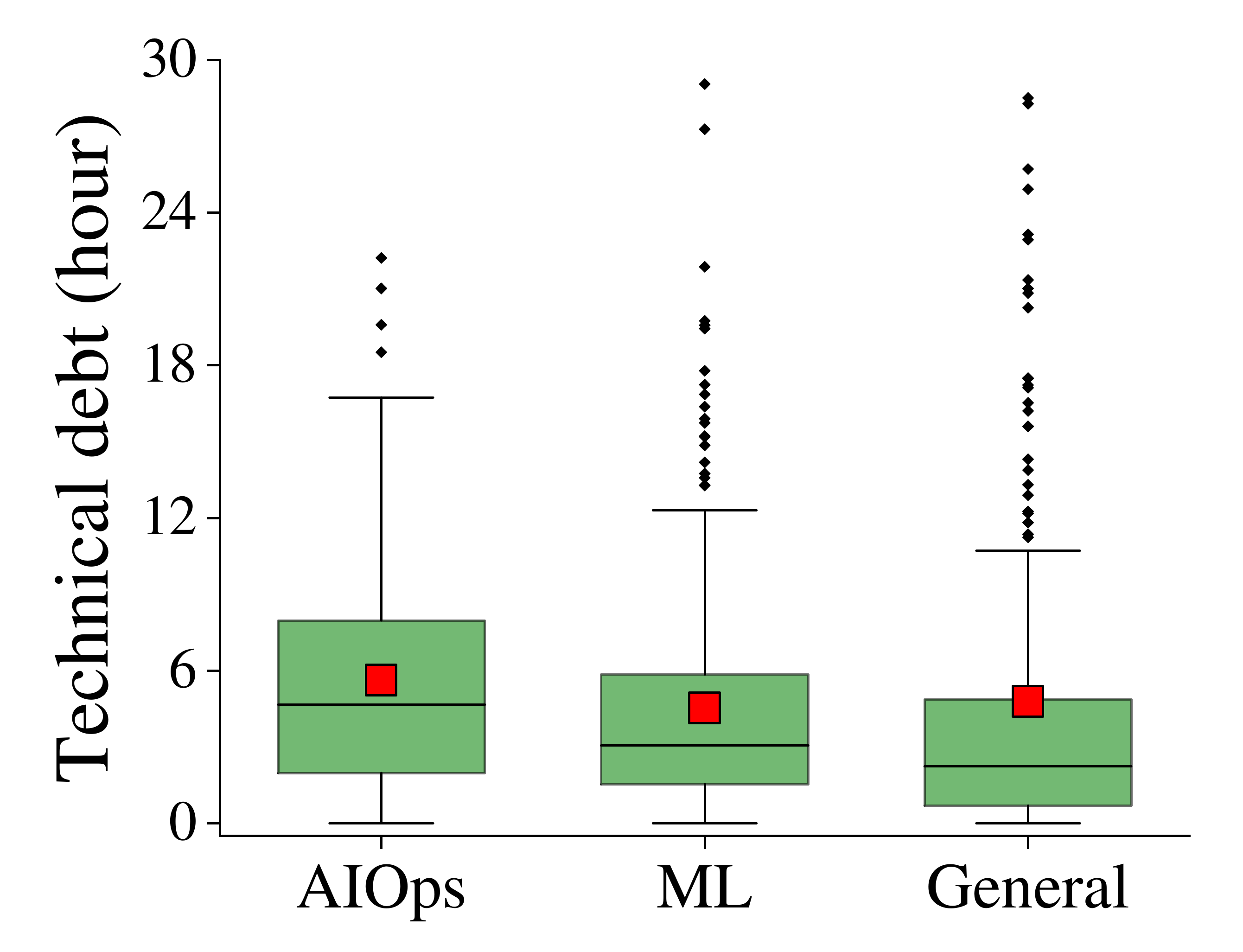}\label{fig:norm_debt}}
  \caption{Box plots of code quality metrics for AIOps and baseline projects. The values are normalized by LOC of each project.}
\label{fig:norm_sonarqube_box_plot}
 \vspace{-1em}
\end{figure*}

\noindent\textbf{AIOps projects have poorer quality than the ML and General baselines.}
Figure~\ref{fig:sonarqube_box_plot} represents the box plots of code quality metrics extracted from SonarQube for AIOps and baseline projects. Table~\ref{tab:sonarqube_metrics_statistics} illustrates the p-value and effect size of the code quality metrics of AIOps projects comparing to the baselines. For bugs, code smells, vulnerabilities, security hotspots, and technical debt, we further report the values normalized by the LOC of each project. We report these normalized values in Figure~\ref{fig:norm_sonarqube_box_plot} to ensure that our findings are robust and not biased by the size of the projects.

\begin{figure*}
\centering
\includegraphics[width=0.65\textwidth]{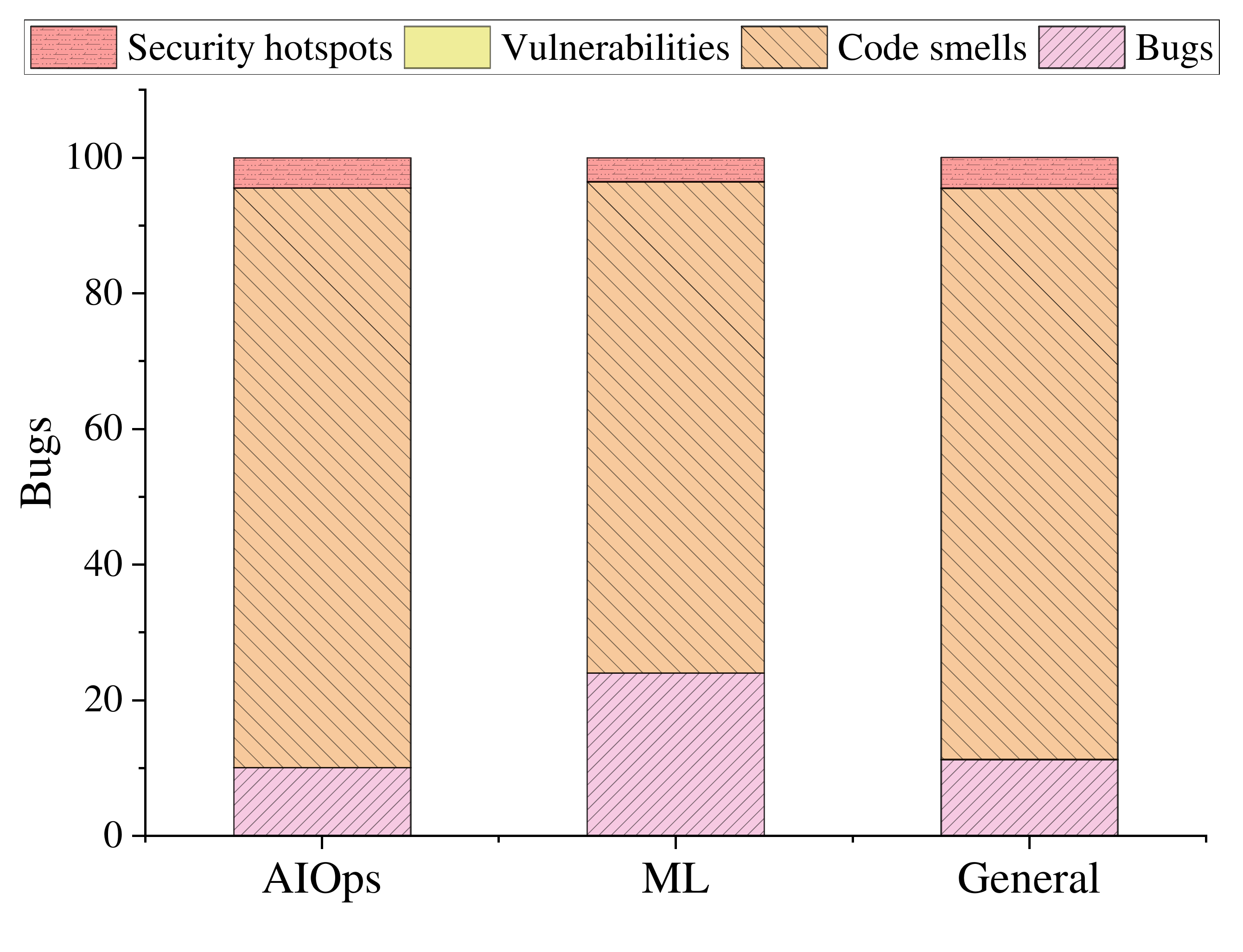}
\caption{The distribution of the quality issues among the AIOps and baseline projects.}
\label{fig:sonarqube_issues}    
\end{figure*}

\begin{table}
\centering
\caption{Detailed results of \textit{Mann–Whitney U} and \textit{Cliff's delta} tests on projects' code quality.}
\begin{threeparttable}
   \begin{tabular}{l|ll|ll}
\hline
  \multirow{2}{*}{Metric} & \multicolumn{2}{c|}{\textbf{AIOps vs. ML}} & \multicolumn{2}{c}{\textbf{AIOps vs. General}} \\ 
  \cline{2-5}
  & \multicolumn{1}{l|}{p-value} & effect size & \multicolumn{1}{l|}{p-value} & effect size \\ \hline
Bugs & \multicolumn{1}{l|}{0.01} & ** & \multicolumn{1}{l|}{0.03} & * \\
Code smells & \multicolumn{1}{l|}{0.00} & ** & \multicolumn{1}{l|}{0.00} & ** \\
Vulnerabilities & \multicolumn{1}{l|}{0.00} & * & \multicolumn{1}{l|}{1.00} & - \\
Security hotspots & \multicolumn{1}{l|}{0.03} & * & \multicolumn{1}{l|}{0.18} & - \\
Technical debt & \multicolumn{1}{l|}{0.00} & ** & \multicolumn{1}{l|}{0.00} & ** \\
LOC & \multicolumn{1}{l|}{0.01} & ** & \multicolumn{1}{l|}{0.02} & * \\
Comment lines & \multicolumn{1}{l|}{0.01} & ** & \multicolumn{1}{l|}{0.00} & *** \\
Comment density & \multicolumn{1}{l|}{0.42} & - & \multicolumn{1}{l|}{0.00} & *** \\ \midrule
\end{tabular}
    \begin{tablenotes}
  \small
  \item \textit{Mann–Whitney U} results are shown in \textit{p-value} columns. If the sets have statistically different distributions, the \textit{Cliff's delta} results are shown in \textit{effect size} columns.

  *: negligible effect
  **: small effect
  ***: medium effect
     \end{tablenotes}
\end{threeparttable}
\label{tab:sonarqube_metrics_statistics}
\end{table}

According to the p-values of~\textit{Mann-Whitney U} test shown in Table~\ref{tab:sonarqube_metrics_statistics}, the difference between most of the code quality metrics of AIOps projects compared to baselines is significant. Considering the number of bugs and code smells, AIOps projects are statistically different from the baselines, and the effect size of their differences is small or negligible. As shown in Figure~\ref{fig:bugs}, AIOps projects have a higher number of bugs in terms of both mean and median, with a median of 3 bugs, while the median is 1 in both baselines. This pattern is also seen in the normalized version in Figure~\ref{fig:norm_bugs}. The number of code smells is also higher in AIOps projects based on both original and normalized versions presented in Figures~\ref{fig:smells} and~\ref{fig:norm_smells}. The median number of code smells in AIOps projects is 51.0, twice the ML baselines with 26.5 and more than 3.5 times the General baseline with 14.0.

Regarding the number of vulnerabilities and security hotspots among the projects, no statistically significant difference is seen between AIOps and General projects, according to Table~\ref{tab:sonarqube_metrics_statistics}. However, AIOps and ML baseline are statistically different, with an effect size of negligible. As Figures~\ref{fig:vulnerabilities} and~\ref{fig:norm_vulnerabilities} show, most of the AIOps and baseline projects do not have any vulnerabilities. However, as Figure~\ref{fig:security_hotspots} represents, the median number of security hotspots in AIOps is 1 while it is 0 in ML baseline; and the mean number of security hotspots in AIOps is 22.4 while it is only 5.1 in ML baseline.

Technical debt is the next analyzed security metric. Table~\ref{tab:sonarqube_metrics_statistics} suggests that there is a statistically significant difference between AIOps and baselines with a small effect size. As Figure~\ref{fig:debt} shows, both the mean and median of AIOps technical debt are higher than the baselines. The mean value of technical debt for the AIOps set is 80.5 hours, while it is 26.8 and 57.3 hours for ML and General baselines, respectively. The median value of technical debt in AIOps (5.4 hours) is 2 times the ML baseline (2.9 hours) and 3 times the General baseline (1.7 hours). As Figure~\ref{fig:norm_debt} suggests, technical debt in AIOps projects is higher than the baselines, even in the normalized version. These findings are in line with the high volume of issues observed in AIOps repositories, as shown in Figure~\ref{fig:github_box_plot}.

\noindent\textbf{AIOps projects have more lines of code than baselines and the same amount of comments as the ML baseline.}

Table~\ref{tab:sonarqube_metrics_statistics} shows that AIOps set is statistically different from the baselines in terms of number of lines of code and number of comment lines. According to~\ref{fig:loc}, the mean value of LOC for AIOps projects is 21.2k, 3 times the ML baseline (with 7.0k) and 2 times the General baseline (10.7k). The median value also indicates a similar pattern, with 1.4k for AIOps projects and 0.8k for the baselines. The amount of comments written in AIOps projects is also higher than the baselines, with a median of 0.3k compared to 0.2k and 0.1k for the ML and General baselines. Figure~\ref{fig:comment_density} shows the comment density between the three sets. The density of comments in AIOps and ML projects are similar, with the medians of 19.0\% and 20.9\%, respectively. The comment density for the General baseline is less than half of the AIOps set, with only 8.4\%.

\noindent\textbf{AIOps projects suffer from code smells more than other types of issues.}
Figure \ref{fig:sonarqube_issues} illustrates the distribution of bugs, code smells, vulnerabilities, and security hotspots among the projects. The percentage of vulnerabilities is much smaller than other types of issues in all the projects. In AIOps projects, the main issue is related to code smells, which account for 85\% of all issues, 13\% more than the ML baseline. After code smells, the most common type of issue is bugs, with 10\%, and then security hotspots, with 5\%.

\begin{table}[t]
\centering
\caption{The top-10 violated SonarQube rules for AIOps projects and the baselines. ``Sev'' indicates the severity of issues, ``W'' represents the weight of rules, and ``N'' is the percentage of projects with that rule. ``Mn'' stands for Minor, ``Mj'' stands for Major, ``Cr'' stands for Critical, and ``Bl'' stands for Blocker.}
    \resizebox{1\textwidth}{!}
    {
    \small
    \begin{tabular}{llll|llll|llll}
        \toprule
        \multicolumn{4}{c|}{\textbf{AIOps}} & \multicolumn{4}{c|}{\textbf{ML}} &
        \multicolumn{4}{c}{\textbf{General}}\\
        Rule & Sev & W(\%) & N(\%) &
        Rule & Sev & W(\%) & N(\%) &
        Rule & Sev & W(\%) & N(\%) \\
        \midrule
        python:S117 & Mn & 17.6 & 59.1 & python:S117 & Mn & 16.8 & 59.3 & python:S117 & Mn & 5.2 & 19.7 \\
        python:S125 & Mj & 14.6 & 66.3 & python:S125 & Mj & 15.5 & 63.8 & python:S125 & Mj & 3.9 & 20.3 \\
        python:S1192 & Cr & 7.8 & 67.3 & python:S905 & Mj & 7.7 & 27.7 & javascript:S1117 & Mj & 3.4 & 19.3 \\
        python:S905 & Mj & 7.0 & 25.5 & python:S1192 & Cr & 7.3 & 55.4 & Web:S5254 & Mj & 3.0 & 20.7 \\
        python:S1481 & Mn & 4.2 & 53.6 & python:S1481 & Mn & 5.9 & 52.5 & python:S1192 & Cr & 2.6 & 19.0 \\
        python:S3776 & Cr & 3.3 & 48.1 & python:S1542 & Mj & 4.0 & 41.0 & javascript:S125 & Mj & 2.3 & 16.7 \\
        python:S1542 & Mj & 3.2 & 44.5 & python:S3776 & Cr & 3.2 & 43.5 & python:S1481 & Mn & 1.9 & 18.0 \\
        python:S2208 & Cr & 2.3 & 25.4 & python:S2320 & Mj & 2.8 & 8.8 & xml:S125 & Mj & 1.8 & 5.2 \\
        python:S5754 & Cr & 2.2 & 35.4 & python:S2208 & Cr & 2.4 & 23.4 & Web:S1827 & Mj & 1.7 & 7.5 \\
        python:S100 & Mn & 2.1 & 20.9 & python:S5754 & Cr & 2.1 & 25.4 & javascript:S2703 & Bl & 1.7 & 12.4
    \end{tabular}
    }
        
    \label{tab:sonarqube_rules}
\end{table}

\begin{table}[t]
\centering
        \caption{The top-10 violated SonarQube tags (rule categories) for AIOps projects and the baselines. ``W'' represents the weight of tags, and ``N'' is the percentage of projects with that tag.
        }
    \resizebox{1\textwidth}{!}
    {
    \small

    \begin{tabular}{lll|lll|lll}
        \toprule
        \multicolumn{3}{c|}{\textbf{AIOps}} & 
        \multicolumn{3}{c|}{\textbf{ML}} &
        \multicolumn{3}{c}{\textbf{General}} \\
        Tag & W(\%) & N(\%) &
        Tag & W(\%) & N(\%) &
        Tag & W(\%) & N(\%) \\

        \midrule
        convention & 25.7 & 76.4 & unused & 26.7 & 83.6 & unused & 15.1 & 62.3 \\
        unused & 24.2 & 93.6 & convention & 21.9 & 67.8 & convention & 10.8 & 40.0 \\
        design & 9.1 & 79.1 & cwe & 6.7 & 56.0 & suspicious & 7.5 & 52.5 \\
        cwe & 8.3 & 66.4 & design & 6.5 & 57.6 & pitfall & 7.5 & 47.9 \\
        suspicious & 5.6 & 66.4 & suspicious & 4.4 & 48.0 & cwe & 6.1 & 50.5 \\
        brain-overload & 4.2 & 60.9 & pitfall & 3.4 & 39.8 & design & 5.1 & 38.4 \\
        pitfall & 3.3 & 42.7 & brain-overload & 3.3 & 49.2 & accessibility & 4.5 & 29.2 \\
        error-handling & 2.1 & 49.1 & obsolete & 2.0 & 13.6 & brain-overload & 3.6 & 43.6 \\
        bad-practice & 2.0 & 51.8 & bad-practice & 2.0 & 37.9 & wcag2-a & 3.5 & 27.5 \\
        clumsy & 1.6 & 36.4 & error-handling & 2.0 & 33.6 & confusing & 3.1 & 38.0
    \end{tabular}
    }

    \label{tab:sonarqube_tags}
\end{table}

\noindent\textbf{AIOps and ML projects have similar types of issues in terms of violated SonarQube rules and rule categories.}
% \heng{add discussions to support this claim: use concrete numbers and examples. For example, out of the top 10 rules, how many overlap; out of the top 10 rule categories, how many overlap; how the top ones are similar... }
% \heng{The text below (including the following few paragraphs should be merged into the approach section}
We report the most violated rules and rule categories in Tables \ref{tab:sonarqube_rules} and \ref{tab:sonarqube_tags}. We also define these violated rules and rule categories in AIOps field (based on SonarQube website\footnote{\url{https://rules.sonarsource.com/}}\footnote{\url{https://docs.sonarqube.org/latest/user-guide/built-in-rule-tags/}}) in Tables~\ref{tab:rules_definition} and~\ref{tab:tags_definition}. As shown in the tables, AIOps and ML projects have similar rule and rule category violations. In terms of violated rules, 9 out of the top-10 violated rules are common between AIOps and ML projects. Furthermore, the first two most violated rules (i.e., python:S117 and python:S125) are the same in AIOps and ML projects, with similar weights (17.6\% and 14.6\% for AIOps, and 16.8\% and 15.5\% for ML projects). Also, in terms of most violated rule categories, 9 out of the top-10 violated rule categories are common between AIOps and ML projects.
% \heng{Briefly mention the reuslts of the general baseline and the difference from the AIOps projects} 

Comparing issues between the AIOps projects and the General baseline, 4 of the top-10 violated rules and 7 of the top-10 violated rule tags are common. However, the weight and percentage of projects having these issues in AIOps projects are much higher. As an example, Python:S117 is the most violated rule in both AIOps and General projects. However, the weight and percentage of projects having this violated rule in AIOps are 17.6\% and 59.1\% but in the General baseline are 5.2\% and 19.7\%, respectively.

We provide the complete list of violated rules and rule categories in AIOps projects and other baselines in our replication package. 

\begin{table}[h!]
\centering
        \caption{The most violated SonarQube rules and their definitions in AIOps projects.
        }
    \resizebox{1\textwidth}{!}
    {
    \small

    \begin{tabular}{|ll|}
        \toprule
        \multicolumn{2}{|c|}{\textbf{Rules}} \\
        Name & Definition \\

        \midrule
        python:S117 & Local variable and function parameter names should \\
        & comply with a naming convention. \\
        python:S125 & Sections of code should not be commented out. \\
        python:S1192 & String literals should not be duplicated \\
        python:S905 & Non-empty statements should change control flow or have \\
        & at least one side-effect. \\
        python:S1481 & Unused local variables should be removed. \\
        python:S3776 & Cognitive Complexity of functions should not be too high. \\
        python:S1542 & Function names should comply with a naming convention. \\
        python:S2208 & Wildcard imports should not be used. \\
        python:S5754 & \textit{SystemExit} exception should be re-raised immediately. \\
        python:S100 & Method names should comply with a naming convention. \\
        \midrule
    \end{tabular}
    }

    \label{tab:rules_definition}
\end{table}

\begin{table}[h!]
\centering
        \caption{The most violated SonarQube tags (rule categories) and their definitions in AIOps projects.}
        
    \resizebox{1\textwidth}{!}
    {
    \small

    \begin{tabular}{|ll|}
        \toprule
        \multicolumn{2}{|c|}{\textbf{Tags}} \\
        Name & Definition \\

        \midrule
        convention & Coding convention - typically formatting, naming, whitespace.  \\
        unused & Unused code; e.g., a private variable that is never used. \\
        design & There is something questionable about the design of the code. \\
        cwe & Relates to a rule in the Common Weakness Enumeration. \\
        & For more information, visit \url{https://cwe.mitre.org/}. \\
        suspicious & It's not guaranteed that this is a bug, but it looks suspiciously \\
        & like one. At the very least, the code should be re-examined \\
        & and likely refactored for clarity. \\
        brain-overload & There is too much to keep in your head at one time. \\
        pitfall & Nothing is wrong yet, but something could go wrong in the \\
        & future; a trap has been set for the next person, and they'll \\
        & probably fall into it and screw up the code. \\
        error-handling & Issues related to handling the errors such as \textit{Exception} methods. \\
        bad-practice & The code likely works as designed, but the way it was designed \\
        & is widely recognized as being a bad idea. \\
        clumsy & Extra steps are used to accomplish something that could be \\
        & done more clearly and concisely.

 \\
        \midrule
    \end{tabular}
    }

    \label{tab:tags_definition}
\end{table}

% We symbolize the projects with \textit{p} and the violated rules or tags with \textit{i}. Therefore, project p\textsubscript{1} can have i\textsubscript{1}, i\textsubscript{2}, ... , i\textsubscript{n}, and each of these issues can have differenct occurence of \textit{n}. First, we calculate the frequency of each issue by dividing its occurrence by the total number of issues on that project. So for the first project, we have}

% \[i_{p_{1}}=\frac{n_{i_{1}}}{n_{p_{1}}}+\frac{n_{i_{2}}}{n_{p_{1}}}+...+\frac{n_{i_{n}}}{n_{p_{1}}}\] \heng{is not this formula always equal to 1?}

% We calculate this frequency for all the issues in all the projects. Next, we sum the obtained frequencies for each issue and divide it by the total number of projects. Hence, our equation for the first issue is}
% \[i_{1}=\tfrac{\frac{n_{i_{1}}}{n_{p_{1}}}+\frac{n_{i_{1}}}{n_{p_{2}}}+...+\frac{n_{i_{1}}}{n_{p_{n}}}}{n_{total}}\] \heng{the calculation is not intuitive, why not just calculating the weight by $w = n_{ij} / n_{pj}$ where $n_{ij}$ is the frequency of issue i in project j and $n_{pj}$ is the total number of issues in project j?}

% \heng{Add a take home message to guide the discussions, for example: \textbf{Naming conventions, unused or commented-out code, complex code or poor design, wildcard imports are among the top code smells of AIOps code, which indicate that developers tend to write AIOps code in an \textit{ad hoc} manner}, which hinders their projects' reusability and maintainability.}

\noindent\textbf{Naming convention, commented-out code, duplicated string literals, high complexity of functions, and wildcard imports are among the top violated rules in AIOps projects.}
Regarding the most violated rules in AIOps, the first three are \textit{python:S117}, \textit{python:S125}, and \textit{python:S1192}. The first one has a minor severity and is about the non compliance of naming convention. Shared naming conventions are vital and allow teams to collaborate effectively. The second one has a major severity and is about commenting out the unused sections of code which reduces readability. Instead, unused code should be deleted. The third one has a critical severity and is about using duplicated string literals. It makes the process of refactoring error-prone because the programmer must be sure to update all occurrences of the string.

In addition to these violated rules, three more rules with critical severity exist in the top 10 violated rules of AIOps projects; \textit{python:S3776}, \textit{python:S2208}, and \textit{python:S5754}. \textit{Python:S3776} is about the cognitive complexity of code. It measures how hard the control flow of a function is to understand. Functions with high cognitive complexity will be difficult to maintain. \textit{Python:S2208} is about using wildcard imports (i.e., from module import *). Importing all public names from a module has multiple disadvantages. It can lead to conflicts between local names and imported ones, or same names between two different packages. It also reduces code readability and may cause confusion about which classes are imported and used. \textit{python:S5754} is about handling exceptions. This rule indicates that \textit{SystemExit} exception should be re-raised immediately.

\noindent\textbf{Naming convention, unused or commented-out code, and poor design are among the top violated rule categories.} Regarding the most violated rule categories in AIOps repositories, \textit{convention}, \textit{unused}, and \textit{design} are the worst issues. \textit{Convention} category is about fulfilling coding conventions, including naming functions and variables, complying white-spaces and indentations. The second most violated rule category is \textit{unused}. This category is about unused code that decreases the performance of the system. Some examples of this category are unused assignments, unused private classes, and empty test cases. The third most ignored rule category, \textit{design}, is about the bad design of the software. Some examples of this group are unstable tests, duplicated string literals, and using randomized data in test cases. 

Having rule category issues such as naming conventions, unused code, and bad design indicate that developers tend to write AIOps code in an \textit{ad hoc} manner, which hinders their project’s reusability and maintainability. Also, looking at the most violated rules and rule categories, we realize paying more attention to details can increase the quality
% \Foutse{this statement may be too strong since violating/fixing naming convention may not be critical in terms of quality} 
of AIOps solutions. For example, 3 out of the top-10 violated rules are related to naming conventions. Only following the naming conventions can reduce the number of issues heavily. Furthermore, since most of the rules (9 of top-10) and rule categories (9 of top-10) are common in AIOps and ML projects, AIOps projects can benefit from the quality assessment and quality assurance tools and techniques that have been developed for ML systems. The above points are discussed in more detail in Section~\ref{sec:discussion}.

\newenvironment{mybox3}[1]{%
    \begin{tcolorbox}[title={Summary of RQ3}]%
    }{
    \end{tcolorbox}
}
\begin{mybox3}{}
%Considering the code quality metrics, we understand that the quality of writing AIOps projects is poorer than the baselines. The number of bugs and security hotspots is higher in AIOps projects, while code smells are the primary issue, counting for 82\% of all the found issues. We further diagnose the most commonly violated rules and tags; may the researchers and developers find practical approaches to oppose them.
Although AIOps projects have an adequate amount of comments compared to the ML and General baselines, AIOps projects exhibit poorer quality than the two baselines in terms of a variety of quality metrics (e.g., bugs, code smells, and technical debt). In particular, code smells are the dominant type of issue in AIOps projects. We also identify the most common issues in AIOps projects: naming convention, unused code, and bad design are the top-3 violated rule categories. Moreover, we observe the similarity of violated rules between AIOps and ML projects. Future efforts are needed to reduce the issues identified in this work and improve the quality of AIOps projects, e.g., by designing tools for fixing bugs/smells or deriving coding guidelines. We also encourage future AIOps projects to reduce the \textit{ad-hoc}ness of their code to improve reusability and maintainability.
% We also identify the most violated issues in AIOps projects and the baselines to understand the most common issues that practitioners make when they are developing their systems.
% \heng{don't put approach here: replace it with the corresponding take-home message (check the comment above)} 
%\Foutse{we should formulate a recommendation for researchers and practitioners to help improve this!}\roozbeh{I added a line but I don't think is enough. Any suggestions what should we recommend?}
\end{mybox3}

\section{Discussion}
\label{sec:discussion}

Based on the findings of our study, in this section, we discuss the state of AIOps in open-source projects. We then discuss the current challenges of developing AIOps applications and foresee possible future directions for AIOps researchers and practitioners. Finally, to assess the robustness of our findings, we add a stricter filtering criterion and report the main results considering only more mature projects. We compare the results obtained from these projects with our main findings to make sure our findings are robust enough.

\subsection{AIOps: Where is it now?}
\noindent\textbf{The number of AIOps applications is growing fast, and they are receiving a lot of attention from the open-source community.}
Regarding the results of Section~\ref{rq1_results}, we find that the speed of growth for AIOps applications is faster than that for general-purposed projects on GitHub. We further find that AIOps applications are receiving more attention regarding GitHub metrics compared to machine learning and general-purposed applications. This attention towards the AIOps area encourages researchers and practitioners to develop new technical innovations in various areas of AIOps. We discuss some of the potential future directions below.

\bigskip 
\noindent\textbf{Monitoring data - especially logs, performance metrics, and network-traffic data - are the most common data types of AIOps applications.}
Regarding the results of Section~\ref{rq2_results}, we find that almost 70\% of the projects use logs, performance metrics, or network-traffic data. The usage of other types of input data (e.g., alarms) is limited in open-source AIOps applications. This finding is in line with Notero et al.~\citeyearpar{notaro2021systematic} that many AIOps-related papers also use event logs and performance metrics as their data sources. Future efforts may be invested to investigate how to optimally process these common data types in various downstream tasks. For example, time-series data representation techniques~\citep{wilson2017data} and time-series segmentation techniques~\citep{lovric2014algoritmic} may be explored to improve data representations of such time-series data.

\bigskip 
\noindent\textbf{Classical machine learning is the most common technique in AIOps applications.}
We find that over half (54\%) of the AIOps applications use classical machine learning as their primary technique, while only 14\% use deep learning techniques. AIOps practitioners also develop AIOps applications using a variety of methods, including time-series and statistical models. However, some techniques (e.g., natural language processing) are not used frequently. We suggest that future AIOps solutions may further utilize more sophisticated techniques of deep learning and natural language processing. However, as suggested by Li et al.~\citeyearpar{li2020predicting} and Lyu et al.~\citeyearpar{lyu2021towards}, AIOps models should be trustable and interpretable; hence, using black-box models without the ability to interpret is not recommended. Therefore, we also suggest that future AIOps solutions should always experiment with classical solutions instead of simply assuming deep learning techniques are the optimal solution.

\bigskip 
\noindent\textbf{Anomaly detection is the most common goal of AIOps applications.}
60\% of open-source AIOps applications’ primary goal is to detect anomalies. Monitoring, anomaly prediction, root cause analysis, and providing AIOps infrastructure are the other common goals of AIOps applications. Only 5\% of projects provide a public AIOps dataset, and only 1\% of projects provide self-healing features. Our results align with the findings of Notaro et al.~\citeyearpar{notaro2021systematic}, where they observe that 62\% of AIOps papers are associated with failure management, including failure detection, failure prediction, and root cause analysis.

\bigskip 
\noindent\textbf{AIOps applications suffer from more code quality issues than machine learning and general-purpose-based projects.}
Considering our results in Section~\ref{sec:rq3_results}, AIOps applications have a higher number of quality issues (e.g., bugs and code smells) compared to the baselines. We list the most common violated issues in AIOps applications in Tables~\ref{tab:sonarqube_rules} and~\ref{tab:sonarqube_tags}. AIOps practitioners may pay attention to these issues in their applications and address them to reduce the risk of future problems.

\subsection{AIOps: Challenges and future directions.}
\textbf{More benchmarking datasets could be designed for AIOps applications.}
We find that only 5\% of the projects aim to design and publish AIOps benchmark datasets. As also mentioned by Bogatinovski et al.~\citeyearpar{bogatinovski2021artificial}, there is a lack of good and public AIOps benchmarks, where different AIOps approaches could be compared. As most of the AIOps applications use logs, performance metrics, and network-traffic data as their input data source, we suggest AIOps researchers design and publish real-world and public datasets of these data types so other AIOps practitioners and researchers could leverage them in their applications. In the future, we also plan to design a benchmarking framework that can help future work generate customized datasets.

\bigskip 
\noindent\textbf{Natural Language Processing (NLP) techniques can receive more attention in AIOps in the coming years.}
Our results in Section~\ref{sec:rq2} indicate that only 2\% of the studied projects use NLP techniques. On the other hand, approximately one-third of the projects use logs as their input data. Prior research has shown that logs can be approximately represented as natural language text since the logs are generated by logging statements in the source code written by humans~\citep{hindle2016naturalness}. Other works also illustrate that logs are even more predictable and repetitive than natural languages, such as English~\citep{tu2014localness, he2018characterizing, yao2020study}. Hence, we suggest that AIOps projects can benefit from leveraging the advances in NLP techniques to analyze and model the input data such as logs. 
% \heng{also briefly mention this point in the results section when discussing the techniques, to make the results-discussion connection smoother. Same for the discussion points below}

\bigskip 
\noindent\textbf{More attention should be paid to increasing automation and reducing human interventions in AIOps solutions.}
Regarding the goal of the studied projects, most projects aim to detect and predict anomalies, monitor, and analyze the root causes of failures. All these systems need human interventions when the goal is reached. For example, when an anomaly is detected, an operator should decide the next required action to handle the situation. As demonstrated in Figure~\ref{fig:goals}, only 1\% of projects aim for self-healing, meaning that the system can make necessary changes when needed without human interventions. We believe AIOps solutions should move toward becoming more automated, detecting/predicting the incidents and resolving them autonomously. Self-healing techniques may get more attention and be integrated with detection/prediction techniques~\citep{ding2012healing, ding2014mining, lou2017experience}.

\bigskip 
\noindent\textbf{Paying more attention to simple details can increase the quality of AIOps solutions.}
We identify the most common issues in AIOps projects in Section~\ref{sec:rq3_results}. When looking at the most violated rules, some of them are challenging and time-consuming to fix and need structural changes in code, such as \textit{python:S3776} (cognitive complexity of functions) and \textit{python:S1192} (duplication of string literals). On the other hand, some are easy to handle. 3 of the 10 most violated rules are related to naming conventions. Only following the naming conventions can reduce the number of issues heavily. Furthermore, rules such as \textit{python:S2208} (using wildcard imports) and \textit{python:S1481} (removing unused variables) are quick to determine and easy to fix. Identifying the most common issues help practitioners and researchers in the field to become aware of them and take measurements to reduce them. 
% \heng{use a short phrase to remind people what it means, same for the next one} 
% \Roozbeh{should we mention the url of our github again (we have mentioned it in intro)}\heng{no need to mention the url again. however, this sentence should be moved to the results section where you discuss the top rules.}

\bigskip 
\noindent\textbf{Most of the tools and techniques used for the quality assurance of machine learning systems can also be used for AIOps solutions.}
Our results indicate that a high proportion of issues in ML and AIOps techniques are the same (90\% of most violated rules and 90\% of most violated rule categories are shared). Besides, the quality assurance of ML systems has received special attention in the previous years, and different approaches have been built to preserve the quality of ML-based systems~\citep{nakajima2018quality, poth2020quality,Nikanjam21,NikanjamDRL,Braiek22,TAMBON2023107129}. Hence, we believe most of the developed tools and techniques for quality assurance in ML systems can be utilized in AIOps techniques. Applying these tools and techniques may reduce the quality issues mentioned in Section~\ref{sec:rq3_results}.

\bigskip 
\noindent\textbf{Open-source AIOps solutions can benefit from broader use-cases and scenarios.}
As described in previous works~\citep{dang2019aiops, gartnerwebsite}, AIOps can encompass a wide range of scenarios, from predicting the future status of systems to improving the productivity of engineers. However, the scope of current AIOps projects available on GitHub appears to be limited, focusing primarily on certain aspects with a skewed distribution. For example, we find that anomaly detection is the dominant goal of AIOps projects with 60\% of use cases. To address this disparity, AIOps researchers and practitioners may pay more attention to other AIOps scenarios, such as minimizing engineers’ tedious tasks and better system automation. This can contribute to advancing this interdisciplinary field and realizing its full potential across diverse use cases and applications.

\subsection{The impact of using a stronger filtering criterion.}
\label{sec:case_study}

\begin{figure*}[]
\centering
\includegraphics[width=0.75\textwidth]{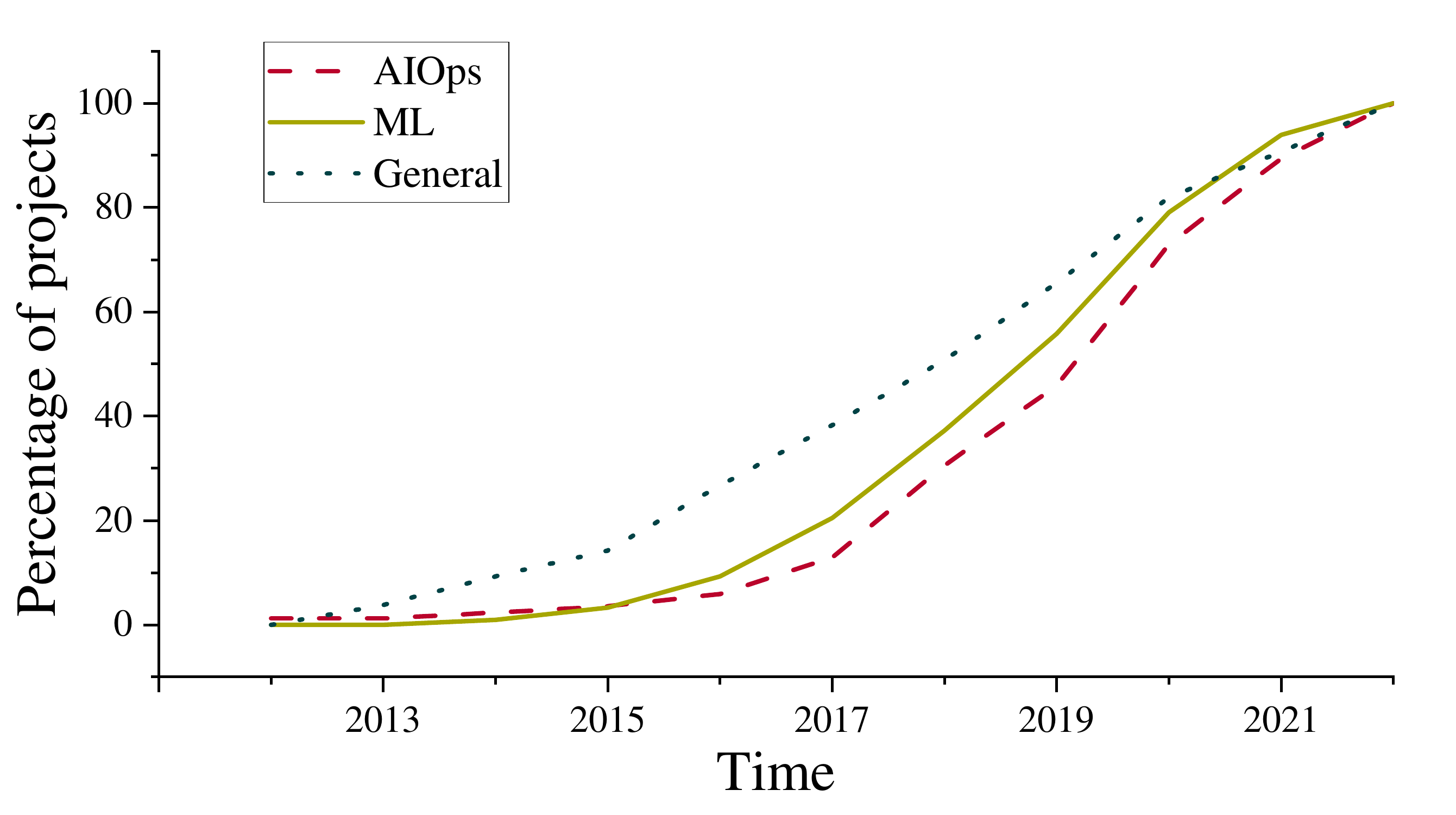}
\caption{The cumulative distribution of the creation time of AIOps and baseline projects with the new filtering criterion.}
\label{fig:d_creation-date}    
\end{figure*}

\begin{table}[]
    \centering
    \caption{The top-5 languages of AIOps and baseline projects with the new filtering criterion.}
    \resizebox{1\textwidth}{!}
    {
    \small
    \begin{tabular}{ll|ll|ll}
        \toprule
        \multicolumn{2}{c|}{\textbf{AIOps}} &  \multicolumn{2}{c|}{\textbf{ML}} &
        \multicolumn{2}{c}{\textbf{General}} \\
        Language & Usage $(\%)$ & 
        Language & Usage $(\%)$ & 
        Language & Usage $(\%)$ \\

        \midrule
        Python & 65.9 & Python & 81.4 & Python & 18.6\\
        Java & 10.6 & C++ & 2.3 & JavaScript & 15.3 \\
        Go & 4.7 & HTML & 2.3 & Java & 7.7 \\
        HTML & 3.5 & JavaScript & 1.9 & Typescript & 6.6 \\
        JavaScript & 2.4 & MATLAB & 1.4 & C & 6.0
    \end{tabular}
    }
    \label{tab:d_languages}
\end{table}

To assess the robustness of our findings, we add an extra filter to both AIOps and baseline projects and discuss the main findings of RQ1 and RQ3. The intention is to confirm that the obtained results are meaningful insights and are robust. To do so, we analyze the distribution of the AIOps projects based on their number of stars. Then, we use the elbow curve to determine a threshold to filter the projects, similar to prior work which uses the elbow curve to filter the AutoML tools to be studied~\citep{majidi2022empirical}. Specifically, we create a line chart of the sorted number of stars of the projects and find the elbow point. The elbow point is the distinct bend or “elbow” in a line chart that indicates a change in the distribution~\citep{kodinariya2013review}. We find that the elbow point in the number of stars of AIOps projects is 5. Hence, we add extra filtering and only consider projects that have the number of stars greater or equal to 5 (stars: >= 5). We also add this filtering criterion to the ML and General baselines to ensure consistency. Adding this criterion reduces the number of AIOps projects from 119 to 85, ML baseline from 383 to 215, and General baseline from 385 to 183.

\begin{table}
\centering
\caption{Detailed results of \textit{Mann–Whitney U} and \textit{Cliff's delta} tests on projects' GitHub metrics with the new filtering criterion.}
\begin{threeparttable}
   \begin{tabular}{l|ll|ll}
\hline
  \multirow{2}{*}{Metric} & \multicolumn{2}{c|}{\textbf{AIOps vs. ML}} & \multicolumn{2}{c}{\textbf{AIOps vs. General}} \\ 
  \cline{2-5}
  & \multicolumn{1}{l|}{p-value} & effect size & \multicolumn{1}{l|}{p-value} & effect size \\ \hline
Stars & \multicolumn{1}{l|}{0.02} & ** & \multicolumn{1}{l|}{0.00} & *** \\
Forks & \multicolumn{1}{l|}{0.01} & ** & \multicolumn{1}{l|}{0.00} & *** \\
Pull requests & \multicolumn{1}{l|}{0.02} & ** & \multicolumn{1}{l|}{0.81} & - \\
Size & \multicolumn{1}{l|}{0.14} & - & \multicolumn{1}{l|}{0.00} & *** \\ \midrule
\end{tabular}
    \begin{tablenotes}
  \small
  \item \textit{Mann–Whitney U} results are shown in \textit{p-value} columns. If the sets have statistically different distributions, the \textit{Cliff's delta} results are shown in \textit{effect size} columns.

  *: negligible effect
  **: small effect
  ***: medium effect
     \end{tablenotes}
\end{threeparttable}
\label{tab:d_github_metrics_statistics}
\end{table}

\begin{figure*}[]
  \centering
  \subfloat[]{\includegraphics[width=0.45\textwidth]{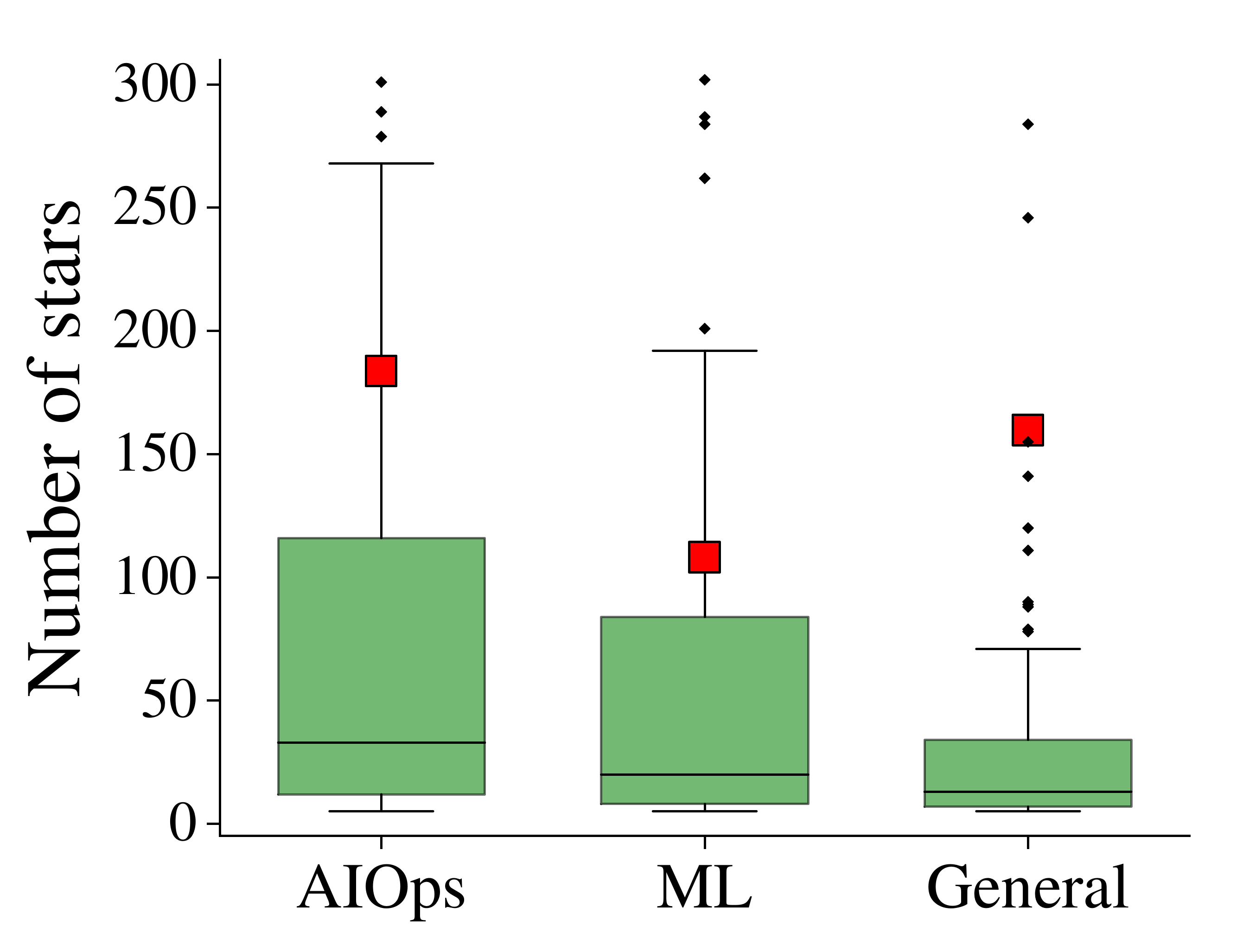}\label{fig:d_stars}}
  \subfloat[]{\includegraphics[width=0.45\textwidth]{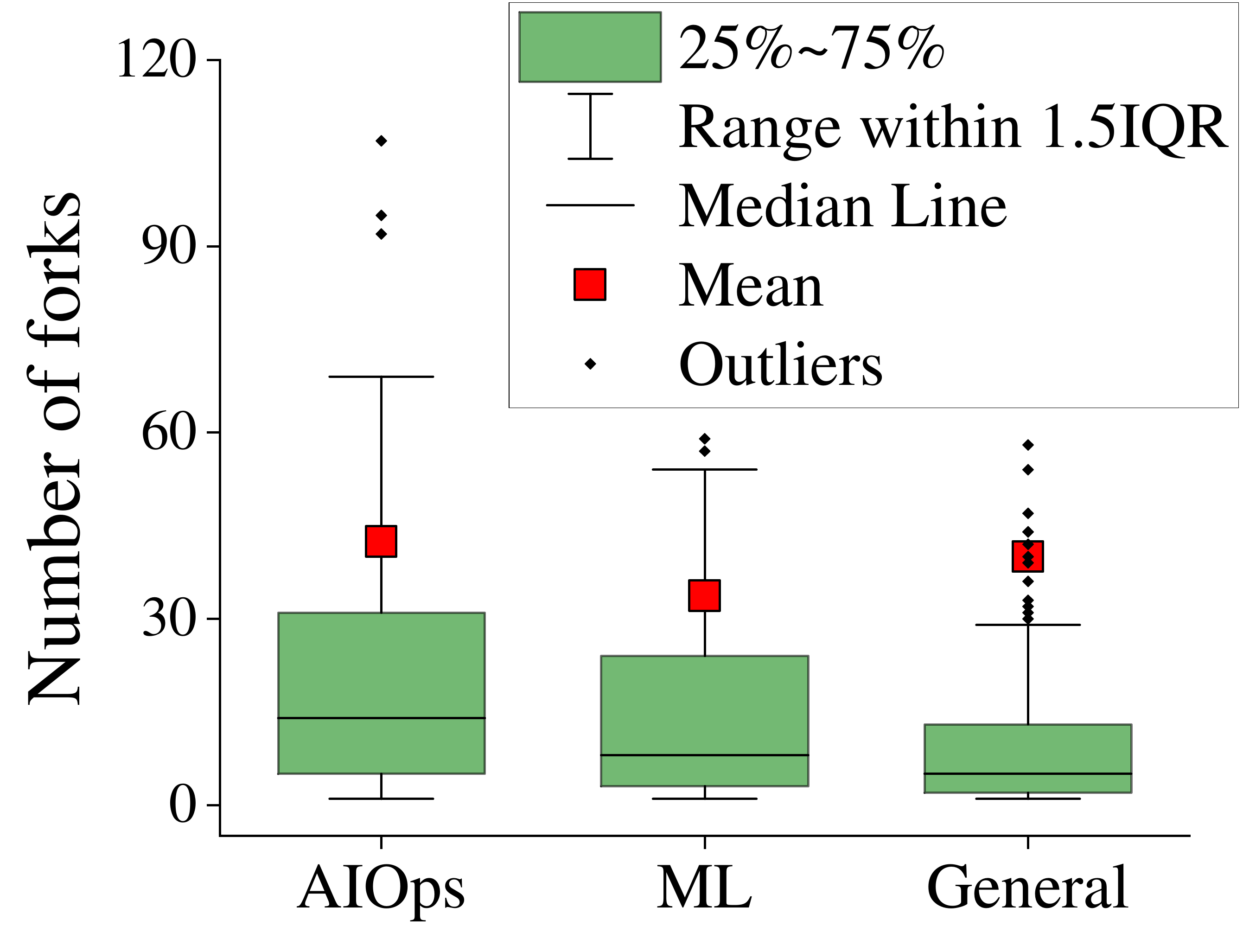}\label{fig:d_forks}} \\
  \subfloat[]{\includegraphics[width=0.45\textwidth]{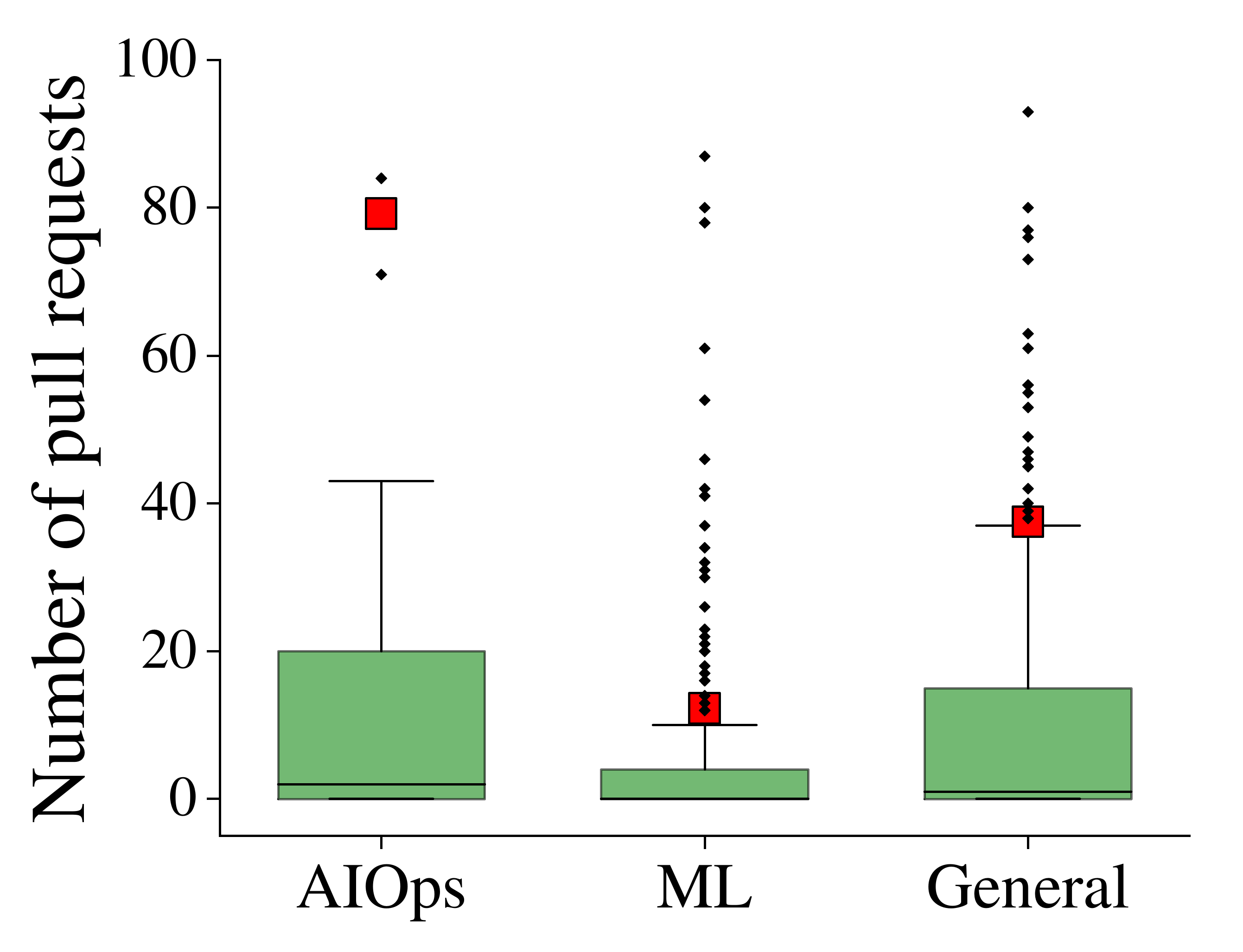}\label{fig:d_pull_requests}}
  \subfloat[]{\includegraphics[width=0.45\textwidth]{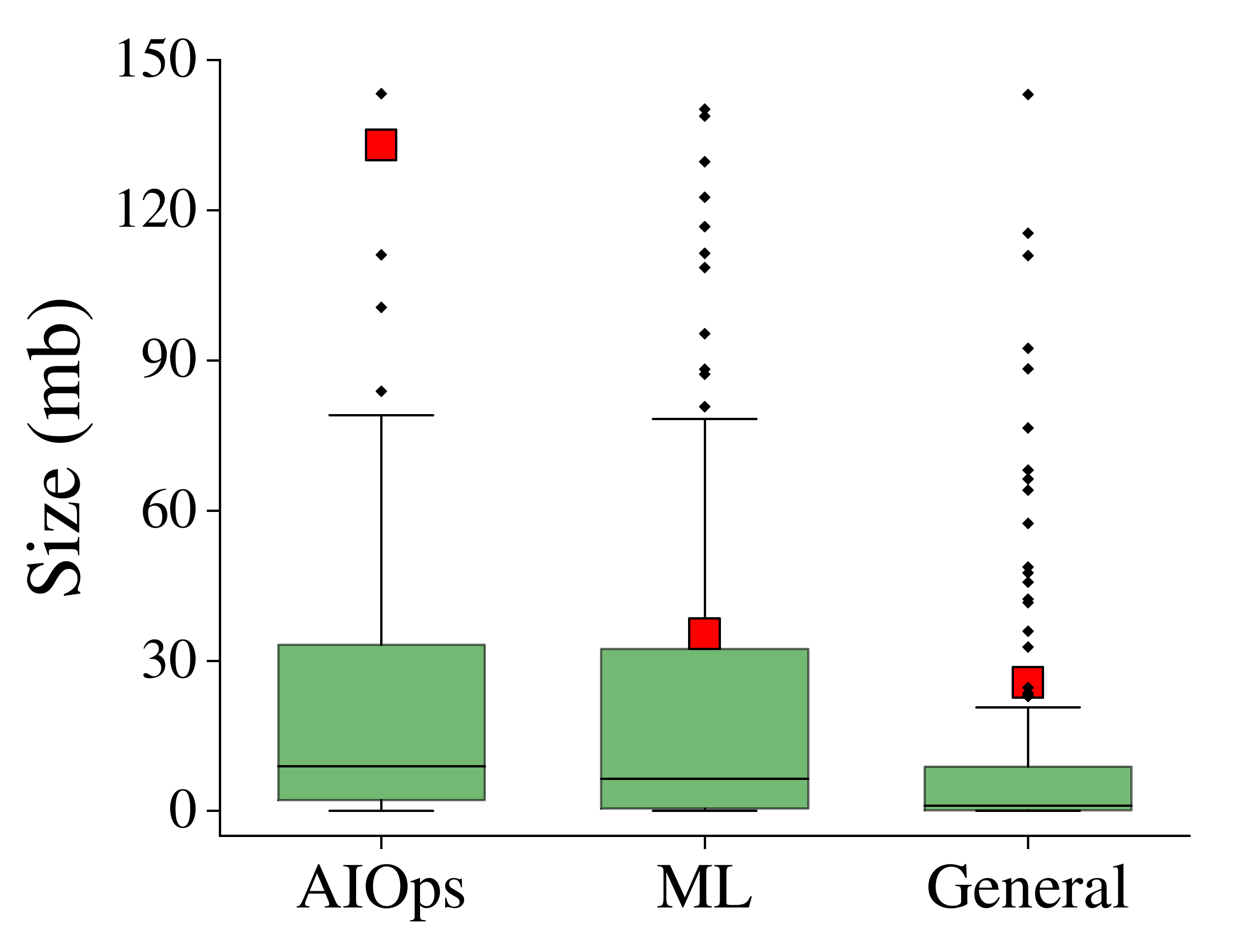}\label{fig:d_size}}
  \caption{Box plots of GitHub metrics for AIOps and baseline projects with the new filtering criterion.}
\label{fig:d_github_box_plot}
 \vspace{-1em}
\end{figure*}

\bigskip 
\noindent\textbf{Results of RQ1 with the new filtering criterion.}
We report the main results of RQ1 in regard to the new filtering criterion. Figure~\ref{fig:d_creation-date} presents the percentage of projects created in and before each year for AIOps and baseline projects. As illustrated, AIOps projects are growing faster than ML and General baselines, which is also the same trend in our initial results (cf. Figure~\ref{fig:creation-date}).

Table~\ref{tab:d_languages} shows the top-5 most common languages in the AIOps and baseline sets. The dominant programming language in AIOps projects is Python, with usage of 65.9\%, followed by Java with 10.6\%. The most used language in the ML projects is Python with 81.4\%, and the most common language in General baseline is Python with 18.6\%, followed by JavaScript with 15.3\%. All these trends are also visible in our initial results (cf. Table~\ref{tab:languages}).

Figure~\ref{fig:d_github_box_plot} depicts the box plots of some of the GitHub metrics for the AIOps and baseline projects. Table~\ref{tab:d_github_metrics_statistics} illustrates the p-value and effect size of the same metrics of AIOps projects compared to the baselines. Regarding the number of stars and forks, AIOps projects demonstrate higher popularity than the baselines, as evidenced by both the median and mean values. The median value of stars in AIOps, ML, and General sets are 33, 20, and 13, respectively. Also, the median value of forks in AIOps, ML, and General sets are 14, 8, and 5, respectively.

Table~\ref{tab:d_github_metrics_statistics} shows that the number of stars and forks have a small effect size compared to the ML baseline and a medium effect size compared to the General baseline. Considering the number of pull requests, as shown in Figure~\ref{fig:d_pull_requests}, AIOps projects experience more of them, with a median value of 2, compared to 0 in ML and 1 in General. The mean value of pull requests in AIOps projects is also much higher than the baselines. Statistical tests suggest that the effect size of pull requests in AIOps compared to the ML baseline is small, and there is not a significant difference between AIOps and the General baseline. According to~\ref{fig:d_size}, AIOps projects have a larger size compared to the baselines. With a median size of 9.0 MB in AIOps projects, they tend to be 9 times larger than General baselines with a median size of 1.1 MB. Statistical tests suggest that the difference between AIOps and General baseline in terms of their size is medium. However, there is no significant difference between AIOps and ML baseline. All the mentioned results and trends are consistent with our initial results (cf. Figure~\ref{fig:github_box_plot} and Table~\ref{tab:github_metrics_statistics}). In summary, when we apply a stricter project filtering criterion, our main findings in RQ1 still hold, indicating the robustness of our findings.

\begin{table}
\centering
\caption{Detailed results of \textit{Mann–Whitney U} and \textit{Cliff's delta} tests on projects' code quality with the new filtering criterion.}
\begin{threeparttable}
   \begin{tabular}{l|ll|ll}
\hline
  \multirow{2}{*}{Metric} & \multicolumn{2}{c|}{\textbf{AIOps vs. ML}} & \multicolumn{2}{c}{\textbf{AIOps vs. General}} \\ 
  \cline{2-5}
  & \multicolumn{1}{l|}{p-value} & effect size & \multicolumn{1}{l|}{p-value} & effect size \\ \hline
Bugs & \multicolumn{1}{l|}{0.01} & ** & \multicolumn{1}{l|}{0.01} & ** \\
Code smells & \multicolumn{1}{l|}{0.00} & ** & \multicolumn{1}{l|}{0.00} & *** \\
Technical debt & \multicolumn{1}{l|}{0.00} & ** & \multicolumn{1}{l|}{0.00} & *** \\
LOC & \multicolumn{1}{l|}{0.02} & ** & \multicolumn{1}{l|}{0.01} & ** \\ \midrule
\end{tabular}
    \begin{tablenotes}
  \small
  \item \textit{Mann–Whitney U} results are shown in \textit{p-value} columns. If the sets have statistically different distributions, the \textit{Cliff's delta} results are shown in \textit{effect size} columns.

  *: negligible effect
  **: small effect
  ***: medium effect
     \end{tablenotes}
\end{threeparttable}
\label{tab:d_sonarqube_metrics_statistics}
\end{table}

\begin{figure*}[]
  \centering
  \subfloat[]{\includegraphics[width=0.45\textwidth]{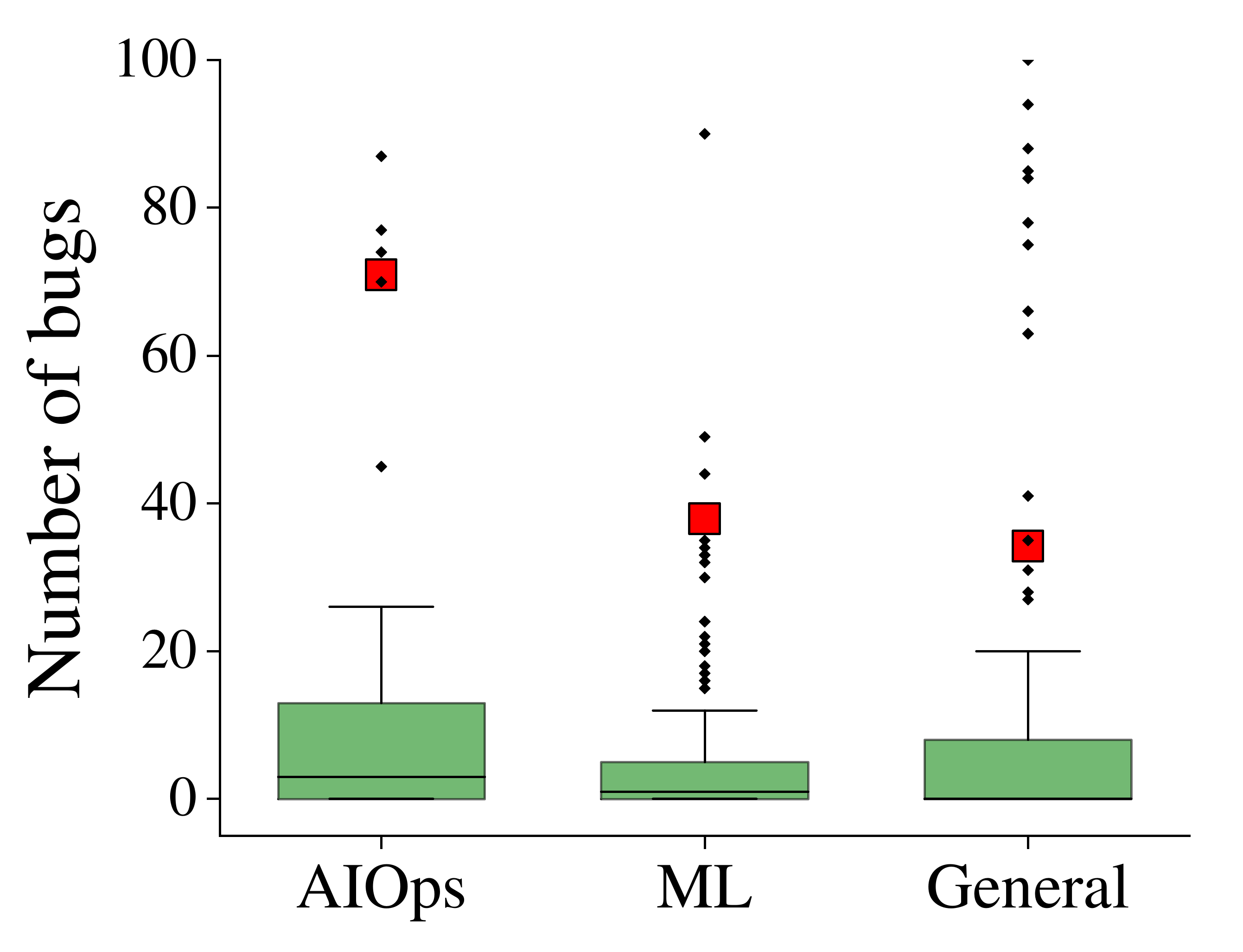}\label{fig:d_bugs}}
  \subfloat[]{\includegraphics[width=0.45\textwidth]{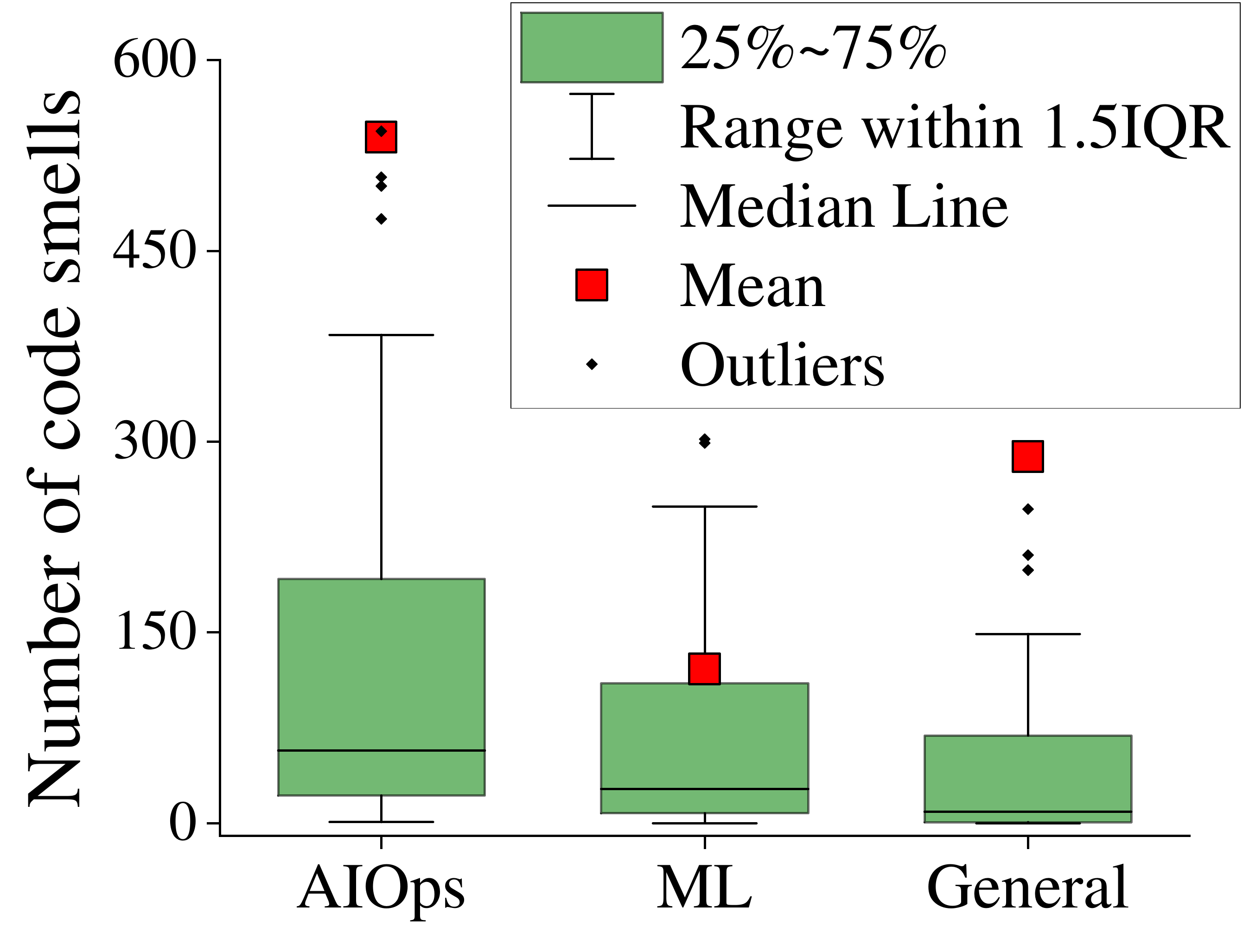}\label{fig:d_smells}}\\
  \subfloat[]{\includegraphics[width=0.45\textwidth]{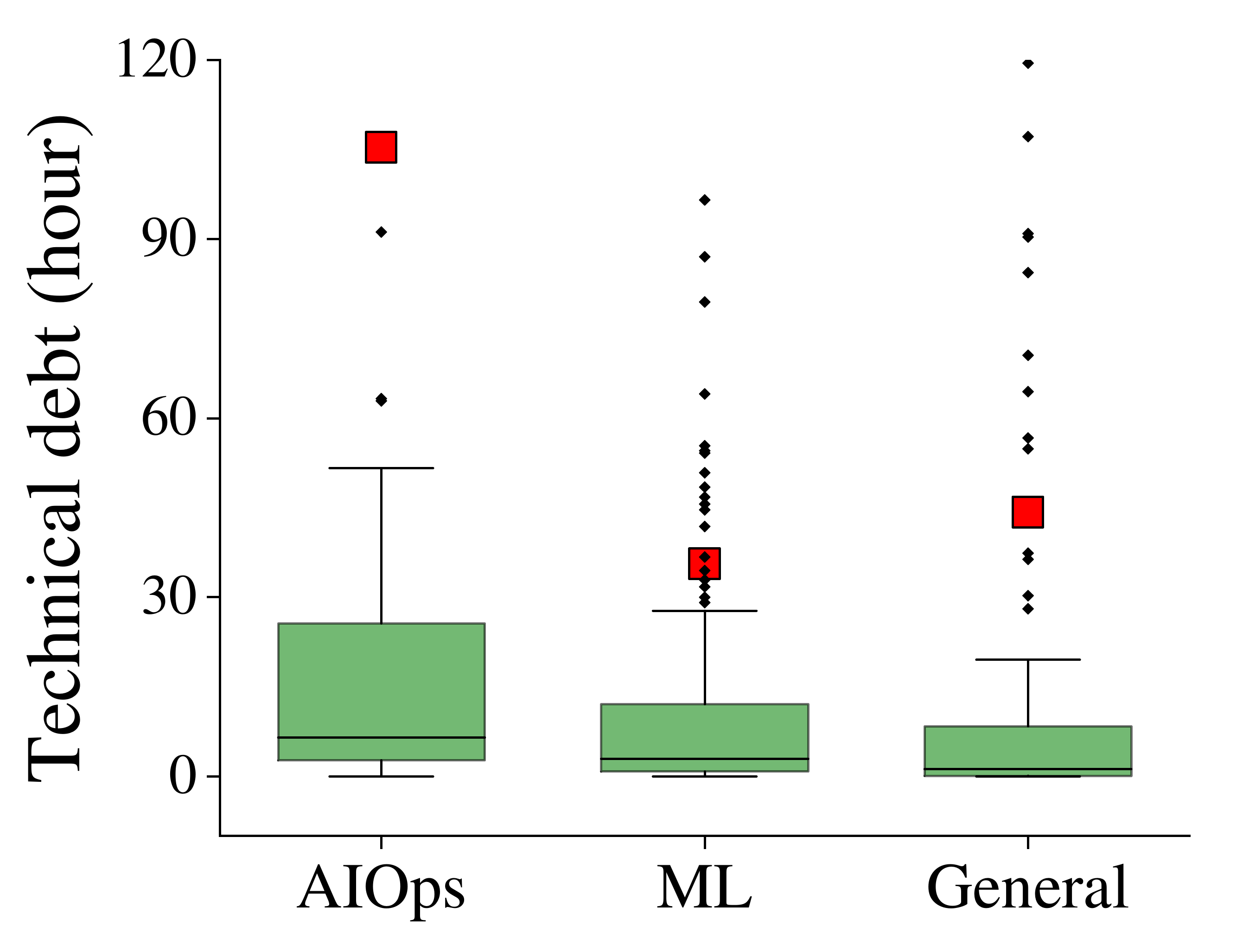}\label{fig:d_debt}}
  \subfloat[]{\includegraphics[width=0.45\textwidth]{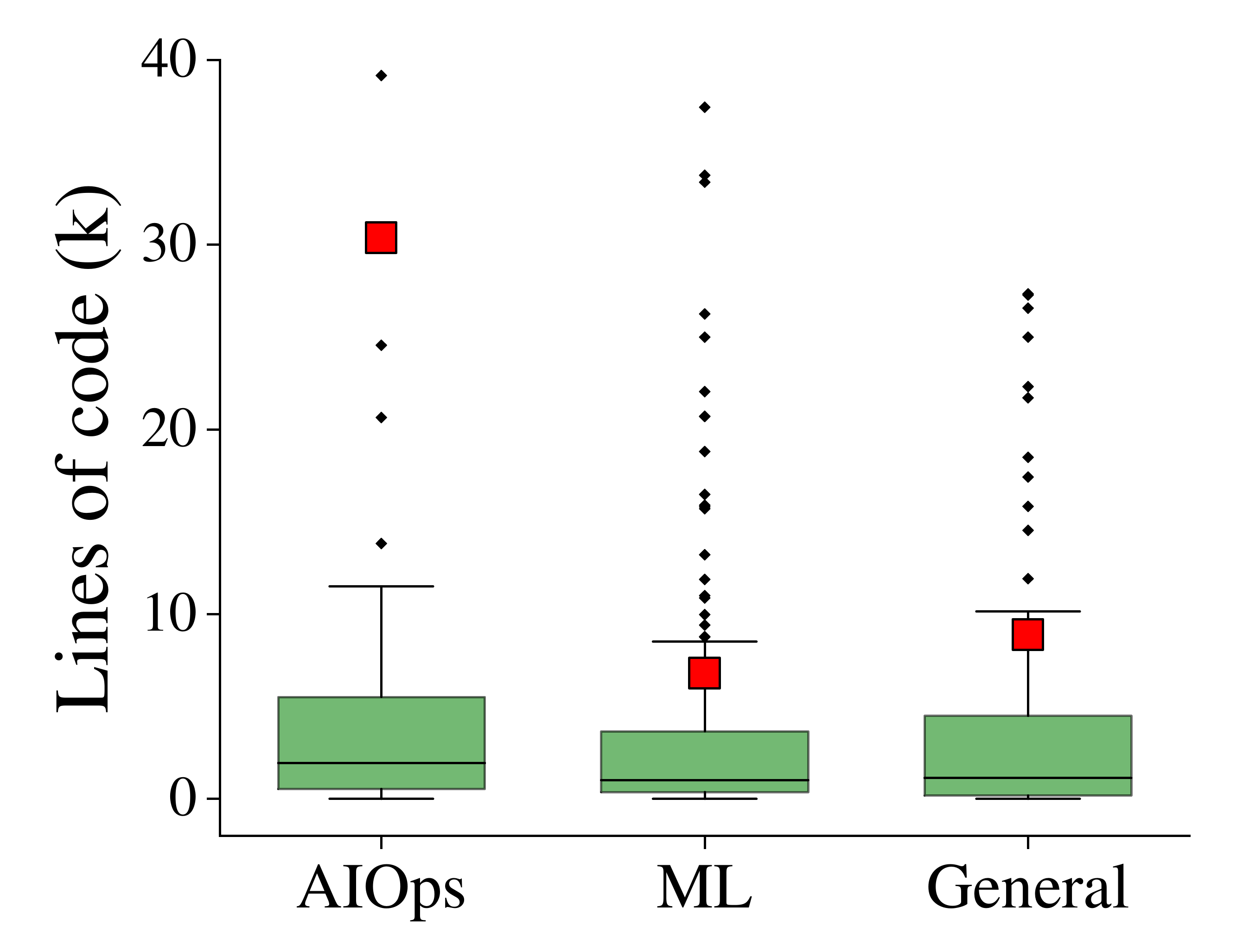}\label{fig:d_loc}}
  \caption{Box plots of code quality metrics for AIOps and baseline projects with the new filtering criterion.}
\label{fig:d_sonarqube_box_plot}
 \vspace{-1em}
\end{figure*}

\bigskip 
\noindent\textbf{Results of RQ3 with the new filtering criterion.}
We also report the main results of RQ3 with the more mature projects that are selected with the stricter filtering criterion. Figure~\ref{fig:d_sonarqube_box_plot} shows the box plots of some of the code quality metrics extracted from SonarQube for AIOps and baseline projects. Please check Table~\ref{tab:sonar_definition} for the definitions of these code quality metrics. Besides, Table~\ref{tab:d_sonarqube_metrics_statistics} presents the p-value and effect size of the code quality metrics of AIOps projects compared to the baselines. As shown in Figure~\ref{fig:d_sonarqube_box_plot}, AIOps projects experience more bugs and code smells than the baselines. The median amount of bugs in AIOps is 3, while it is 1 in ML and 0 in General baselines. Besides, the median amount of code smells is 57 in AIOps, while it is 27 in ML and 9 in General baselines. Furthermore, the amount of technical debt in AIOps projects is much higher than the baselines (the median value of 6.5 hours for the AIOps set, 2.9 hours for the ML set, and 1.2 hours for the General set). The LOC in AIOps projects is also higher than the baselines.

While all these trends are also visible in the initial results (cf. Figure~\ref{fig:sonarqube_box_plot}), it seems that the more mature AIOps projects (i.e., the projects with the new filtering criterion) have more LOC and thus more issues in terms of bugs, code smells, and technical debt.

\begin{table}[t]
    \centering
    \caption{The top-5 violated SonarQube rules and tags (rule categories) for AIOps projects and the baselines with the new filtering criterion. ``W'' represents the weight of tags, and ``N'' is the percentage of projects with that tag.}
    \resizebox{1\textwidth}{!}{
        \small
        \begin{tabular}{lll|lll|lll}
            \toprule
            \multicolumn{3}{c|}{\textbf{AIOps}} &
            \multicolumn{3}{c|}{\textbf{ML}} &
            \multicolumn{3}{c}{\textbf{General}} \\
            Rule & W(\%) & N(\%) &
            Rule & W(\%) & N(\%) &
            Rule & W(\%) & N(\%) \\
            \midrule
            python:S117 & 17.1 & 58.7 & python:S117 & 17.5 & 60.7 & python:S117 & 3.7 & 15.3 \\
            python:S125 & 12.7 & 66.7 & python:S125 & 14.4 & 64.2 & Web:S5254 & 3.2 & 24.7 \\
            python:S1192 & 6.7 & 65.3 & python:S1192 & 7.8 & 60.2 & python:S125 & 2.9 & 18.0 \\
            python:S905 & 4.8 & 22.7 & python:S1481 & 6.3 & 57.7 & javascript:S1117 & 2.8 & 18.0 \\
            python:S1481 & 3.9 & 60.0 & python:S905 & 5.5 & 23.9 & Web:S1827 & 2.7 & 10.0 \\
            \bottomrule
            % Add a small space between the two sets of data
            \addlinespace[0.5em]
            % Next set of data starts here
            % Adjust the rows below to represent the new data
            Tag & W(\%) & N(\%) &
            Tag & W(\%) & N(\%) &
            Tag & W(\%) & N(\%) \\
            \midrule
            % Add your new data rows here
            convention & 25.7 & 76.4 & unused & 25.0 & 85.1 & unused & 12.1 & 58.0 \\
            unused & 24.2 & 93.6 & convention & 22.2 & 69.2 & convention & 8.9 & 34.7 \\
            design & 9.1 & 79.1 & design & 7.1 & 63.2 & pitfall & 7.3 & 44.7 \\
            cwe & 8.3 & 66.4 & cwe & 5.8 & 59.2 & suspicious & 7.0 & 50.0 \\
            suspicious & 5.6 & 66.4 & suspicious & 5.1 & 55.7 & accessibility & 4.8 & 32.7 \\
            % Add more rows as needed for the new data
            \bottomrule
        \end{tabular}
    }

    \label{tab:d_sonarqube_rules_tags}
\end{table}

Considering the statistical tests presented in Table~\ref{tab:d_sonarqube_metrics_statistics}, we can observe that AIOps projects are statistically different from both baselines in the mentioned quality metrics (i.e., number of bugs, number of code smells, technical debt, and lines of code). Besides, the effect sizes of the differences are small or medium. Compared to the initial results presented in Table~\ref{tab:sonarqube_metrics_statistics}, a larger effect size can be seen while comparing AIOps and General baseline.

We also report the most violated rules and rule categories in Table~\ref{tab:d_sonarqube_rules_tags}. These rules and rule categories are defined in Tables~\ref{tab:rules_definition} and~\ref{tab:tags_definition}, respectively. Comparing the new criterion (Table~\ref{tab:d_sonarqube_rules_tags}) with the initial criterion (Tables~\ref{tab:sonarqube_rules} and~\ref{tab:sonarqube_tags}), no change is seen for the AIOps projects. The top-5 rules and rule categories remain the same, with similar weight (i.e., W(\%) column in Table~\ref{tab:d_sonarqube_rules_tags}) and usage (i.e., N(\%) column in Table~\ref{tab:d_sonarqube_rules_tags}). Another finding (alike to the initial results) is the similarity between AIOps and ML projects in terms of their quality issues. Overall, the results of our stricter filtering criterion show that our initial findings are robust, as the trends and main findings remain the same.
\section{Threats to Validity} 
\label{sec:threats}
This section discusses threats to the validity of our results.

\noindent\textbf{External validity.}
In this work, we identified and studied a set of AIOps projects on GitHub. These projects may not cover all AIOps projects on GitHub, those hosted on other platforms, or private projects. Besides, as ``AIOps’’ is a new terminology, not all AIOps projects mention the keyword of ``AIOps’’ in their repositories. Future work examining other sources of AIOps projects (e.g., those published in the literature or closed-source projects) can complement our results. To broaden the generalizability of our work and maximize the coverage of AIOps projects, we follow a process that combines automated search (two rounds), keyword expansion, manual verification, and filtering. Two authors of the paper carefully examined each of the candidate projects to select the ones for our study. The third author steps in to resolve any disagreement. 
Nevertheless, future work can leverage our replication package and extend our study by analyzing more AIOps projects.

To select the projects, we consider repositories with stars and forks greater than or equal to 1. We selected this criteria to have a balance between filtering the low-bar projects and having a proper portion of projects to study. Having stricter filtering criteria would heavily reduce the number of AIOps projects, as there are not many AIOps projects available on GitHub. However, having low filtering criteria could potentially lead to biased results and lack of robustness in the findings. To address this concern and enhance the robustness of our findings, we conducted a sensitivity analysis (cf. ~\ref{sec:case_study}). In this sensitivity analysis, we validated our results using a stricter filtering criterion, providing additional evidence to support the reliability of our findings. The analysis shows that our results remain consistent and robust even with stricter filtering criteria.

We also collect the ML baseline based on two keywords of ``machine learning'' and ``deep learning''. However, having these two keywords may not include all machine learning repositories on GitHub. Future work can expand our results by analyzing more projects extracted from more diverse keywords.

% \heng{Follow the example above to discuss other threats: first admit the threat (also give the context, like in which RQ), then discuss what we did to mitigate the threat, finally suggest future work (optional).}

\noindent\textbf{Internal validity.} We study the code quality metrics of the AIOps projects using a set of metrics (e.g., code smells). Nevertheless, these metrics may not accurately represent the quality of the projects. To reduce the effect of this threat, we leveraged a variety of metrics that represent different aspects of each project. These metrics have also been used in other articles to measure code quality~\citep{tan2018towards, businge2019studying, lenarduzzi2019diffuseness}.

We rely on the code issues detected by SonarQube to answer our RQ3. We choose SonarQube since it is one of the most used tools for analyzing code quality~\citep{lenarduzzi2017analyzing, lenarduzzi2018survey}. We are aware that SonarQube might have false positives or false negatives. In order to evaluate the accuracy of code quality metrics extracted through SonarQube, we conduct an manual verification. We select the 10 most violated SonarQube rules in AIOps projects presented in Table~\ref{tab:rules_definition}. For each of the rules, we randomly select 20 issues from 20 different projects (from all the AIOps and baseline projects) where that rule violation is present. Therefore, for each of the 10 rules,20 instances are analyzed, and totally we analyze 200 issues. In our analysis, we look for false positives (i.e., any issue that SonarQube has detected but is not actually an issue). We did not find any false positives in the analyzed issues. Our results align with SonarQube’s claim that it has zero false-positives for code smells and bugs\footnote{\url{https://docs.sonarqube.org/latest/user-guide/rules/}}.

% \heng{Talk about threat: In RQ3, we study the code quality of the AIOps projects using a set of metrics (e.g., code smells). These metrics may not accurately represent the quality of the projects}.
\noindent\textbf{Construct validity.} To expand our set of AIOps projects, we perform pattern mining and choose 4 out of 194 pairs of keywords. We choose these 4 pairs of keywords after a systematic process and based on the discussions between the authors, in order to reduce the bias. %We did not choose these keywords subjectively, and 
All of the keywords are among the most frequently used keywords.

To answer RQ2, we use qualitative analysis to categorize the input data, analysis techniques, and goals of each AIOps project. Our results may be biased by personal deductions like any other qualitative study. In order to mitigate this threat, two authors of the paper performed the qualitative analysis carefully and followed a 5-step process to deduce the categories. We achieve a Cohen's kappa value of 0.84 which shows a strong and reliable agreement. 
\section{Related Work}
\label{sec:related_work}

%Our study aims to understand the current state of AIOps approaches. We extract our desired AIOps projects from GitHub and analyze their GitHub metrics. We also analyze the source code of these projects to discover their quality. 
This work studies AIOps projects on GitHub and analyzes the quality of the projects using SonarQube. Thus, we discuss the related work on the following three aspects.

% \heng{Each related work subsection can be condensed by further grouping the studies and using one or two examples for each group. The first and second subsections are a bit long.}

\subsection{AIOps solutions.}
In recent years and with the emergence of AIOps, more and more studies have been conducted in this field. Prior works have come up with different solutions for various problems. 
Anomaly detection \citep[e.g.,][]{brown2018recurrent, nedelkoski2019anomaly}, failure prediction~\citep[e.g.,][]{notaro2021survey, zhao2021predicting}, ticket management \citep[e.g.,][]{xue2016managing, xue2018spatial}, self-healing \citep[e.g.,][]{lou2017experience, lou2013software, ding2014mining}, and issue diagnosis \citep[e.g.,][]{luo2014correlating} are among the most studied topics in AIOps solutions. We divide AIOps papers by their goals into five sections and discuss them in more detail.

\noindent \textbf{Anomaly Detection.}
Anomaly detection is one of the most common tasks in AIOps \citep{bogatinovski2021artificial}. Previous studies perform anomaly detection using different approaches; for example, Fu et al.~\citeyearpar{fu2009execution} and Sharma et al.~\citeyearpar{sharma2013cloudpd} use clustering to find the anomalies, Liu et al.~\citeyearpar{liu2015opprentice} and Xu et al.~\citeyearpar{xu2009detecting} use tree-based models, and Brown et al.~\citeyearpar{brown2018recurrent} and Du et al.~\citeyearpar{du2017deeplog} use neural networks to perform the task. 
To perform anomaly detection, different data sources have been used. Even though event logs \citep{beschastnikh2014inferring, brown2018recurrent, du2017deeplog, fu2009execution} are the most common data source, other sources such as performance metrics \citep[e.g.,][]{sharma2013cloudpd, su2019robust}, network traffic \citep[e.g.,][]{lakhina2004diagnosing, lakhina2005mining}, and traces \citep[e.g.,][]{nedelkoski2019anomaly2, nedelkoski2019anomaly} have also been employed.

\noindent \textbf{Failure Prediction.}
Predicting failures is also one of the most common tasks of AIOps \citep{notaro2021survey}. Failure prediction can be divided into two groups of hardware failures and system failures. Compared to anomaly detection tasks where various data sources have been employed, in failure prediction, usually only the performance metrics of the systems are used. Performance metrics are measurements that aid in identifying and analyzing system bottlenecks and diagnosing issues. Some of the most popular performance metrics include CPU utilization, memory utilization, response time, throughput, I/O, and network latency.
Zhao et al.~\citeyearpar{zhao2021predicting} use different system metrics, such as CPU usage and the number of live threads of the system, to predict performance anomalies at run-time. Lin et al.~\citeyearpar{lin2018predicting} leverage temporal data (e.g., CPU and memory utilization metrics, alerts) and spatial data (e.g., rack locations) and construct MING, a deep learning-based approach. MING is able to rank the faulty nodes of the cloud system. MING is applied to the maintenance of one of the cloud service systems in Microsoft. Li et al.~\citeyearpar{li2020predicting} try to enhance the performance of MING, e.g., by enriching the data representing node failures through a novel oversampling approach. They also discuss some criteria for the successful adoption of AIOps solutions, including trust, interpretation, maintenance, scalability, and in-context evaluation.

\noindent \textbf{Root Cause Analysis.}
Root cause analysis is the approach of identifying the underlying causes of problems or issues in the system. The process involves identifying and analyzing the factors that contributed to the problem, determining the root cause, and developing solutions to address the underlying issue. 
Ding et al.~\citeyearpar{ding2021ttercl} leverage the generated alarms of a software system and find the root causes of the issues in an online manner. Wang et al.~\citeyearpar{wang2021groot} implement a graph-based algorithm and construct GROOT to perform root cause analysis. Their approach uses a combination of different event data (i.e., performance metrics, logs, and developer activities). GROOT is deployed in the production services of eBay. Zhang et al.~\citeyearpar{zhang2021cloudrca} propose a root cause analysis framework that also leverages different data sources: Key Performance Indicators (KPIs), logs, and topology data. Their primary model is a hierarchical Bayesian network that helps handle novel types of root causes.

\noindent \textbf{Incident Management.}
Incident management refers to the process of identifying, analyzing, and resolving incidents that occur within a software system. Although the majority of organizations have established unique methods for handling incidents, critical service incidents still happen unexpectedly, and the incident system fails to mitigate them. 
Chen et al.~\citeyearpar{chen2020towards} provide an overview of incident management. They analyze the incident management practices at Microsoft over two years and identify the distribution of incident severities for each section (e.g., network, database). Saha et Hoi.~\citeyearpar{saha2022mining} leverage past incident root cause analysis reports to manage the incidents. They present an incident causation analysis engine that extracts information from previous root cause analysis reports and use that knowledge for new incidents. They employ pre-trained NLP models over reports collected over a few years at Salesforce. Li et al.~\citeyearpar{li2022intelligent} propose an AIOps framework for incident detection. Their framework consists of four main parts: multi-aspect detection (where it is able to automatically identify the combinations of different data types such as logs and performance metrics), proactive detection (where it can proactively search for future hardware and software failures), incident refinement (where it can provide a global view of the incident and prioritize high-impact incidents), and incident enrichment (where it can locate the faulty scope).

\noindent \textbf{AIOps Literature Reviews.}
There exist two articles that survey existing AIOps studies. Notaro et al.~\citeyearpar{notaro2021systematic} conduct a systematic mapping study to identify past research in AIOps. They provide a taxonomy of AIOps papers in order to investigate the trends and also a comparison of AIOps papers for specific problems. Their findings demonstrate an on-growing research interest in the field of AIOps, particularly for downstream tasks such as anomaly detection and root cause analysis. According to their paper, the majority of studies (more than 60\%) are related to failure management (e.g., failure prediction and failure detection). Resource management and scheduling are two other popular areas in AIOps. To understand the trends in AIOps articles, similarly, Rijal et al.~\citeyearpar{rijal2022aiops} perform a literature review about AIOps works. Their findings support the growing interest in AIOps \citep{notaro2021systematic}. Their results count several benefits of AIOps: better monitoring of IT work, efficient time saving, improved human-AI collaborations, proactive IT work, and faster Mean Time To Resolve (MTTR). The development of AIOps solutions also faces a number of difficulties, including doubt about the efficiency of AI and ML, low-quality data, identifying the proper use cases, and traditional engineering approaches. Their paper concludes the need for further research to improve human and AIOps interaction in order to enhance human productivity. The results of our work can complement these literature reviews.

However, none of the existing work studies real-world AIOps projects. Different from these studies, in this work, we perform an empirical study to understand the practices and characteristics of real-world AIOps projects on GitHub.

% For example, Gao et al.~\citep{gao2020task} attempt to predict the task failures in cloud data centers. They use Google cluster trace \citep{reiss2012heterogeneity} and can achieve a 93\% accuracy in task failure prediction. Similarly, Islam et al.~\citep{islam2017predicting} characterize the failures of the Google cluster trace and then uses LSTM (Long Short-Term Memory) to predict the task failures. 
% There also exists studies focusing on hardware failures~\citep{pinheiro2007failure, schroeder2007understanding, mahdisoltani2017proactive, xu2018improving, botezatu2016predicting}, 
% %and~\citep{botezatu2016predicting}, 
% where they perform large-scale studies on disk failures. 
% Different from these studies, in this work, we perform an empirical study to understand the practices and characteristics of real-world AIOps projects on GitHub.

% \heng{do not use a citation number as a subject in a sentence, instead, use Gao \emph{et al.}~\citep{gao2020task} attempt to... Fix other places as well} 

% \heng{Briefly explain our difference from these studies.}

\subsection{Characterizing GitHub projects.}
As the biggest hosting service for open-source software, GitHub has been studied remarkably in the past years. Some studies focus on the best approaches for finding the most prominent repositories on GitHub~\citep{kalliamvakou2014promises, kalliamvakou2016depth, dabic2021sampling}. Many studies also have used GitHub as a source for mining software repositories~\citep{vadlamani2020studying, subramanian2020analyzing, horschig2018java, guzman2014sentiment, manes2021studying, lopes2017dejavu, kallis2021predicting, coelho2018identifying, businge2019studying, wessel2018power}. 
% As an example of the first group, \citep{kalliamvakou2014promises} performs an experimental study to categorize the most critical dangers of mining GitHub. It also provides suggestions for the ways researchers can find suitable projects. In their further research \citep{kalliamvakou2016depth}, they expand their findings by comparing the projects’ source code over time. \citep{dabic2021sampling} proposes GitHub search, a dataset containing quantitative metrics researchers might need. This tool is built to overcome GitHub limitations such as the API rate limit. It also provides some metrics that GitHub API does not directly provide, such as the number of commits.
For example, Vadlamani et  al.~\citep{vadlamani2020studying}, along with  Subramanian et al.~\citep{subramanian2020analyzing} and Horschig et al.~\citep{horschig2018java} try to characterize the developers of open-source software. As another example, Kallis et al.~\citep{kallis2021predicting} try to predict the issue types of projects.
More similar to our study, Ghrairi et al.~\citep{ghrairi2018state} study the state of Virtual Reality (VR) projects extracted from GitHub. In another work, Coppola et al.~\citep{coppola2019characterizing} characterize the popularity of Kotlin in Android projects by analyzing 1,232 applications on GitHub. 
However, as far as we know, no studies have been conducted yet on the subject of AIOps. Our work is the first study that uses GitHub to characterize and analyze AIOps projects.

% In another work, \citep{businge2019studying} studies the popularity of Android applications by analyzing the GitHub and Google Play Store metrics.

% In another work, \citep{wessel2018power} samples 351 popular projects on GitHub and measures the utilization of bots in these projects. 

% \heng{Briefly explain our work's relationship with these studies.}

\subsection{Analyzing code quality using SonarQube.}
SonarQube is one of the most popular static code analyses used both in academia~\citep{lenarduzzi2017analyzing, lenarduzzi2018survey} and industry~\citep{vassallo2020developers}. Previous research has conducted many analyses on the code quality of open-source software projects using SonarQube. Businge et al.~\citep{businge2019studying} use SonarQube and analyses the source code of 119 applications to find their number of bugs. In another work, Tan et al.~\citep{tan2018towards} study 9 Apache software systems written in Python to investigate their technical debt. Lenarduzzi et al.~\citep{lenarduzzi2019diffuseness} also investigate the technical debt of 33 Apache systems written in Java. They find that the amount of code smell is much higher than bugs or vulnerabilities in their set of projects. They also report that the code smells with the severity level \textsc{major} take the longest time to get fixed. 
Compared to the mentioned studies, our paper analyzes the quality of 6,101 projects, which is much higher than in the previous studies, making our results more confident and robust.

% \heng{Briefly explain our work's relationship with these studies.}

\section{Conclusion} 
\label{sec:conclusion}
%Let's conclude everything
% \heng{First briefly summarize the results, then focus on the implications and future work.}

This work studies the characteristics of AIOps projects on GitHub and conducts a comparative analysis with two baseline sets (ML and General projects). We combine both quantitative and qualitative analyses to understand the current state of AIOps solutions. Specifically, we illustrate the state of AIOps projects on GitHub in RQ1, observing that they are relatively new and have been growing rapidly in recent years. We determine the most common input data, techniques, and goals of AIOps solutions in RQ2. By uncovering these patterns,  researchers and practitioners can learn from successful approaches and adopt optimal AIOps solutions for their specific application scenarios. We then investigate the quality of AIOps projects in terms of code quality and compare them to the baselines in RQ3. We observe that the quality of AIOps projects is poorer compared to the baselines. 
Our findings highlight the need for future efforts to enhance AIOps practices. For example, addressing weak aspects such as self-healing and reducing the \textit{ad-hoc}ness of coding can lead to improved AIOps quality.

Our study could be extended in different ways. First, we study the AIOps projects publicly available on GitHub. However, there might be differences between open-source projects and closed-source projects which are implemented for private datasets in companies. Thus, one extension to our study could be to gather and analyze the industrial AIOps solutions and report any similarities and differences with AIOps projects on GitHub. Second, based on our results regarding the inputs, techniques, and goals of AIOps solutions, future work may develop a pipeline to automate or augment the development of AIOps solutions. Third, we discussed the potential of integrating existing quality assurance techniques from machine learning systems into AIOps solutions. Hence, future research could implement these techniques for AIOps solutions, examining their impact on enhancing AIOps code quality. Finally, conducting a future study that includes interviews with AIOps researchers and practitioners could serve to validate and extend the findings of our work.

\bigskip

\noindent \textbf{Code Availability} Our datasets, code and analysis are publicly available on \url{https://github.com/roozbehaghili/studying_aiops_github}.

\bigskip

\noindent \textbf{Conflict of Interest} The authors have no conflicts of interest to declare that are relevant to the content of this article.

\bibliography{bibliography}
\bibliographystyle{spbasic}

\end{document}